\documentclass[a4paper,11pt]{article}
\pdfoutput=1 
\usepackage{physics}
\usepackage[utf8]{inputenc}
\usepackage{jheppub} 
\usepackage[T1]{fontenc} 
\usepackage{slashed,float}
\usepackage{tikz}

\newcommand{\be}{\begin{equation}}
\newcommand{\ee}{\end{equation}}

\usepackage{subcaption}
\usepackage{hyperref}
\usepackage{amsmath}
\usepackage{amsfonts}
\usepackage{comment}
\usepackage{amssymb}
\usepackage{array}
\usepackage{epsfig}
\usepackage{slashed}
\usepackage{lipsum}
\usetikzlibrary{arrows.meta,decorations.text,positioning, calc,quotes,angles}

\usepackage{hyperref}
\hypersetup{
	colorlinks=true,
	linkcolor=blue,
	filecolor=red,
	urlcolor=blue,
	citecolor=red
} 

\title{Constraining momentum space CFT correlators with consistent position space OPE limit and the collider bound }

\author{Sachin Jain,}
\author{ Renjan Rajan John,}
\author{ Abhishek Mehta,}
\author{Dhruva K. S}

\affiliation{Indian Institute of Science Education and Research, Homi Bhabha Road, Pashan, Pune 411 008, India}
\emailAdd{sachin.jain@iiserpune.ac.in}
\emailAdd{renjan.john@acads.iiserpune.ac.in}
\emailAdd{abhishek.mehta@students.iiserpune.ac.in}
\emailAdd{k.s.dhruva@students.iiserpune.ac.in}

\abstract{Consistency with position space OPE limit requires momentum space CFT correlators to have only total energy singularity. We show that this requirement gives a  simple proof of the known result that the parity-odd structure cannot exist for  three-point correlators of  exactly conserved currents with spins $s_i,s_j,s_k$, when triangle inequality $s_i\le s_j+s_k$ is violated. We also show that even for parity even correlation functions the properties are different inside and outside the triangle inequality. It was previously shown that if we allow for weakly broken higher spin symmetry, parity-odd correlators can  exist even when triangle inequality is violated. In this paper we establish a relation between non-conservation Ward-Takahashi (WT) identities for weakly broken currents using known WT identities for exactly conserved currents. This allows us to calculate the parity violating results outside the triangle inequality using parity-even free bosonic and free fermionic results.

In general, there is one parity-odd structure and two parity-even structures for three-point functions. It can be shown that the coefficient of one of the parity-even and odd  parts can be combined into a complex parameter $c$ when correlators are expressed in spinor-helicity variables. When this complex parameter takes real value $c=\pm 1$ it corresponds to either the free boson or free fermion theory. When $c$ is a  pure phase, it corresponds to Chern-Simons matter theories.  Furthermore, re-expressing known results for conformal collider bounds we see that $|c| \le 1$
for generic 3d CFTs and $|c| \le f(\Delta_{gap})$ for holographic CFTs.}

\begin{document}
	
\maketitle

\raggedbottom

\section{Introduction}
Understanding correlation functions in  Conformal Field Theory (CFT) is important and is explored in position space mostly. In three dimensions, both parity-even and parity-odd structures are possible. Detailed studies were performed for exactly conserved currents in \cite{Giombi:2011rz,Costa:2011mg,Costa:2011dw,Maldacena:2011jn}. For weakly broken higher spin symmetry three-point functions were also constrained and calculated, see \cite{Maldacena:2012sf,Giombi:2016zwa}. In general, the parity-even structures are generated by the free fermionic and free bosonic theory, whereas the parity violating structure can be explicitly obtained in Chern-Simons matter theories \cite{Giombi:2011kc,Aharony:2011jz,Maldacena:2011jn,Maldacena:2012sf,Giombi:2016zwa,Gerasimenko:2021sxj,Skvortsov:2018uru}. For explicit results for correlation functions in Chern-Simons matter theories including their supersymmetric extensions, see \cite{Aharony:2012nh,GurAri:2012is,Bedhotiya:2015uga,Jain:2020puw,Aharony:2019mbc,Inbasekar:2019wdw,Jain:2020rmw,Aharony:2018pjn,Li:2019twz,Silva:2021ece,Kalloor:2019xjb}. See \cite{Vasiliev:1992av,Vasiliev:1995dn,Vasiliev:1999ba,Vasiliev:2003ev,Skvortsov:2018uru,Giombi:2010vg,Giombi:2016ejx} for discussion on derivation of CFT correlators from dual Vasiliev theory perspective.

Even though CFT is well studied in position space, only very recently has the momentum space aspects been explored \cite{Bzowski:2013sza, Coriano:2013jba,sissathesis,Bzowski:2015pba,Bzowski:2017poo,Bzowski:2018fql,Isono:2019ihz,Gillioz:2019lgs,Bzowski:2019kwd,Bzowski:2020kfw,Jain:2020rmw,Jain:2020puw,Jain:2021wyn,Jain:2021qcl,Coriano:2020ees,Mata:2012bx,Maldacena:2011nz,Gillioz:2021sce}. In three dimensions, momentum space analysis has already given a lot of insights into the structure of CFT correlation functions. The double copy structure \cite{Farrow:2018yni,Lipstein:2019mpu,Jain:2021qcl} of CFT correlation functions becomes transparent only in momentum space \footnote{In position space, it is very difficult to establish such relations. One can do a Fourier transform of the momentum space double copy relation which will result in a very ugly integral representation.  In Spinor Helicity variables it becomes even more clear.}. Momentum space analysis has also led to an interesting relation between parity-even and odd parts of correlation functions \cite{Caron-Huot:2021kjy,Jain:2021gwa,Gandhi:2021gwn}. In momentum space in light cone gauge in light cone coordinates in specific kinematic regions relations between the parity even and parity odd parts of a correlator can be observed from the results quoted in \cite{Aharony:2012nh,GurAri:2012is,Skvortsov:2018uru}. 
In particular, in \cite{Skvortsov:2018uru} from the Vasiliev theory perspective it was shown that the parity-breaking parameter $\theta$ results from certain EM duality.
Another attractive feature of momentum space correlation functions is their relation to flat space amplitudes and cosmological correlation functions \cite{Maldacena:2011nz,Mata:2012bx,Arkani-Hamed:2018kmz,Baumann:2019oyu,Baumann:2020dch,Kundu:2015xta,Kundu:2014gxa,Ghosh:2014kba,Pajer:2020wxk,Jazayeri:2021fvk,Bonifacio:2021azc}.

In this work we use consistency with position space OPE limit which implies that the only singularity structure that is allowed is the total energy singularity. More precisely, for a generic correlator of the form $\langle J_{s_1}(k_1)J_{s_2}(k_2)J_{s_3}(k_3)\rangle$, the only pole structure that is allowed is in $E=k_1+k_2+k_3$. Other pole structures, for example poles  at $k_{1}=k_{2}+k_3$ or in general poles at $k_i=k_j+k_k$ are not consistent with position space OPE limit. We use it to show that for exactly conserved currents, when triangle inequality $s_i\le s_j+s_k$ is violated, the parity-odd structure cannot exist. This was shown in position space in Appendix B of \cite{Giombi:2016zwa}.
However, in  \cite{Giombi:2016zwa} it was shown that for weakly broken higher spin symmetry, it is possible to have consistent parity-odd correlation functions. Interestingly, it turns out that, as will be discussed in this paper the Ward-Takahashi (WT) identity induced by non-conservation of higher spin currents can be derived from WT identity of exactly conserved current by what we call an epsilon transform in the paper. This gives us a general prescription to calculate the parity-odd structure outside the triangle inequality. We also discuss that in spinor helicity variables all the known conformal collider bounds that were derived for generic 3d CFT and holographic CFT have a very natural and simple interpretation. We discuss the correspondence with flat space amplitudes and the supersymmetric extension briefly.

This paper is organised as follows. In Section \ref{CCSHV} we introduce the notation and convention we use in the paper and review some known results in the literature. We also discuss the requirement imposed by the consistency with the position space OPE limit. In Section \ref{SectionECC} we discuss the structure of three-point correlation functions of exactly conserved currents. 
In Section \ref{noodd} we show that the there is no parity-odd contribution outside the triangle inequality for exactly conserved currents when all the spins are non-zero. In Section \ref{SectionWBC} we extend our analysis to the structure of three-point functions of weakly broken conserved currents. In Section \ref{susy} we comment on the structure of correlation functions in supersymmetric theories. In Section \ref{ccb} we illustrate how the conformal collider bound takes a natural form in our way of expressing correlation functions. We end with a summary and discussion in Section \ref{summary}. We collect a few technical details in the Appendices. In Appendix \ref{theory1} we discuss the list of theories that we study in this paper. In Appendix \ref{WT-identity} we give a derivation of the Ward-Takahashi (WT) identity for a general symmetry transformation. In Appendix \ref{countingapp} we discuss the WT identities that correlators in the fermionic and bosonic theories satisfy when the spins of the operator insertions satisfy and as well as violate triangle inequality. In Appendix \ref{2pt1} we show that conformal invariance and the WT identity together implies a relation between 2-point function coefficients which hint towards higher spin symmetry. In Appendix \ref{hotngap} we give details of the computation encountered in Section \ref{hotng}, related to correlators that lie outside the triangle inequality. In Appendix \ref{cfba} we give details of the conformal bound computation of Section \ref{ccb}. 
\section{Notation, conventions and review of some known results}
\label{CCSHV}
We study three-point functions of symmetric traceless conserved currents. The ansatz for a general correlator in spinor-helicity variables is given by
\begin{align}\label{generalansatz}
\langle J_{s_1}^{h_1}(k_1)J_{s_2}^{h_2}(k_2)J_{s_3}^{h_3}(k_3) \rangle&=(c_1\,F^{h_1h_2h_3}_1(k_1, k_2, k_3) +i\,c_2\,F^{h_1h_2h_3}_2(k_1, k_2, k_3) )\nonumber\\[5pt]
&\hspace{1cm}\langle 12 \rangle^{h_3-h_1-h_2} \langle 23 \rangle^{h_1-h_2-h_3} \langle 31 \rangle^{h_2-h_3-h_1}
\end{align}
where $F_1(k_1, k_2, k_3)$ and $F_2(k_1, k_2, k_3)$ are form-factors that we will determine by imposing dilatation and special conformal invariance. The form factor $F_2(k_1,k_2,k_3)$ corresponds to the parity-odd contribution to the correlator. 

In our analysis we will split the three-point function into homogeneous (h) and non-homogeneous (nh) pieces 
\begin{align}
\langle J_{s_1} J_{s_2}J_{s_3}\rangle=\langle J_{s_1} J_{s_2}J_{s_3}\rangle_{\text{h}}+\langle J_{s_1} J_{s_2}J_{s_3}\rangle_{\text{nh}}
\end{align}
where $\langle J_{s_1} J_{s_2}J_{s_3}\rangle_{\text{h}}$ solves the conformal Ward identity 
\begin{align}\label{Kkappah}
\widetilde K^\kappa\left\langle\frac{J_{s_1}}{k_1^{s_{1}-1}}\frac{J_{s_2}}{k_2^{s_{2}-1}}\frac{J_{s_3}}{k_3^{s_{3}-1}}\right\rangle_{\text{h}}=0
\end{align}
and $\langle J_{s_1} J_{s_2}J_{s_3}\rangle_{\text{nh}}$ is a solution of %
\begin{align}\label{nhpiece}
\widetilde{K}^{\kappa} \left\langle\frac{J_{s_1}}{k_1^{s_{1}-1}}\frac{J_{s_2}}{k_2^{s_{2}-1}}\frac{J_{s_3}}{k_3^{s_{3}-1}}\right\rangle_{\text{nh}} &= 2\left(z_1^{-\kappa}\frac{k_{1\mu}}{k_1^2}\langle J^{\mu} J^- J^- \rangle+z_2^{-\kappa}\frac{k_{2\mu}}{k_2^2}\langle J^- J^{\mu} J^- \rangle+z_3^{-\kappa}\frac{k_{3\mu}}{k_3^2}\langle J^- J^- J^{\mu} \rangle\right)
\end{align}
where $\widetilde K^\kappa$ is the generator of special conformal transformations and spin indices have been suppressed for brevity. For more details on the distinction between homogeneous and non-homogeneous contributions, see \cite{Jain:2021vrv}. 
 One can further separate the homogeneous and non-homogeneous pieces into parity-even and parity-odd parts as follows
 \begin{align}\label{hnhoe}
  \langle J_{s_1} J_{s_2}J_{s_3}\rangle_{\text{h}}& = \langle J_{s_1} J_{s_2}J_{s_3}\rangle_{\text{h,even}}+ \langle J_{s_1} J_{s_2}J_{s_3}\rangle_{\text{h,odd}}\nonumber\\   
 \langle J_{s_1} J_{s_2}J_{s_3}\rangle_{\text{nh}}& = \langle J_{s_1} J_{s_2}J_{s_3}\rangle_{\text{nh,even}}+ \langle J_{s_1} J_{s_2}J_{s_3}\rangle_{\text{nh,odd}}.
 \end{align}
 
 \subsection{Consistency with position space OPE limit}
 In this section we review the important fact that consistency with position space OPE imposes the restriction that the only singularity structure that can exist in momentum space correlators is in the  total energy $E=k_1+k_2+k_3$ \cite{Maldacena:2011nz,Mata:2012bx}. To illustrate this we consider the particularly simple example of the three-point function of the  scalar operator $O$ with dimension $\Delta=2$. Here we reproduce the argument presented in \cite{Maldacena:2011nz}.  One can show that there exist four solutions to the conformal Ward identity. This gives the correlator to be of the form 
 \begin{align}
     \langle OOO\rangle =a_1 f(k_1+k_2+k_3)+a_2 f(-k_1+k_2+k_3)+a_3 f(k_1-k_2+k_3)+a_4 f(k_1+k_2-k_3)
 \end{align}
 with $f(k_1+k_2+k_3)=\ln \left(k_1+k_2+k_3\right)$. Permutation symmetry implies $a_2=a_3=a_4$. In the limit $x_{23}\rightarrow 0,$ the OPE expansion in position space implies $\langle OOO\rangle \sim \frac{1}{x_{23}^2}\frac{1}{x_{12}^4}$. This in momentum space gives
$\langle OOO\rangle \sim\frac{k_1}{k_3}$ as ${\vec{k}}_1\rightarrow 0$ and $k_2\approx k_3 \gg k_1$.  It can be checked that only $\ln\left(k_1+k_2+k_3\right)$ reproduces the correct OPE limit. We conclude that a singularity of the form $f(k_i-k_j+k_k)$ is not allowed and that the only singularity structure that is allowed is in $E=k_1+k_2+k_3\rightarrow 0$.

 \subsection{Triangle inequality}
Throughout this paper we will distinguish between correlators that satisfy the triangle inequality and those that do not. The inequality is given by
 \begin{equation}\label{tangle}
     s_i \le s_j + s_k,\quad\quad\forall i,j,k\in\{1,2,3\}
 \end{equation} 
 We now classify correlators based on whether the spins of the operator insertions  satisfy or violate  \eqref{tangle}.
 
 \subsection*{Inside the triangle inequality}
Examples of correlation functions that satisfy triangle inequality are
 \begin{align}\label{IT}
     &\langle J_{s_1} J_{s_2}J_{s}\rangle~~~~ \rm{where} ~~ |s_1-s_2| \le  s \le s_1 +s_2\notag\\
    & \langle J_{s} J_{s}O_{\Delta}\rangle,~~\langle O_{\Delta_1}O_{\Delta_2}O_{\Delta_3}\rangle\,.
 \end{align}

 \subsection*{Outside the triangle inequality}
Examples of correlation functions that violate triangle inequality are
 \begin{align}\label{OT}
    & \langle J_{s_1} J_{s_2}J_{s}\rangle~~~~ \rm{where}~~    s > s_1 +s_2~~ \rm{or}~~ s < |s_1-s_2|\notag\\
    &\langle J_{s_1} J_{s_2}O_{\Delta}\rangle ~~~~ \rm{where}~~    s_1 \ne s_2 \notag\\
  & \langle J_{s_1} O_{\Delta_1} O_{\Delta_2}\rangle ~~~~ \rm{where}~~    s_1 \ne 0\,.
 \end{align}
 \subsection{Summary of the known results in position space}
 In this section we summarize the results for three-point functions involving  exactly conserved and weakly broken higher spin currents. A detailed position space analysis of these correlators in three dimensions was performed in \cite{Giombi:2011rz}, see also \cite{Maldacena:2011jn,Maldacena:2012sf,Costa:2011mg,Costa:2011dw}. It was shown that for conserved currents $J_{s_i}$, correlation functions of the form  $\langle J_{s_i}O_{\Delta}O_\Delta\rangle$ have one parity-even structure. Correlators of the form $\langle J_{s_1}J_{s_2}O_\Delta\rangle$ with $s_1=s_2\neq0$ have one parity-even and one parity-odd structure. When $s_1\neq s_2\neq0$, these correlators are non-zero only for  $\Delta=1$ and $\Delta=2$ and they have one parity-even and one parity-odd structure. When the scalar operator has dimension $\Delta=1,2,$ the parity-even and the parity-odd correlators can be obtained suitably using free bosonic, critical bosonic, free fermionic or critical fermionic theories. It was also shown in \cite{Giombi:2011rz},\cite{Maldacena:2012sf,Giombi:2016zwa} that 
correlators of the form $\langle J_{s_1}J_{s_2}J_{s_3}\rangle$ where $s_i\ge 1$ have two parity-even and one parity-odd structure when the spins satisfy triangle inequality \eqref{tangle}. When the sum of spins is even, parity-even contributions arise from the free boson and the free fermion theories. When the sum of spins is odd, parity-even contributions arise from correlators of non-abelian currents in the free theory of multiple scalars or fermions. 
When the spins violate triangle inequality, it was shown that the parity-odd structure vanishes leaving only two parity-even structures \cite{Giombi:2011rz,Maldacena:2011jn}. For a proof see Appendix B of \cite{Giombi:2016zwa}. 

In \cite{Giombi:2016zwa}, it was shown  that parity-odd structures can arise even when triangle inequality is violated for theories with weakly broken higher spin symmetry. In this paper we shall show that these parity-odd structures can be obtained from the free theory parity-even results.

\begin{center}
\begin{table}[h!]
    \begin{tabular}{ | l | l | l |}
    \hline
   Correlators  & Parity-even & Parity-odd \\ \hline
    $\langle J_sO_{\Delta}O_{\Delta}\rangle$ & 1 & 0  \\ \hline
    $\langle J_{s_1}J_{s_2}O_{\Delta}\rangle$ $s_1=s_2\ne 0$ & 1 & 1\\
    \hline
    $\langle J_{s_1}J_{s_2}O_{\Delta}\rangle$ $s_1\ne s_2\ne 0$ and $\Delta=1,2$ & 1 & 1\\ \hline
    $\langle J_{s_1}J_{s_2}O_{\Delta}\rangle$ $s_1\ne s_2\ne 0$ and $\Delta\notin \{1,2\}$ & 0 & 0\\ \hline
    $\langle J_{s_1}J_{s_2}J_{s_3}\rangle$ $s_i\le s_j+s_k$ for $i,j,k\in\{1,2,3\}$  & 2 & 1\\ \hline
     $\langle J_{s_1}J_{s_2}J_{s_3}\rangle$ $s_i> s_j+s_k$ for any $i,j,k\in\{1,2,3\}$, $J_{s_i}$ are exactly conserved  & 2 & 0\\ \hline
          $\langle J_{s_1}J_{s_2}J_{s_3}\rangle$ $s_i> s_j+s_k$ for any $i,j,k\in\{1,2,3\}$, $J_{s_i}$ are weakly broken  & 2 & 1\\ \hline
    \end{tabular}
     \caption{\label{tab:table-name}In this table we summarise the number of parity-even and parity-odd structures for various 3-point correlators. We refer the reader to \cite{Giombi:2011rz,Maldacena:2011jn,Giombi:2016zwa} for details.}
    \end{table}
\end{center}

 \subsection{Contact and Semilocal terms}\label{smlcon}
 In this section we will briefly discuss contact terms and semi-local contributions to correlation functions. Contributions to three-point functions of the following form 
 \begin{align}
 \langle J_{s_1} J_{s_2}J_{s_3}\rangle\propto \delta^3(x_1-x_2)  \delta^3(x_2-x_3) 
 \end{align}
are called contact terms. In momentum space they correspond to even polynomials of absolute value of momentum $f(p_1^2,p_2^2,p_3^2)$. For example, let us consider the following
\begin{align}\label{mcont}
  X(p_1,p_2,p_3)=(z_1\cdot z_2)^a (z_1\cdot z_3)^b (z_2\cdot z_3)^c (z_1\cdot p_2)^d (z_2\cdot p_3)^e (z_3\cdot p_1)^g f(p_1^2,p_2^2,p_3^2)
\end{align}
where $z_i$ are transverse polarization tensors.
When converted to position space this takes the form
\begin{align}\label{cntct1}
 X(x_1,x_2,x_3)&\propto (z_1\cdot z_2)^a (z_1\cdot z_3)^b (z_2\cdot z_3)^c (z_1\cdot \partial_{x_2})^d (z_2\cdot \partial_{x_3})^e (z_3\cdot \partial_{x_1})^g\cr
 &\hspace{1.5cm}f(\Box_{x_1},\Box_{x_2},\Box_{x_3}) \delta^{3}(x_1-x_2)\delta^{3}(x_2-x_3)
\end{align}
three-point functions of the form 
 \begin{align}
 \langle J_{s_1} J_{s_2}J_{s_3}\rangle\propto \delta^3(x_i-x_j ) 
 \end{align} with $i\ne j$, where $i$ and $j$ can take values $1,2$ and $3$ are called semilocal terms.
For example, let us consider 
\begin{align}
 Y(p_1,p_2,p_3)=(z_1\cdot z_2)^a (z_1\cdot z_3)^b (z_2\cdot z_3)^c (z_1\cdot p_2)^d (z_2\cdot p_3)^e (z_3\cdot p_1)^g f(p_1^2,p_2^2,p_3^2)~ p_3
\end{align}
which when converted to position space takes the form
\begin{align}
 Y(x_1,x_2,x_3)&\propto (z_1\cdot z_2)^a (z_1\cdot z_3)^b (z_2\cdot z_3)^c (z_1\cdot\partial_{x_2})^d (z_2\cdot\partial_{x_3})^e (z_3\cdot\partial_{x_1})^g\cr
 &\hspace{1.5cm}f(\Box_{x_1},\Box_{x_2},\Box_{x_3})\frac{1}{|x_1-x_3|^4} \delta^{3}(x_1-x_2)\,.
\end{align}

\subsection{Parity-odd two-point function is a contact term}
In this section we will discuss the parity-odd contribution to the two-point function of the spin-$s$ operator $J_s$. This term is a contact term and to illustrate it we consider the simplest example of spin-one current operator.
The two-point function is given by
\begin{align}
    \langle J_{\mu}(k_1) J_{\nu}(-k_1) \rangle = \left( \delta_{\mu\nu}-\frac{k_{1\mu} k_{1\nu}}{k_1^2}\right)k_1 + \epsilon_{\mu\nu\rho}k_1^{\rho}.
\end{align}
In position space the parity-odd contribution takes the following form
\begin{align}
  \langle J_{\mu}(x_1) J_{\nu}(x_2) \rangle  = \epsilon_{\mu\nu\rho}\partial_{x_{1\rho}} \delta^3(x_1-x_2).
\end{align}
One can show in general that
\begin{align}\label{podcon}
    \langle J_{s}(x_1) J_{s}(x_2) \rangle  \propto f(x_1,\partial_{x_{1}} )\delta^3(x_1-x_2) 
\end{align}
where we have not written down the explicit form of $f$. 

\subsection{Parity odd non-homogeneous three-point function is a contact term}
In \cite{Jain:2021vrv}, it was shown that the parity-odd non-homogeneous part of a CFT correlation function is a contact term. The basic argument is that the WT identity for these contributions is purely a contact term, which implies that the parity-odd non-homogeneous term is always a contact or at best a semilocal term. For correlators such as $\langle TTT \rangle$ and $\langle JJJ \rangle$ the non-homogeneous parity-odd term was indeed shown to be a contact term in \cite{Jain:2021vrv}.

\section{Correlation function for exactly conserved currents}
\label{SectionECC}
In this section we discuss properties of correlations functions for exactly conserved currents. First we discuss correlators inside the triangle inequality where both parity-even and parity-odd structures are allowed. We then discuss correlators outside the triangle inequality and finally we discuss certain correlators with no parity-odd structure.
\subsection{Inside the triangle inequality}
In this section, we analyse correlation functions  of the form given in \eqref{IT}. The simplest of these, namely $\langle O_{\Delta_1}O_{\Delta_2}O_{\Delta_3}\rangle$ has only one structure. 
We analyse the other  cases in detail in the following sub-sections. 

\subsubsection{$\langle J_{s} J_{s}O_{\Delta}\rangle$}
Let us consider correlation functions of the form $\langle J_{s} J_{s}O_{\Delta}\rangle$. They can always be defined in such a way that they are completely transverse i.e. they do not have a local piece \cite{Bzowski:2018fql,Jain:2021vrv}. 
The action of the generator of special conformal transformations  in spinor-helicity variables then takes the form \eqref{nhpiece}
\begin{align}\label{Kkappah}
\widetilde K^\kappa \langle J_{s} J_{s}O_{\Delta}\rangle=0
\end{align}
which implies that the non-homogeneous part of the correlator is zero
\begin{equation}
    \langle J_{s} J_{s}O_{\Delta}\rangle_{\text{nh}} =0.
\end{equation}
Thus the correlator is entirely given by its homogeneous part 
\begin{align}
\langle J_{s} J_{s}O_{\Delta}\rangle = \langle J_{s} J_{s}O_{\Delta}\rangle_{\text{h}}
\end{align}
One can show by explicit computation that for the free fermion and free boson theories this is indeed the case, up to a semilocal term \cite{Bzowski:2018fql,Jain:2021wyn}. 
\subsection*{Homogeneous contribution: parity-even and parity-odd}
Now that we have shown that the correlator gets a contribution only from the homogeneous part, let us separate it into its parity-even and parity-odd parts
 \begin{equation}
     \langle J_{s} J_{s}O_{\Delta}\rangle_{\text{h}} =\langle J_{s} J_{s}O_{\Delta}\rangle_{\text{h},\text{even}}+ \langle J_{s} J_{s}O_{\Delta}\rangle_{\text{h},\text{odd}}
 \end{equation}
Plugging the ansatz for the correlator using \eqref{generalansatz} into  \eqref{Kkappah}, it was shown in \cite{Jain:2021vrv} that the equations that $F_{\text{even},\text{h}}$ and $F_{\text{odd},\text{h}}$ satisfy are identical and we have
 \begin{equation}
   F_{\text{even},\text{h}}(k_1,k_2,k_3) \propto F_{\text{odd},\text{h}}(k_1,k_2,k_3).
 \end{equation}
 In fact it can be shown that 
  the correlator is given by \cite{Jain:2021vrv,Gandhi:2021gwn} 
 \begin{align}
    & \langle J_{s}^{-} J_{s}^{-}O_{\Delta}\rangle_{\text{h}} = \left(c_1+ i c_2\right)F_h(k_1,k_2,k_3)\nonumber\\
    & \langle J_{s}^{+} J_{s}^{+}O_{\Delta}\rangle_{\text{h}} = \left(c_1- i c_2\right)F_h(k_1,k_2,k_3)\nonumber\\
    & \langle J_{s}^{-} J_{s}^{+}O_{\Delta}\rangle_{\text{h}} = 0.
 \end{align}
In spinor-helicity variables, the only non-zero components are those with ${--}$ and ${++}$ helicities. For explicit expressions we refer the reader to \cite{Jain:2021vrv}. 

We notice an interesting fact that in spinor-helicity variables there is a complex number which behaves like an OPE coefficient. However, when we convert the result back into momentum space, the term proportional to $c_1$ contributes to the parity-even part and the term proportional to $c_2$ contributes to the parity-odd part of the correlation function and they look completely different as is the case in position space. For an explicit expression in momentum space see \cite{Bzowski:2013sza,Bzowski:2018fql,Jain:2021wyn,Jain:2021vrv}. We conclude that the homogeneous piece contributes to two structures, one parity-even and one parity-odd. 
As an example let us consider $\langle T T O\rangle$.  When the scalar operator $O$ has a dimension $\Delta$ then the correlator is given by the following \cite{Jain:2021vrv}
\begin{align}
\label{ttodeltasph}
\begin{split}
\langle T^- T^+ O_{\Delta} \rangle &= 0\\
\langle T^-T^- O_{\Delta} \rangle&=\langle T^-T^- O_{\Delta} \rangle_{\text{even}}+\langle T^-T^- O_{\Delta} \rangle_{\text{odd}}=\left(c_1+ i c_2\right)\,k_1 k_2 I_{\frac{9}{2}\{\frac{1}{2},\frac{1}{2},\Delta-\frac{3}{2}\}} \langle 12 \rangle^4\\[5 pt]
\langle T^+T^+ O_{\Delta} \rangle&=\langle T^+ T^+ O_{\Delta} \rangle_{\text{even}}+\langle T^+ T^+ O_{\Delta} \rangle_{\text{odd}}=\left(c_1 - i c_2\right)\,k_1 k_2 I_{\frac{9}{2}\{\frac{1}{2},\frac{1}{2},\Delta-\frac{3}{2}\}} \langle \bar 1\bar 2 \rangle^4\\[5 pt]
\end{split}
\end{align}
When $\Delta=1$ this takes the form 
\begin{align}
\begin{split}
&\langle T^-T^-O_1 \rangle_{\text{even}} = c_1 k_1 k_2\frac{1}{k_3(k_1+k_2+k_3)^4}\langle 12 \rangle^4\\[5 pt]
&\langle T^-T^-O_1\rangle_{\text{odd}} = ic_2 k_1k _2\frac{1}{k_3(k_1+k_2+k_3)^4}\langle 12 \rangle^4
\end{split}
\begin{split}
&\langle T^-T^+O_1 \rangle_{\text{even}} =0\\[5 pt]
&\langle T^-T^+O_1 \rangle_{\text{odd}} =0
\end{split}
\end{align}
We emphasize that the only pole that exists is when $E=p_1+p_2+p_3 \rightarrow 0$ and in this limit we obtain the flat space scattering amplitude as the residue of the pole.

\subsubsection{$\langle J_{s_1} J_{s_2} J_{s_3}\rangle$}
Correlation functions of the form $\langle J_{s_1} J_{s_2} J_{s_3}\rangle$ have both homogeneous and non-homogeneous contributions \cite{Jain:2021vrv}. In the following we first analyse the homogeneous part and then the non-homogeneous piece.

\subsection*{Homogeneous contribution: parity-even and parity-odd}
 The homogeneous piece has parity even and parity odd contributions 
 \begin{equation}
     \langle J_{s_1} J_{s_2}J_{s_3}\rangle_{\text{h}} =\langle J_{s_1} J_{s_2}J_{s_3}\rangle_{\text{h},\text{even}}+ \langle J_{s_1} J_{s_2}J_{s_3}\rangle_{\text{h},\text{odd}}
 \end{equation}
 In spinor-helicity variables we can write
 \begin{equation}
     \langle J_{s_1} J_{s_2}J_{s_3}\rangle_{\text{h}} = \left(c_1 F_{\text{h,even}}(k_1,k_2,k_3)+ i c_2 F_{\text{h,odd}}(k_1,k_2,k_3)\right)\langle 12 \rangle^{h_3-h_1-h_2} \langle 23 \rangle^{h_1-h_2-h_3} \langle 31 \rangle^{h_2-h_3-h_1}
 \end{equation}
 It was shown in \cite{Jain:2021vrv} that 
 \begin{equation}
     F_{\text{h,even},\text{h}}(k_1,k_2,k_3) =F_{\text{h,odd},\text{h}}(k_1,k_2,k_3)=F_{\text{h}}(k_1,k_2,k_3)= \frac{1}{E^{s_1+s_2+s_3}} 
 \end{equation}
 This implies that the homogeneous piece is given by
 \begin{equation}
     \langle J_{s_1} J_{s_2}J_{s_3}\rangle_{\text{h}} = \left(c_1+ i c_2\right)F_{\text{h}}(k_1,k_2,k_3)\langle 12 \rangle^{h_3-h_1-h_2} \langle 23 \rangle^{h_1-h_2-h_3} \langle 31 \rangle^{h_2-h_3-h_1}
 \end{equation}
 
\subsection*{Non-homogeneous contribution: parity-even and parity-odd} 
Since the conformal Ward identity takes the following form 
\begin{align}\label{nhpiece1}
\widetilde K^\kappa\left\langle\frac{J_{s_1}}{k_1^{s_{1}-1}}\frac{J_{s_2}}{k_2^{s_{2}-1}}\frac{J_{s_3}}{k_3^{s_{3}-1}}\right\rangle_{\text{nh}}=\text{transverse Ward identity terms}.
\end{align} 
we need to carefully analyse the WT identity to calculate the non-homogeneous contributions to the correlator. Before dealing with the most general case, let us consider the example of the three-point function of stress tensor.

\subsection*{$\langle T(k_1) T(k_2) T(k_3) \rangle$}
 The transverse Ward identity that the correlator obeys is given by \cite{Baumann:2020dch}
\begin{align}\label{wtidttt1}
\begin{split}
z_{1\mu}k_{1\nu}&\langle T^{\mu\nu}(k_1) T(k_2) T(k_3) \rangle \\
&= -(z_1 \cdot k_2) \langle T(k_1 + k_2) T(k_3) \rangle +2(z_1 \cdot z_2)k_{2\mu}z_{\nu}\langle T^{\mu\nu}(k_1+k_2) T(k_3) \rangle\\[5 pt]
&\hspace{0.25cm}-(z_1 \cdot k_3)\langle  T(k_1+k_3) T(k_2) \rangle + 2(z_1 \cdot z_3)k_{3\mu}z_{3\nu}\langle T^{\mu\nu}(k_1+k_3) T(k_2) \rangle\\[5 pt]
&\hspace{0.25cm}+(k_1 \cdot z_2)z_{1\mu}z_{2\nu}\langle T^{\mu\nu}(k_1+k_2) T(k_3) \rangle+(z_1 \cdot z_2)k_{1\mu}z_{2\nu}\langle T^{\mu\nu}(k_1+k_2) T(k_3) \rangle\\[5 pt]
&\hspace{0.25cm}+(k_1 \cdot z_3)z_{1\mu}z_{3\nu}\langle T^{\mu\nu}(k_1+k_3) T(k_2) \rangle+(z_1 \cdot z_3)k_{1\mu}z_{3\nu}\langle T^{\mu\nu}(k_1+k_3) T(k_2) \rangle
\end{split}
\end{align}
If we take the parity-even part of $\langle T T\rangle$, this gives rise to the parity-even part of the non-homogeneous solution, see Section 4.3.3 of \cite{Baumann:2020dch}. 

Now let us turn our attention to the parity-odd part. As was discussed in \cite{Jain:2021vrv}, if we take the parity-odd part of $\langle T T\rangle$, the non-homogeneous pieces is a contact term and is given by
 \begin{align}\label{tttnewbasis}
&\langle TTT \rangle_{\text{odd},\text{nh}} \nonumber\\
&= \bigg[ \epsilon^{k_1 z_1 z_2} (z_1 \cdot z_2) (k_1 \cdot z_3)^2 - (k_1 \leftrightarrow k_2)\epsilon^{k_2 z_1 z_2} (z_1 \cdot z_2) (k_1 \cdot z_3)^2\bigg. \nonumber\\[5 pt]
&\bigg.+\left(2 k_1^2 +\frac{7}{2}k_2^2+\frac{7}{2}k_3^2\right) \epsilon^{k_1 z_1 z_2} (z_1 \cdot z_3) (z_2 \cdot z_3)-\left(2 k_2^2 +\frac{7}{2}k_1^2+\frac{7}{2}k_3^2\right)\epsilon^{k_2 z_1 z_2} (z_1 \cdot z_3) (z_2 \cdot z_3)\bigg]  \nonumber\\&+ \text{cyclic permutations}
\end{align}
The RHS of \eqref{tttnewbasis} is of exactly the same form as given in \eqref{mcont} and when converted to position space is  a contact term \eqref{cntct1}. The same conclusion holds  for other three-point functions involving spinning operators with  $0<s_i \le 2$ \cite{Jain:2021vrv}. Thus for correlators of the form $\langle J_{s_1}J_{s_2}J_{s_3}\rangle$ where $0<s_i \le 2$, the only contribution to the non-homogeneous piece comes from the parity-even part. 
\subsection*{$\langle J_{s_1}J_{s_2}J_{s_3} \rangle$}
Let us now analyse the case involving general spins $s_1,s_2$ and $s_3$. We will show that non-homogeneous piece has only one parity-even contribution, and that the  parity-odd contribution is always a contact term. To understand this, we need to write down the WT identity for the correlator. Solving the conformal Ward identity involving general spins is quite complicated. However, for our purposes we do not need to solve the Ward identity. 

The generic structure of WT identity in position space is given by
\begin{align}\label{WThgnspin}
 \partial_{x_{1\mu_1}} \langle J^{\mu_1\mu_2\cdots \mu_{s_1}}_{s_1}(x_1) J_{s_2}(x_2)J_{s_3}(x_3)\rangle \propto & f_1(\partial_{x_i}) \langle J_{s_3}(x_2)J_{s_3}(x_3)\rangle \delta^3(x_1-x_2)\nonumber\\
 &\hspace{.2cm}+ f_2(\partial_{x_i}) \langle J_{s_2}(x_2)J_{s_2}(x_3)\rangle \delta^3(x_1-x_3)
\end{align}
where the functions $f_1$ and $f_2$ need to be derived for each correlator. In general they can be quite complicated. 
\subsection*{Parity odd contribution}
Let us first analyse the parity-odd term. 
The RHS of \eqref{WThgnspin} using \eqref{podcon} gives 
\begin{align}\label{WThgnspinod}
 \partial_{x_{1\mu_1}} \langle J^{\mu_1\mu_2\cdots \mu_{s_1}}_{s_1}(x_1) J_{s_2}(x_2)J_{s_3}(x_3)\rangle \propto  f(\partial_{x_i})  \delta^3(x_1-x_3)  \delta^3(x_1-x_2)
\end{align}
which is a contact term. One can also check that the corresponding contribution to the three-point function is a contact term or at best a semilocal term \cite{Jain:2021vrv}. This explains the claim that the parity-odd non-homogeneous contribution to the correlator is a contact term.

\subsection*{Parity even contribution}
Let us now analyse the parity-even contribution. For this purpose let us first write down the conformal Ward identity \ref{nhpiece} in spinor-helicity variables
\begin{align}
\sum_{a=1}^{3}\widetilde{K}^i_a\frac{\langle J^{h_1}_{s_1}J^{h_2}_{s_2}J^{h_3}_{s_3}\rangle}{k^{s_1-1}_1k^{s_2-1}_2k^{s_3-1}_3} &\sim \rm{WT~identity}\nonumber\\
& \sim c_{s_1} f_1(k_1,k_2,k_3)\langle J_{s_1} J_{s_1}\rangle +c_{s_2} f_2(k_1,k_2,k_3) \langle J_{s_2} J_{s_2}\rangle +c_{s_3} f_3(k_1,k_2,k_3)\langle J_{s_3} J_{s_3}\rangle. 
\end{align}
The right hand side of the above equation has three independent two-point function coefficients  $c_{s_1},c_{s_2},c_{s_3}$. One would then conclude that there are three independent structures that are allowed which is in contradiction with \cite{Maldacena:2011jn,Maldacena:2012sf,Giombi:2011rz}. The apparent contradiction is resolved by the fact that if we also demand exact conservation of higher-spin currents, then they satisfy higher-spin algebra. Using higher-spin algebra it is easy to show  that $c_{s_i} = c_{s_j} =c_2$. In other words all two-point function coefficients are related to each other. We refer the reader to Appendix \ref{2pt1} for an alternate understanding of this fact using conformal Ward identity. This immediately implies that there is only one parity-even non-homogeneous contribution to the correlation function.

\subsubsection{Summary of the section}
We summarise our findings from this section in the following tables. 

\begin{table}[h!]
\centering
    \begin{tabular}{ | l | l | l |}
    \hline
    $\langle J_sJ_sO_{\Delta}\rangle$ & Parity-even & Parity-odd \\ \hline
    Homogeneous & 1 & 1  \\ \hline
    Non-homogeneous & 0 & 0\\
    \hline
    \end{tabular}
     \caption{\label{tab:table-name}In this table we summarise the number of parity-even and parity-odd contributions to homogeneous and non-homogeneous terms to correlators of the form $\langle J_sJ_sO_{\Delta}\rangle$.}
    \end{table}

\begin{table}[h!]
\centering
    \begin{tabular}{ | l | l | l |}
    \hline
    $\langle J_{s_1}J_{s_2}J_{s_3}\rangle$ & Parity-even & Parity-odd \\ \hline
    Homogeneous & 1 & 1  \\ \hline
    Non-homogeneous & 1 & 0~(contact term)\\
    \hline
    \end{tabular}
    \caption{\label{tab:table-name}In this table we summarise the number of parity-even and parity-odd contributions to homogeneous and non-homogeneous terms to correlators of the form $\langle J_{s_1}J_{s_2}J_{s_3}\rangle$ when the spins satisfy triangle inequality.}
    \end{table}
%
 We see that this is perfectly consistent with the structure of three-point functions in position space as studied in \cite{Maldacena:2011jn,Maldacena:2012sf,Giombi:2011rz}. Since the parity-even structures can be obtained from free bosonic and fermionic theories
 we conclude that for $s_1\ne 0,s_2\ne 0,s_3\ne 0$ we have 
 \begin{equation}\label{fff1}
   \langle J_{s_1} J_{s_2}J_{s_3}\rangle = n_s  \langle J_{s_1} J_{s_2}J_{s_3}\rangle_{FF}+ n_f  \langle J_{s_1} J_{s_2}J_{s_3}\rangle_{FB}+ n_{odd}  \langle J_{s_1} J_{s_2}J_{s_3}\rangle_{odd}
 \end{equation}
 where $FF$ and $BB$ correspond to free bosonic and free fermionic theory contributions. Correlation functions involving scalar operator $O$ with dimension $\Delta$ can be written in terms of free theory correlators only when $\Delta =1,2$ \footnote{For example, consider $\langle JJO_2\rangle$. This has one parity-even and one parity-odd structure. The parity-odd structure is given by the free fermion theory whereas the parity-even structure can be obtained by shadow transforming the free bosonic theory answer.}. For $\Delta=1,$ we have the free bosonic theory which gives the parity-even answer and the critical fermionic theory or Gross-Neveu model which gives the parity-odd result. For  $\Delta=2$, we have the critical-bosonic theory which gives the parity-even answer and the free fermionic theory which gives the parity-odd result.

Equation \eqref{fff1} can also be written in
 the following interesting basis with appropriate normalization \cite{Gandhi:2021gwn}
 \begin{equation}\label{fff2}
   \langle J_{s_1} J_{s_2}J_{s_3}\rangle = \left(n_s+n_f\right)  \langle J_{s_1} J_{s_2}J_{s_3}\rangle_{\text{nh}}+ \left(n_s-n_f\right)  \langle J_{s_1} J_{s_2}J_{s_3}\rangle_{h,even}+ n_{odd}  \langle J_{s_1} J_{s_2}J_{s_3}\rangle_{odd}
 \end{equation} where subscripts $h$ and $nh$ stand for homogeneous and non-homogeneous contributions respectively. One can establish using WT identity that $n_s+ n_f= c_s$, the two-point function coefficient.
In spinor-helicity variables, the homogeneous parity-even and parity-odd parts in \eqref{fff2} can be combined to obtain  \cite{Gandhi:2021gwn}
\begin{align}\label{fffsh}
   \langle J_{s_1}^{-} J_{s_2}^{-}J_{s_3}^{-}\rangle &= \left(n_s- n_f + i ~n_{odd}\right) \langle J_{s_1}^{-} J_{s_2}^{-}J_{s_3}^{-}\rangle_{{\bf{h}}}= |c| e^{i\theta}\langle J_{s_1}^{-} J_{s_2}^{-}J_{s_3}^{-}\rangle_{{\bf{h}}}\nonumber\\
   \langle J_{s_1}^{+} J_{s_2}^{+}J_{s_3}^{+}\rangle 
   &= |c| e^{-i\theta}\langle J_{s_1}^{+} J_{s_2}^{+}J_{s_3}^{+}\rangle_{{\bf{h}}}
 \end{align}
 where we have written only the homogeneous part  and introduced the notation $n_s- n_f + i ~n_{odd}\equiv c=|c| e^{i\theta}$ where $c$ is some complex number. This implies, we can combine the two OPE coefficients into one complex number. This is interesting as we can use this representation to do the bootstrapping \cite{Caron-Huot:2021kjy}.

\subsection{Outside the triangle inequality}\label{hotng}
We now turn our attention to correlation function outside the triangle inequality listed in \eqref{OT}.
As will shown below, the homogeneous contribution is not physically acceptable as it has bad poles, see Appendix \ref{outTH} for details of the calculation. So the only contribution to the correlation function is non-homogeneous. 
Let us first illustrate the non existence of homogeneous solution outside the triangle inequality. 

\bigskip
\subsubsection{Non existence of  homogeneous solution outside the triangle inequality}
We start with simplest correlation function $\langle J_{s} O_{\Delta_1} O_{\Delta_2}\rangle$.
\subsubsection*{$\langle J_{s} O_{\Delta_1} O_{\Delta_2}\rangle$}
In \cite{Bzowski:2013sza,Bzowski:2018fql,Jain:2021vrv}, it was shown that this correlator is zero if $\Delta_1 \ne \Delta_2$. 
The homogeneous solution for this correlation function is calculated in details in Appendix \ref{outTHjsoo}. The result is given by 
\begin{align}
\langle J^{-}_{s_1}O_{\Delta}O_{\Delta}\rangle = \frac{k^{s_1-1}_1k^{\Delta-2}_2k^{\Delta-2}_3}{E^{s_1}}\left(\frac{\langle 12\rangle\langle 31\rangle}{\langle 23\rangle}\right)^{s_1} = \frac{k^{s_1-1}_1k^{\Delta-1}_2k^{\Delta-1}_3}{E^{s_1}(E-2k_1)^{-h_1s_1}}\left(\langle 12\rangle\langle\bar{2}1\rangle\right)^{s_1}  
\end{align}
and other helicity can be obtained by complex conjugating.
We can also rewrite the above in momentum space 
\begin{align}\label{jsoo11}
\langle J(k_1)O_{\Delta}(k_2)O_{\Delta}(k_3)\rangle  = \frac{k^{s_1-1}_1k^{\Delta-1}_2k^{\Delta-1}_3}{E^{s_1}(E-2k_1)^{s_1}}(z_1\cdot k_2 )^{s_1} 
\end{align}
This has a bad pole and not acceptable as a CFT correlator. This implies that the homogeneous contribution is zero. It has only one non-homogeneous parity-even contribution. \\

\noindent {\bf{Flat space limit}}\\
Interestingly, if we take the flat space limit which is $E=k_1+k_2+k_3\rightarrow 0,$ we get
\begin{align}\label{jsooflt}
\lim_{E\rightarrow 0}\langle J(k_1)O_{\Delta_1}(k_2)O_{\Delta_2}(k_3)\rangle  = \frac{k^{-1}_1k^{\Delta_1-1}_2k^{\Delta_2-1}_3}{E^{s_1}}(z_1\cdot k_2 )^{s_1} \quad \rm{with}\,\Delta_i = 1~\rm{or}~2
\end{align}
which gives the amplitude to be 
\begin{equation}
    A\propto \left(z_1\cdot k_2 \right)^{s_1}.
\end{equation}
Surprisingly, even though CFT homogeneous correlation function is not acceptable as a valid result in this case, its flat space limit reproduces correct flat space amplitude result. 

\subsubsection*{$\langle J_{s_1} J_{s_2} O_{\Delta}\rangle$ with $s_1 \neq s_2$}
Unlike in the case of $s_1=s_2,$ it can be shown that only non-zero contribution to correlation function for $s_1\neq s_2$ is when $\Delta=1,2$. See Appendix \ref{Js1Js2Odeltaapp} for details. In Appendix \ref{Js1Js2Odeltaapp}, we also show that the homogeneous solution can be written as 
\begin{align}\label{s1s2o}
&\langle J^{h_1}_{s_1}J^{h_2}_{s_2}O_{\Delta}\rangle = \frac{k^{s_1-1}_1k^{s_2-1}_2k^{\Delta-2}_3}{E^{s_1+s_2}}\langle12\rangle^{-h_1s_1-h_2s_2}\langle23\rangle^{h_1s_1-h_2s_2}\langle31\rangle^{h_2s_2-h_1s_1}
\end{align}
where $h_i$ denotes helicity which can be $\mp$. Since, $s_1\neq s_2$, one of the angle brackets $\langle ij\rangle$ will end up in the denominator, which would lead to a bad pole and needs to be discarded. Below we look at an example of this class of correlators for $s_1 \neq s_2$ which illustrates the point.\\

\noindent {\bf{Example:  $\langle J_3J_1O\rangle$}}\\
Homogeneous part of the correlation function $\langle J_3J_1O_{\Delta}\rangle$ with $\Delta=1,2$ is given by
\begin{align}
\langle J^{-}_3 J^{-}_1 O_{\Delta}\rangle_h = \frac{k^2_1 k_3^{\Delta-2}}{E^4 } \frac{\langle 12\rangle^4\langle 31\rangle^2}{\langle 23\rangle^2} = \frac{k^2_1 k_3^{\Delta-2}}{E^6(k_1-k_3-k_2)^2 } \langle \bar{2}\bar{3}\rangle^2\langle 12\rangle^4\langle 31\rangle^2\label{hgpsh}
\end{align}
which has a has an unphysical pole at $k_1= k_2  + k_3$. Similar conclusion can be drawn for $++$ helicity as well. For mixed helicity the homogeneous part is always zero.  This unphysical pole shows up in the momentum space variables as well
\begin{align}\label{h310}
    &\langle J_3(z_1, k_1) J_1(z_2, k_2) O(k_3)\rangle_h = -\frac{k^4_1 k_3^{\Delta-2}}{E^4(E-2k_1)^2}(z_1\cdot k_2)^2\left(2z_2\cdot k_1  z_1\cdot k_2 +E(E-2k_3)z_1\cdot z_2\right).
\end{align}
This unphysical pole is not acceptable as a valid CFT correlation function and should be discarded.\\

\noindent {\bf{Flat space limit}}\\
In the flat space limit using \eqref{h310}, we get the amplitude of the form
\begin{equation}\label{310hfl1}
    A\propto \left(z_1\cdot k_2 \right)^3 z_2\cdot k_1 
\end{equation}
which is again correct flat space amplitude for the spin-3, spin-1 and a scalar vertex.
\subsubsection*{$\langle J_{s_1} J_{s_2}J_{s_3}\rangle$}
 The homogeneous contribution in this case is given by
\begin{align}
&\langle J^{h_1}_{s_1}J^{h_2}_{s_2}J^{h_3}_{s_3}\rangle = \frac{1}{E^{s_1+s_2+s_3}}\langle12\rangle^{h_3s_3-h_1s_1-h_2s_2}\langle23\rangle^{h_1s_1-h_2s_2-h_3s_3}\langle31\rangle^{h_2s_2-h_3s_3-h_1s_1}
\end{align}
Again, we see that when triangle inequality is violated, one of the brackets $\langle ij\rangle$ will end up in the denominator leading to a bad pole. We take an example of $\langle J_4J_1J_1\rangle$ to illustrate this point.\\

\noindent {\bf{Example:  $\langle J_4J_1J_1\rangle$}}\\
Let us consider $\langle J_4J_1J_1\rangle$ for simplicity. The homogeneous part is given by
\begin{align}
\langle  J^{-}_4 J^{-}_1J^{-}_1\rangle_h = \frac{k^3_1}{E^6} \frac{\langle 12\rangle^4\langle 31\rangle^4}{\langle 23\rangle^2} = \frac{k^3_1}{E^8(k_1-k_3-k_2)^2} \langle \bar{2}\bar{3}\rangle^2\langle 12\rangle^4\langle 31\rangle^4\label{hgpsh2}
\end{align}
which has a has an unphysical pole at $k_1= k_2  + k_3$. Similar conclusion can be drawn for $+++$ helicity as well. For mixed helicity the homogeneous part is always zero.  This unphysical pole shows up in the momentum space variables as well
\begin{align}\label{h411a}
    \langle J_4J_1J_1\rangle_{h} = -\frac{(z_1\cdot k_2 )^2k^5_1\left(2z_1\cdot k_2  z_3\cdot k_1 -z_1\cdot z_3E(E-2k_2)\right)\left(2 z_1\cdot k_2  z_2\cdot k_1 -z_1\cdot z_2 E(E-2k_3)\right)}{256 E^6 (E-2k_1)^2},
\end{align}
This unphysical pole is not acceptable as valid CFT correlation function and should be discarded. We conclude that outside the triangle inequality, the homogeneous solution can not be considered as a valid CFT correlation function.\\

\noindent {\bf{Flat space limit}}\\
Again taking flat space limit of \eqref{h411a} we get 
\begin{equation}\label{hflt411}
    A= \left(z_1\cdot k_2  \right)^4 z_2\cdot k_2 z_3\cdot k_1 .
\end{equation}
which turns out to be the correct flat space limit. Thus we observe that even though we have a homogeneous solution which is not a good CFT correlator, it has good flat space limit\footnote{In \cite{Baumann:2021fxj} it was discussed, given a  flat space amplitude how to get a CFT correlator or $dS_4$ amplitude. As we will see later, the non-homogeneous piece as well gives correct flat space amplitude. So given a flat space amplitude, the corresponding CFT correlator is not unique. However acceptable CFT correlator is  unique and for that we need to carefully distinguish between homogeneous or non-homogeneous nature of the correlator. For example, outside the triangle all the CFT correlators are non-homogeneous, so before using the prescription in \cite{Baumann:2021fxj}, we need to specify the WT identity carefully before computing it. Also we would like to emphasize, even though homogeneous piece in \ref{hflt411} gives non-minimal amplitude in the flat space limit, this homogeneous solution solution is not the correct CFT correlator. }. 

\subsubsection{The non-homogeneous solution outside the triangle inequality}
We start with simplest correlation function $\langle J_{s} O_{\Delta_1} O_{\Delta_1}\rangle$.
\subsubsection*{$\langle J_{s} O_{\Delta_1} O_{\Delta_2}\rangle$}\label{jsoonh}
For non-homogeneous solution we need to write down the Ward Takahashi identity first.
The WT identity schematically can be written as
\begin{align}\label{WTsoo}
k_1\cdot\langle J(k_1)O_{\Delta_1}(k_2)O_{\Delta_2}(k_3)\rangle  \propto \langle O_{\Delta_1} O_{\Delta_2} \rangle 
\end{align}
where we have neglected other momentum dependence in RHS. Note that \eqref{WTsoo} 
 is proportional to scalar two-point function which does not have any parity-odd contribution to it. Hence the non-homogeneous part will not have any parity-odd part to it. Also note that RHS of \eqref{WTsoo} is only non zero for $\Delta_1=\Delta_2$. Using \eqref{nhpiece1} and \eqref{WTsoo} we conclude that only possible solution is a non-homogeneous parity-even solution.
In \cite{Jain:2021vrv,Baumann:2020dch} explicit results  can be found.
\subsubsection*{$\langle J_{s_1} J_{s_2} O_{\Delta}\rangle$ with $s_1\neq s_2$}\label{js1js2oonh}
To evaluate this correlation function in general, we need to know the WT identity. One can use, in principle, higher-Spin algebra to derive WT identity for free boson and free fermionic theory. However, for our purposes, it is sufficient to consider  an examples which illustrates the basic physics. In this section, we compute $\langle J_3 J_1 O \rangle$ for free bosonic and free fermionic theory. The aim of this section is to show that non-homogeneous solutions do not have bad poles that appeared in homogeneous solutions. 
Let us first consider the free bosonic theory. \\

\noindent {\bf{Example: Bosonic $\langle J_3J_1O\rangle$}}\\
The result in momentum space is given by 
\begin{equation}\label{310anst}
     \langle J_3(z_1, k_1)J_1(z_2, k_2)O(k_3)\rangle = A (z_1\cdot k_2 )^3(z_2\cdot k_1 )+B (z_1\cdot k_2 )^2(z_1\cdot z_2)
\end{equation}
where $A$ and $B$ are given by,
\begin{align}\label{res1}
A = -\frac{E^2 + 2k_1(E+k_1)}{E^4 k_3},\quad B = \frac{3}{E}+\frac{k_1(3E + 2k_1)}{E^3}-\frac{5}{2k_3}.
\end{align}
The  WT identity corresponding to this correlator is given by,
\begin{align}\label{WTfb}
   \left\langle k_1\cdot J_3\left(z_1, k_{1}\right) J_1\left(z_2, k_{2}\right) O\left(k_{3}\right)\right\rangle &= \frac{5 z_1\cdot k_2 \left(2 (z_1\cdot k_2) (z_2\cdot k_1)+\left(k_1^2-5 k_2^2-k_3^2+6 k_2 k_3\right) z_1\cdot z_2\right)}{k_3}\nonumber\\
    k_{2 \rho}\left\langle J_3\left(z_1, k_{1}\right) J_1^{\rho}\left(z_2, k_{2}\right) O\left(k_{3}\right)\right\rangle &=0
\end{align}
where we have used the notation $k_1\cdot J_3= k_{1\mu} J_3^{\mu\alpha\beta}$. 
Converting \eqref{310anst},
\eqref{res1} to spinor-helicity variables we obtain
\begin{align}\label{310spab} 
   &\langle J^-_3J^-_1O\rangle = f^{-~-}(k_1, k_2, k_3)\langle 12\rangle^4\langle 31\rangle^2\langle \bar{2}\bar{3}\rangle^2,~~~
     \langle J^-_3J^+_1O\rangle = f^{-+}(k_1, k_2, k_3)\langle 12\rangle^2\langle 31\rangle^4\langle \bar{2}\bar{3}\rangle^4
\end{align}
where,
\begin{align}
    f^{-~-}(k_1, k_2, k_3) =\frac{2 E^3+3 E^2 k_1+4 E k_1^2+4 k_1^3}{128 E^6 k_1^3 k_3},~~~
    f^{-~+}(k_1, k_2, k_3) =\frac{2E+k_1}{E^6 k_1^3 k_3}.
\end{align}
The other helicity configurations can be obtained by complex conjugating the ones we have listed.\\
\\
{\bf{Flat space limit}}\\
In the flat space limit \eqref{310anst} with \eqref{res1} gives 
\begin{equation}
    A\propto \left(z_1\cdot k_2 \right)^3 z_2\cdot k_1 
\end{equation}
which is again the correct flat space amplitude. This is precisely what we obtained in \eqref{310hfl1}. So in conclusion, both the homogeneous solution in \eqref{h310} and non-homogeneous solution in \eqref{h411a} give same flat space amplitude.\\

\noindent {\bf{Scalar with $\Delta=2$}}\\
For this case, the answer can be obtained from free bosonic answers by doing a Legendre transform.  More precisely, 
\begin{equation}\label{LTsc}
    \langle J_3J_1 O_{\Delta=2}\rangle= k_3\langle J_3J_1 O_{\Delta=1}\rangle
    =k_3\langle J_3J_1 O\rangle_{FB}.
\end{equation}
This in momentum space gives
\begin{equation}
    \langle J_3(z_1, k_1)J_1(z_2, k_2)O_{\Delta=2}(k_3)\rangle ={\widetilde {A}} (z_1\cdot k_2 )^3(z_2\cdot k_1 )+{\widetilde {B}} (z_1\cdot k_2 )^2(z_1\cdot z_2)
\end{equation}
with 
\begin{equation}
    {\widetilde {A}} =k_3  A,~~~ {\widetilde {B}}=k_3 B 
\end{equation}
with $A,B$ gives by \eqref{res1}.
In spinor-helicity variables \eqref{310spab} we have
\begin{align}
   f^{-~-}_{\Delta=2}(k_1, k_2, k_3) =k_3  f^{-~-}_{\Delta=1}(k_1, k_2, k_3),~~~ f^{-~+}_{\Delta=2}(k_1, k_2, k_3) =k_3  f^{-~+}_{\Delta=1}(k_1, k_2, k_3)
\end{align}

The other helicity configurations are obtained by complex conjugating the ones that are listed.
The WT identity for this case can be computed using \eqref{WTfb} and is given by
 \begin{align}\label{WTcb}
  k_{2 \rho}\left\langle J_3\left(z_1, k_{1}\right) J_1^{\rho}\left(z_2, k_{2}\right) O\left(k_{3}\right)\right\rangle &=0,\nonumber\\
   \left\langle k_1\cdot J_3\left(z_1, k_{1}\right) J_1\left(z_2, k_{2}\right) O\left(k_{3}\right)\right\rangle &= 5 z_1\cdot k_2 \left(2 (z_1\cdot k_2) (z_2\cdot k_1)+\left(k_1^2-5 k_2^2-k_3^2+6 k_2 k_3\right) z_1\cdot z_2\right)\nonumber\\
   &=30 k_2 k_3 z_1\cdot k_2 z_1\cdot z_2+ \left(2 (z_1\cdot k_2) (z_2\cdot k_1)+\left(k_1^2-5 k_2^2-k_3^2\right) z_1\cdot z_2\right) \nonumber\\
   &=30 k_2 k_3 z_1\cdot k_2 z_1\cdot z_2+ (\rm{terms~that ~contribute~to~contact~term })\nonumber\\
   &\sim 30 k_2 k_3 z_1\cdot k_2 z_1\cdot z_2
\end{align}
where in the penultimate line we have used the fact that upon Fourier transforming we get
\begin{align}
FT: & \left(2 (z_1\cdot k_2) (z_2\cdot k_1)+\left(k_1^2-5 k_2^2-k_3^2\right) z_1\cdot z_2\right)\nonumber\\
&= \left(2 (z_1\cdot\partial_{x_2}) (z_2\cdot \partial_{x_1})+\left(\Box_{x_1}-5 \Box_{x_2}-\Box_{x_3}\right) z_1\cdot z_2\right)\delta^3(x_1-x_2) \delta^3(x_2-x-3)
\end{align}
which will contribute to the purely contact term or semilocal terms in correlation function, see section \ref{smlcon} for more discussion.
\\

\noindent{\bf{Example: Fermionic $\langle J_3 J_1 O_2\rangle$}}\\
The $J_3$ higher spin current is defined as
\begin{align}
    J_3(z, x) = 3\bar{\psi}(z\cdot\gamma) (z\cdot\partial)^2\psi+ 10 (z\cdot\partial\bar{\psi})(z\cdot\gamma)(z\cdot\partial)\psi + 3(z\cdot\partial)^2\bar{\psi}(z\cdot\gamma) \psi
\end{align}
which in momentum space is given by
\begin{align}
    J_3(z, k) = \int d^3l (z\cdot l)^3\bar{\psi}(l)(z\cdot\gamma)\phi(l-k)(z\cdot l)^3
\end{align}
Unlike in the free bosonic case discussed in the previous section, the free fermionic $\langle J_3 J_1 O_2\rangle$ is parity-odd. A direct calculation gives
\begin{align}
    \langle J_3(z_1, k_1)J_1(z_2, k_2)O_2(k_3)\rangle = A (z_1\cdot k_2 )^2\epsilon^{z_1 z_2 k_1}+B (z_1\cdot k_2 )^2\epsilon^{z_1 z_2 k_2}
\end{align}
where the form factors $A, B$ are given by
\begin{align}
    A = \frac{k_2}{E^2}+\frac{2k_1 k_2}{E^3}+\frac{2k_1^2k_2}{E^4},\quad B = -\frac{2}{E}-\frac{2k_1}{E^2}-\frac{2k_1^2}{E^3}-\frac{2k^3_1}{E^4}
\end{align}
The correlator $\langle J_3 J_1O_2 \rangle$ obeys the following Ward identity
\begin{align}\label{WTFf1}
  \left\langle k_1\cdot J_3\left(z_1, k_{1}\right) J_1\left(z_2, k_{2}\right) O_2\left(k_{3}\right)\right\rangle &= -10 z_1\cdot k_2  \left(k_2\epsilon^{z_1 z_2 k_1}+3(k_2-k_3)\epsilon^{z_1 z_2 k_2}\right)\nonumber\\
  k_{2 \rho}\left\langle J_3\left(z_1, k_{1}\right) J_1^{\rho}\left(z_2, k_{2}\right) O_2\left(k_{3}\right)\right\rangle &=0\,.
  \end{align}
In spinor-helicity variables with the ansatz  in \eqref{310spab} we have 
\begin{align}\label{ff31}
   &f_{-~-}(k_1, k_2, k_3) =-i\left(\frac{2 E^3+3 E^2 k_1+4 E k_1^2+4 k_1^3}{128 E^6 k_1^3 }\right)\notag\\
    &f_{-~+}(k_1, k_2, k_3) =i\frac{2E+k_1}{E^6 k_1^3 }
\end{align}
Other helicity configurations are obtained via complex conjugation of the listed ones.
Once again, we notice that neither the momentum space expression nor the spinor-helicity expressions have any unphysical poles.

    It is interesting to note that, in spinor-helicity variables free fermion result \eqref{ff31} and bosonic result with $\Delta=2$ \eqref{310spab} and \eqref{LTsc} are identical upto some factors of $i$ and possible signs for complex conjugate helicity components. Furthermore one can relate these results also in momentum space as follows \cite{Jain:2021gwa}
\begin{equation}
    \langle J_3 J_1 O_2\rangle_{even}=\frac{1}{k_2}\epsilon^{\alpha z_2 k_2}\langle J_3 J_1^\alpha O_2\rangle_{FF}.
\end{equation}
We can extend this result to arbitrary spin. We will discuss related issues in later sections.  What we conclude is that for the correlators of the form $\langle J_{s_1} J_{s_2} O_{\Delta}\rangle$ with $s_1\neq s_2,$ only non zero answer is obtained for $\Delta=1,2$ and  has one parity-even and one parity-odd non-homogeneous contribution.


\subsubsection*{$\langle J_{s_1} J_{s_2}J_{s_3}\rangle$}
In this subsection we calculate the non-homogeneous contribution to correlators of the kind $\langle J_{s_1} J_{s_2}J_{s_3}\rangle$. To determine non-homogeneous contribution we need to first find out the WT identity. Interestingly, unlike the cases inside the triangle inequality, there can be multiple WT identity. It is interesting to note that  WT identity for inside the triangle inequality  for free boson and free fermion correlators are the same. See Appendix \ref{countingapp} for details. However for correlation function outside the triangle inequality, it turns out that the WT identity for boson and fermion are not the same. Again see \ref{countingapp} for details. 
 We take an example of $\langle J_4J_1J_1\rangle$ to illustrate this point.\\
 
\noindent {\bf{Example: Bosonic $\langle J_4 J_1 J_1\rangle$}}\\
The $J_4$ current is given by
\begin{align}
    J_4(z, x) = \frac{1}{128}\left(\bar{\phi}(z\cdot\partial)^4\phi -28 (z\cdot\partial)\bar{\phi}(z\cdot\partial)^3\phi+70 (z\cdot\partial)^2\bar{\phi}(z\cdot\partial)^2\phi+28 (z\cdot\partial)^3\bar{\phi}(z\cdot\partial)\phi+(z\cdot\partial)^4\bar{\phi} \phi \right)
\end{align}
which in momentum space is given by
\begin{align}
    J_4(z, k) = \int d^3l \bar{\phi}(l)\phi(l-k)(z\cdot l)^4
\end{align}
We now write an ansatz for the correlator $\langle J_4 J_1 J_1\rangle$ after appropriate contractions with transverse polarizations
\begin{align}
   &\langle J_4(z_1, k_1)J_1(z_2, k_2)J_1(z_3, k_3)\rangle = A (z_1\cdot k_2 )^4 z_2\cdot z_3+ C (z_1\cdot k_2 )^3(z_1\cdot z_2)(z_3\cdot k_1 )\notag\\&- C(k_2 \leftrightarrow k_3)(z_1\cdot k_2 )^3(z_1\cdot z_3)(z_2\cdot k_1 )+ D (z_1\cdot k_2 )^2(z_1\cdot z_2)(z_1\cdot z_3) 
\end{align}
 Momentum space answers are cumbersome and can be found in \eqref{res1bans} and \eqref{res1b}.

The WT identity is given by
\begin{align}
\label{WIEx2b}
 & \left\langle k_1\cdot J_4\left(z_1, k_{1}\right) J_1\left(z_2, k_{2}\right) J_1\left(z_3, k_{3}\right)\right\rangle = \frac{7}{512}z_1\cdot k_2 \left[7 (z_1\cdot k_2 )^2 z_2\cdot z_3 (k_2-k_3)\right.\notag\\&\left.+z_1\cdot z_2 z_1\cdot z_3 (k_2-k_3)(-k^2_1+k^2_2+8k_2k_3+k^2_3)+z_1\cdot k_2(z_1\cdot z_3 z_3\cdot k_1(2k_2-7k_3)\right.\cr
 &\left.+z_1\cdot z_3 z_2\cdot k_1(-7k_2+2k_3))\right]\nonumber\\
   & k_{2 \rho}\left\langle J_{4}\left(z_1, k_{1}\right) J_1^{\rho}\left( k_{2}\right) J_1\left(z_3, k_{3}\right)\right\rangle=0.
\end{align}
Let us express above answer in spinor-helicity variables. In spinor-helicity variables we have 
\begin{align}
    &\langle J^-_4 J^-_1J^-_1\rangle = f_{---}(k_1, k_2, k_3)\langle 12\rangle^4\langle 31\rangle^4\langle \bar{2}\bar{3}\rangle^2\label{WIC2b}\notag\\
    & \langle J^-_4 J^-_1J^+_1\rangle = f_{--+}(k_1, k_2, k_3)\langle 12\rangle^6\langle 31\rangle^2\langle \bar{2}\bar{3}\rangle^4\\
    & \langle J^+_4 J^-_1J^-_1\rangle = f_{+--}(k_1, k_2, k_3)\langle23\rangle^6\langle\bar{1}\bar{2}\rangle^4\langle\bar{3}\bar{1}\rangle^4\notag
\end{align}
with
\begin{align}\label{411bhlc}
    f_{-~-~-}(k_1, k_2, k_3) &= \frac{3 E^5+5 E^4 k_1+8 E^3 k_1^2+12 E^2 k_1^3+16 E k_1^4+16 k_1^5}{1048576 E^8 k_1^4}\notag\\
    f_{-~-~+}(k_1, k_2, k_3) &=\frac{5 E^3+5 E^2 k_1+4 E k_1^2+2 k_1^3}{524288 E^8 k_1^4}\\
    f_{+~-~-}(k_1, k_2, k_3) &= \frac{3E+k_1}{1048576 E^8 k_1^4}\notag
\end{align}
The other helicity configurations are obtained either by complex conjugation or by symmetry.
\\

\noindent {\bf{Fermionic $\langle J_4 J_1 J_1\rangle$}}\\
The $J_4$ current is given by
\begin{align}
    J_4(z, x) =\left( -\bar{\psi}(z\cdot\gamma)(z\cdot\partial)^3\psi+7(z\cdot\partial)\bar{\psi}(z\cdot\gamma)(z\cdot\partial)^2\psi-7(z\cdot\partial)^2\bar{\psi}(z\cdot\gamma)(z\cdot\partial)\psi+(z\cdot\partial)^3\bar{\psi}(z\cdot\gamma)\psi\right)
\end{align}
which in momentum space is given by
\begin{align}
    J_4(z, k) = \int d^3l\, \bar{\psi}(l)(z\cdot\gamma)\psi(l-k)(z\cdot l)^3
\end{align}
The ansatz for the correlator $\langle J_4 J_1 J_1\rangle$ is exactly the same as \eqref{411anst1}. Explicit results can be found in Appendix \eqref{res1f}. 
In momentum space there is no bad pole.
Let us re-express the answer in spinor-helicity variables. The ansatz again is given by
\eqref{WIC2b} with 
\begin{align}\label{411fhlc}
   f_{-~-~-}(k_1, k_2, k_3) &= -\frac{3 E^5+5 E^4 k_1+8 E^3 k_1^2+12 E^2 k_1^3+16 E k_1^4+16 k_1^5}{1048576 E^8 k_1^4}\notag\\
    f_{-~-~+}(k_1, k_2, k_3) &=\frac{5 E^3+5 E^2 k_1+4 E k_1^2+2 k_1^3}{524288 E^8 k_1^4}\\
    f_{+~-~-}(k_1, k_2, k_3) &= -\frac{3E+k_1}{1048576 E^8 k_1^4}\notag
\end{align}
The other helicity configurations are obtained either by complex conjugation or symmetry.

Even though free fermionic \eqref{res1f}  and free bosonic results \eqref{res1b}  in momentum space are quite different, we notice that remarkably in spinor-helicity variables the free boson \eqref{411bhlc} and free fermion \eqref{411fhlc} correlators are essentially the same upto a sign in some helicities such as for  $f_{---}$. To exploit this fact, let us consider following combinations.
\subsection*{$\langle J_4J_1J_1\rangle$: Free Boson (FB) $-$ Free Fermion (FF)}
In spinor-helicity variables we have,
\begin{align}
    \langle J_4^{-} J_1^{-} J_1^{-}\rangle_{FB-FF}&=\langle 12\rangle^4\langle 31\rangle^4\langle \bar{2}\bar{3}\rangle^2\frac{3 E^5+5 E^4 k_1+8 E^3 k_1^2+12 E^2 k_1^3+16 E k_1^4+16 k_1^5}{524288 E^8 k_1^4} \nonumber\\
 \langle J_4^{+} J_1^{-} J_1^{-}\rangle_{FB-FF}&=\langle 23\rangle^6\langle \bar{1}\bar{2}\rangle^4\langle \bar{3}\bar{1}\rangle^4\frac{3E+k_1}{524288 E^8 k_1^4},~~~ \langle J_4^{-} J_1^{-} J_1^{+}\rangle_{FB-FF}=0
\end{align}
This combination in flat space limit corresponds to a non-minimal vertex with six derivatives and is discussed below.

\subsection*{$\langle J_4J_1J_1 \rangle$: Free Boson $+$ Free Fermion}

In spinor-helicity variables we have
\begin{align}
    \langle J_4^{-} J_1^{-} J_1^{-}\rangle_{FB+FF}&=0,~~~\langle J_4^{+} J_1^{-} J_1^{-}\rangle_{FB+FF}=0\nonumber\\
    \langle J_4^{-} J_1^{-} J_1^{+}\rangle_{FB+FF}&=\langle 12\rangle^6\langle 31\rangle^2\langle \bar{2}\bar{3}\rangle^4 \frac{5 E^3+5 E^2 k_1+4 E k_1^2+2 k_1^3}{262144 E^8 k_1^4}
\end{align}
This combination in flat space limit corresponds to a non-minimal vertex with six derivatives and is discussed below.\\

\noindent {\bf{Flat space limit}}\\
In the flat-space limit of  $E \to 0$ we have 
\begin{align}
    \langle J_4 J_1 J_1 \rangle_{FB+FF} \sim \frac{1}{E^4}\left((z_1\cdot k_2 )^4 z_2\cdot z_3 + (z_1\cdot k_2)^3 z_1\cdot z_2 z_3\cdot k_1 - (z_1\cdot k_2)^3 z_1\cdot z_3 z_2\cdot k_1\right)+ \mathcal{O}\left(\frac{1}{E^3}\right).
\end{align}
We can identify the amplitude to be 
\begin{align}\label{amina}
    A_{min} = \left((z_1\cdot k_2 )^4 z_2\cdot z_3 + (z_1\cdot k_2)^3 z_1\cdot z_2 z_3\cdot k_1 - (z_1\cdot k_2)^3 z_1\cdot z_3 z_2\cdot k_1\right)
\end{align}
which is the correct flat space minimal amplitude \cite{Conde:2016izb}. 
Again the tensor structure is gauge invariant. Now, we look at  the flat space limit of
\begin{align}
  \lim_{E\rightarrow 0} \langle J_4J_1J_1\rangle_{FB-FF} \sim -\frac{k_1^3}{128 E^6}(z_1\cdot k_2 )^4 z_2\cdot k_3 z_3\cdot k_1  + \mathcal{O}\left(\frac{1}{E^5}\right)\,.
\end{align}
Using above equation, we can identify the non-minimal flat space amplitude to be given by
\begin{align}
    A_{non-min}= (z_1\cdot k_2 )^4 z_2\cdot k_3 z_3\cdot k_1 
\end{align}
which is again the correct flat space non-minimal amplitude and exactly same as \eqref{hflt411} which was obtained from \eqref{h411a}.

\section{No parity-odd contribution to correlation function of exactly conserved currents outside triangle inequality}
\label{noodd}
In this section we show that the there is no parity-odd contribution outside the triangle inequality for exactly conserved currents when all the spins are non-zero. For this purpose, let us start with non-homogeneous contribution. 

\subsection{Parity odd non-homogeneous solution} Let us start with simplest example $\langle J_s OO\rangle$. In subsection \ref{jsoonh} we explicitly argued that for this case there is no parity-odd non-homogeneous solution. For correlators of the form $\langle J_{s_1} J_{s_2} O \rangle $ with $s_1\neq s_2$, however there can be a parity-odd non-homogeneous solution coming from free fermion or critical fermionic theory consistent with \cite{Giombi:2011rz}. For generic correlator of the form $\langle J_{s_1} J_{s_2} J_{s_3} \rangle $ with all non-zero spin, however the parity-odd non-homogeneous contribution is always a contact or a semilocal term. The argument is exactly the same as described in \eqref{WThgnspinod}. So we conclude that, in general parity-odd non-homogeneous term is not present except in the case of $\langle J_{s_1} J_{s_2} O \rangle $ with $s_1\neq s_2$ where the contribution comes from free fermionic theory or critical fermionic theory.
We now turn our attention to parity-odd homogeneous solution.
\subsection{Parity odd homogeneous correlation function}
In spinor-helicity variables we can write
 \begin{equation}
     \langle J_{s_1} J_{s_2}J_{s_3}\rangle_{\text{h}} = \left(c_1 F_{\text{h,even}}(k_1,k_2,k_3)+ i c_2 F_{\text{h,odd}}(k_1,k_2,k_3)\right)\langle 12 \rangle^{h_3-h_1-h_2} \langle 23 \rangle^{h_1-h_2-h_3} \langle 31 \rangle^{h_2-h_3-h_1}
 \end{equation}
 with $s_i$ taking arbitrary value including $0$ such that triangle inequality is violated.
 It is easy to show that
 \begin{equation}\label{heodr}
     F_{\text{h,even}}(k_1,k_2,k_3) =F_{\text{h,odd}}(k_1,k_2,k_3)=F_{\text{h}}(k_1,k_2,k_3) 
 \end{equation}
 In section \ref{hotng} we found that there exist no acceptable homogeneous solution for CFT correlator consistent with OPE limit,  with spins violating triangle inequality. Then \eqref{heodr} immediately implies that there cannot be any parity-odd homogeneous solution. 

We observed that in $---$ and $+++$ helicity, there is a bad pole and is not acceptable as a good solution. One might ask what happens to other helicity components. It is easy to show that, homogeneous contribution coming from other helicity components have even worse pole structure. For details see Appendix \ref{countingapp}. So we conclude we can not have any parity odd homogeneous term either. 

\subsubsection*{Summary so far}
We now summarise the results when triangle inequality is violated. We conclude following counting of structures
\begin{center}
    \begin{tabular}{ | l | l | l |}
    \hline
    Correlator & Parity-even & Parity-odd \\ \hline
    $\langle J_s O_{\Delta}O_{\Delta}\rangle$ & 1 & $\cross$  \\ \hline
    $\langle J_{s_1} J_{s_2}O_{\Delta}\rangle$ & 1 & 1 \\ \hline
    $\langle J_{s_1} J_{s_2}J_{s_3}\rangle$ & 2 & $\cross (\text{up ~to ~contact ~terms})$\\
    \hline
    \end{tabular}
\end{center}
For all the correlators we have non-homogeneous solution.
 So we have
 \begin{align}\label{fff}
   \langle J_{s_1} J_{s_2}J_{s_3}\rangle &= n_f  \langle J_{s_1} J_{s_2}J_{s_3}\rangle_{FF}+ n_{b}  \langle J_{s_1} J_{s_2}J_{s_3}\rangle_{FB}
 \end{align}up to contact terms.
\section{Correlation functions of slightly broken higher spin currents}
\label{SectionWBC}
In the previous subsection, we have analysed the cases when the currents are exactly conserved. Here we shall consider situations where the higher spin currents are not exactly conserved. Examples of such theories include, quasi-bosonic and quasi-fermionic theory\cite{Aharony:2011jz,Giombi:2011kc,Maldacena:2012sf}, see appendix \ref{theory1} for more details. Quasi-bosonic theory corresponds to dual pair of theories which include Chern-Simons gauge field coupled to bosonic theory or critical fermionic theory. Quasi-fermionic theory corresponds to dual pair of theories which include Chern-Simons gauge field coupled to critical bosonic theory or  fermionic theory. The explicit results of non conservation can be found in \cite{Giombi:2016zwa}.  
We can divide the non-conservation in two separate pieces, one involving spin zero and another involving only spinning operators. Non-conservation equation involving spin zero operator, we have two different results, one for quasi bosons and one for quasi fermion. 
For quasi bosons in light cone coordinates we have 
\begin{align}\label{noncs0}
  \left.\partial \cdot J_{s_1}\right|_{s_{2}, 0}=\sum_{n=0}^{s_1-s_2} \epsilon_{\mu \nu-}\left(b_{n} \partial_{-}^{n-1} \partial^{\nu} J_{s_{2}}^{\mu} \partial_{-}^{p-n} J_{0}+f_{n} \partial_{-}^{n} J_{s_{2}}^{\mu} \partial^{\nu} \partial_{-}^{p-n-1} J_{0}\right)
\end{align}
where $b_n,f_n$ can be found explicitly computed in \cite{Giombi:2016zwa}. 
Below we shall use the notation $\partial \cdot J_{s_1} =[J_{s_2} ] [J_{0} ]$.
For quasi fermion
\begin{align}\label{noncs0f}
   \left.\partial \cdot J_{s_1}\right|_{s_{1}, 0}=\sum_{m=0}^{p} c_{m} \partial_{-}^{m} J_{s_{2}} \partial_{-}^{p-m} \widetilde{J}_{0}
\end{align}
where $p = s_1-s_2$ with $p>0$ and $J_0$ and $\widetilde{J}_{0}$ corresponds to scalar operator with scaling dimension $1$ and $2$ respectively. For later purposes it is useful to keep in mind that the spin of the spinning operator that appears on the right hand side is always less than the spin on the left hand side of \eqref{noncs0} and \eqref{noncs0f}. Another possibility of non-conservation which is same for both quasi boson and quasi fermion is given by\footnote{Below we shall use the notation $\partial \cdot J_{s_1} =[J_{s_2} ] [J_{s_3} ]$.} 
\begin{align}\label{noncss1s2}
   \left.\partial \cdot J_{s_1}\right|_{s_{2}, s_{3}}=\sum_{n=0}^{s_1-s_2-s_3} \epsilon_{\mu \nu-}\left(a_{n} \partial_{-}^{n} J_{s_{2}}^{\mu} \partial_{-}^{p-n} J_{s_{3}}^{\nu}+b_{n} \partial_{-}^{n-1} \partial^{\nu} J_{s_{2}}^{\mu} \partial_{-}^{p-n} J_{s_{3}}+e_{n} \partial_{-}^{n} J_{s_{2}} \partial^{\mu} \partial_{-}^{p-n-1} J_{s_{3}}^{\nu}\right) 
\end{align}
where $p = s_1-s_2-s_3$ with $p>0$. Here we would like to emphasize again that sum total of spin of the operators that appears on the right hand side is always less than that appears on the left hand side. This in the language of previous section $p>0$ is basically the violation of triangle inequality relation.

We show below that when we are restricted to inside the triangle inequality, the non conservation does not change any conclusion. However, when we are outside the triangle, we get one parity-odd contribution. In \cite{Giombi:2016zwa}, light cone components of some of the correlators were explicitly computed in position space. 

\subsection{Inside the triangle inequality}
Let us first consider effect of non-conservation Ward identity for correlation function with spin inside the triangle inequality.
We start by considering simplest situation first $\langle J_s J_s O \rangle$.
Using the non conservation equation \eqref{noncs0} we get
\begin{align}
  \langle \partial\cdot J_{s} J_{s}O_{\Delta}\rangle  &= \langle(\sum [J_{s_i}][J_{0}]) J_{s}O_{\Delta}\rangle \nonumber\\
  &=\langle J_{s_i} J_{s}\rangle \langle O_{\Delta} O_{\Delta}\rangle\nonumber\\
  &=0
\end{align}
where in the last line we have used the fact that $s_i < s$ and hence the two-point function is zero. We also have used large $N$ factorization. Also in the first line of the of the last equation we have used $[J_{s_i}][J_{0}]$ symbol as a shorthand for. R.H.S. of \eqref{noncs0}. 
Now turning our attention to the case with all the non-zero spin, using \eqref{noncss1s2} we obtain 
\begin{align}
  \langle \partial.J_{s_1} J_{s_2}J_{s_3}\rangle  &= \langle(\sum [J_{s_i}][J_{s_j}]) J_{s_2}J_{s_3}\rangle \nonumber\\
  &\sim \langle J_{s_i} J_{s_1} \rangle \langle J_{s_j} J_{s_2} \rangle + \langle J_{s_j} J_{s_1} \rangle \langle J_{s_i} J_{s_2} \rangle  \nonumber\\
  &=0.
\end{align}
where in the last line we made use of the fact that non-conservation equation states that $s_1-s_i-s_j> 0,$ however inside the triangle inequality $s_1-s_2-s_3 \le 0$. This implies it can never  be the case that both $s_i$ and $s_j$ be equal to $s_1$ or $s_2$.  This implies we won't ever get non-trivial contribution. Considering conservation with respect to other spins also will not contribute. So we conclude, non-conservation does not affect correlation functions within the triangle inequality.

\subsection{Outside the triangle inequality}
Let us first consider simplest example $\langle J_s O O\rangle$.
{\bf{\subsection*{$\langle J_s O O \rangle $} }}
In this case it is easy to show that non-conservation of higher spin does not effect the correlation function. For the case of correlators with scalar operators we are going to use \eqref{noncs0}. Using \eqref{noncs0}, if we compute
 $\langle\partial.J_s J_0 J_0\rangle$,  we immediately see that\footnote{Here we present the results schematically. For the argument exact details are not essential.}
\begin{align}
    \langle \partial_1.J_s(x_1) J_0(x_2)J_0(x_3)\rangle &= \sum_{m = 0}^p c_m\langle \partial^m_1 J_{s_1}(x_1) \partial^{p-m}_1 J_0(x_1) J_0(x_2) J_0(x_3)\rangle\nonumber\\
    &= \sum_{m = 0}^{p}\partial^m_1\langle J_{s_1}(x_1) J_0(x_2)\rangle  \partial^{p-m}_1\langle J_0(x_1) J_0(x_3)\rangle + (2\leftrightarrow 3)\nonumber\\
    &=0.
\end{align}
In the second equality, we have used the factorization at leading order in large $N$. Since, at leading order in large $N$, the slightly broken higher spin symmetry leaves  the ward identity is unchanged and hence, the correlator also remains unchanged.
    \subsection*{$\langle J_{s_1} J_{s_2} \widetilde{J}_{0} \rangle_{QF}$ with $s_1 \neq s_2$} 
In this case it is easy to show that non-conservation of higher spin can lead to non-trivial contribution. For the case of correlators with scalar operators we are going to use \eqref{noncs0f}. Let us for simplicity use $s_1>s_2$. Also note that, since for $s_1+s_2=odd,$ the correlator $\langle J_{s_1} J_{s_2} \widetilde{J}_{0} \rangle =0,$ we do require $s_1-s_2=2n$ that is even integers. Using \eqref{noncs0f} we immediately see that
\begin{align}
    \langle \partial_1.J_{s_1}(x_1) J_{s_2}(x_2)\widetilde{J}_{0}(x_3)\rangle &= \sum_{m = 0}^{s_1-s_2-1} c_m\langle \partial^m_1 J_{s_2}(x_1) \partial^{p-m}_1 \widetilde{J}_{0}(x_1) J_{s_2}(x_2) \widetilde{J}_{0}(x_3)\rangle\nonumber\\
    &= \sum_{m = 0}^{s_1-s_2-1}\partial^m_1\langle J_{s_2}(x_1) J_{s_2}(x_2)\rangle  \partial^{p-m}_1\langle \widetilde{J}_{0}(x_1) \widetilde{J}_{0}(x_3)\rangle\nonumber\\
    &\neq 0. \label{ncqf}
\end{align}
Here we have used with out loss of generality\footnote{If we instead take $s_2>s_1$ then non conservation of $s_1$ will not involve $s_2$.} that  $s_1> s_2$ which implies that both $s_2$ and spin zero operator can be present in RHS of \eqref{noncs0f}. We have also again used the factorization at leading order in large $N$.  This shows that outside the triangle inequality the non-conservation of currents due to the slightly broken higher spin symmetry leads to a contribution to the ward identity. This will give rise to non vanishing parity-even non-homogeneous contribution to the correlation function. Let us note that, for the case for free fermion $\langle J_{s_1} J_{s_2} O \rangle_f$ is parity-odd naturally for free theory. The non-conservation then contributes to non-homogeneous parity-even part to the correlation function. This parity-even part of the correlator is identical to critical bosonic theory results as will be shown in the next section, is consistent with duality. 
\bigskip
 \subsection*{Example: $\langle J_{3} J_{1} \widetilde{J}_{0} \rangle_{QF}$} 
It is easy to write down the covariant\footnote{Let us note that \eqref{noncs0f} is given only in light cone coordinates.} non conservation of $J_3$  in the QF theory and is given by
\begin{equation}
     z_{1\nu}z_{1\rho}\partial_\mu J_{3QF}^{\mu \nu \rho}=z_{1\nu}z_{1\rho}(3 J_1^{\nu}(\partial^{\rho} \widetilde{J}_{0})-2(\partial^{\nu} J_1^{\rho})\widetilde{J}_{0})
\end{equation}
where we have used transverse polarization vectors such that $z_i^2=0$.
Thus we have,
\begin{align}
    &z_{1\nu}z_{1\rho}z_{2\alpha}\partial_\mu \langle J_3^{\mu \nu \rho}(x_1) J_1^\alpha(x_2) O_2(x_3) \rangle_{QF}\notag\\=&z_{1\nu}z_{1\rho}z_{2\alpha}\bigg(3 \langle J_1^{\nu}(x_1)\partial^{\rho} \widetilde{J}_{0}(x_1)J_1^\alpha(x_2)\widetilde{J}_{0}(x_3)\rangle-2\langle\partial^{\nu} J_1^{\rho}(x_1)\widetilde{J}_{0}(x_1)J_1^\alpha(x_2)\widetilde{J}_{0}(x_3)\rangle\bigg)
\end{align}
In momentum space this equals,
\begin{equation}
     z_{1\nu}z_{1\rho}z_{2\alpha}\bigg(3 k_3^\rho \langle J_1^{\nu}(-k_2)J_1^\alpha(k_2)\rangle\langle \widetilde{J}_{0}(-k_3)\widetilde{J}_{0}(k_3)\rangle-2k_2^\nu\langle J_1^{\rho}(-k_2)J_1^\alpha(k_2)\rangle\langle \widetilde{J}_{0}(-k_3)\widetilde{J}_{0}(k_3)\rangle\bigg).
\end{equation}
Evaluating this explicitly using \small$\langle J_1^\mu(p) J_1^\nu(-p)\rangle=p \pi^{\mu\nu}(p)$ and $\langle \widetilde{J}_{0}(p) \widetilde{J}_{0}(-p)\rangle=p$\normalsize, we get
\begin{equation}\label{cb1}
     \langle k_1\cdot J_3(k_1,z_1)J_1(k_2,z_2) O_2(k_3)\rangle_{QF}\propto k_2k_3(z_1\cdot k_2)(z_1\cdot z_2).
\end{equation}
RHS of the above equation is precisely the WT identity for critical bosonic theory. This is exactly what we obtained by Legendre transforming free bosonic WT in \eqref{WTcb}. 

 \subsection*{$\langle J_{s_1} J_{s_2} J_{0} \rangle_{QB}$ with $s_1 \neq s_2$}  
For quasi-bosonic theory argument is exactly same as quasi-fermionic theory. 
 For this case we have
\begin{align}
    \langle \partial_1.J_s(x_1) J_{s_2}(x_2)J_0(x_3)\rangle &= \sum_{m = 0}^{s_1-s_2} \epsilon_{\mu\nu-}[b_n\langle \partial^{m-1}_1\partial^{\nu} J^{\mu}_{s_2}(x_1)J_{s_2}(x_2)\rangle \partial^{p-m}_1 \langle J_0(x_1) J_0(x_3) \rangle\notag\\& + f_n\partial^{m}_1\langle  J^{\mu}_{s_2}(x_1)J_{s_2}(x_2)\rangle \partial^{p-m-1}_1 \partial^{\nu}\langle J_0(x_1) J_0(x_3) \rangle\label{ncqb}
\end{align}
similar to the fermionic case, there is a nontrivial contribution at leading order in large $N$. 

When $s_1 \neq s_2$, for the free bosonic theory, the correlator is parity-even.  We find that the RHS of (\ref{ncqb}) has a epsilon term and hence if we consider the parity-even contribution from the two-point function $\langle J J \rangle,$ this will contribute to parity-odd WT identity. This in turn will give non-homogeneous parity-odd correlation function. 
\bigskip
 \subsection*{Example: $\langle J_{3} J_{1} J_{0} \rangle_{QB}$} 
The covariant non conservation of $J_3$ in the QB  theory can be shown to be given by
\begin{align}
    z_{1\nu}z_{1\rho}\partial_\mu J_{3~QB}^{\mu \nu \rho}=2\,z_{1\rho}\epsilon^{a b z_1}\bigg(-4\partial_b J_{1a}\partial^\rho J_{0}+(\partial^\rho \partial_b J_{1a})J_{0}+\frac{3}{2} J_{1a}\partial_b\partial^\rho J_{0}-\partial^\rho J_{1a}\partial_b J_{0}\bigg)
\end{align}
Thus we have,
\begin{align}
    &z_{1\nu}z_{1\rho}z_{2\alpha}\partial_\mu \langle J_3^{\mu \nu \rho}(x_1)J_1^\alpha(x_2) J_{0}(x_3)\rangle_{QB}\notag\\=&2\epsilon^{a b z_1}z_{1\rho}z_{2\alpha}\bigg(-4\langle\partial_b J_{1a}(x_1)\partial^\rho J_{0}(x_1)J_1^\alpha(x_2)J_{0}(x_3)\rangle+\langle\partial^\rho \partial_b J_{1a}(x_1)J_{0}(x_1)J_1^\alpha(x_2)J_{0}(x_3)\rangle\notag\\+&\frac{3}{2} \langle J_{1a}(x_1)\partial_b\partial^\rho J_{0}(x_1)J_1^\alpha(x_2)J_{0}(x_3)\rangle-\langle \partial^\rho J_{1a}(x_1)\partial_b J_{0}(x_1)J_1^\alpha(x_2)J_{0}(x_3)\rangle\bigg)
\end{align}
Using large $N$ factorization and converting the result into momentum space we get,
\begin{align}
       &\langle k_1\cdot J_3(k_1,z_1)J_1(k_2,z_2) J_{0}(k_3)\rangle_{QB}\notag\\=&-2\epsilon^{a b z_1}z_{1\rho}z_{2\alpha}\bigg(-4k_{2b}k_3^\rho\langle J_{1a}(-k_2)J_1^\alpha(k_2)\rangle\langle J_{0}(-k_3)J_{0}(k_3)\rangle+k_2^\rho k_{2b}\langle J_{1a}(-k_2)J_1^\alpha(k_2)\rangle\langle J_{0}(-k_3)J_{0}(k_3)\rangle\notag\\&+\frac{3}{2}k_3^\rho k_{3b} \langle J_{1a}(-k_2)J_1^\alpha(k_2)\rangle\langle J_{0}(-k_3)J_{0}(k_3)\rangle-k_2^\rho k_{3b}\langle  J_{1a}(-k_2)J_1^\alpha(k_2)\rangle\langle J_{0}(-k_3)J_{0}(k_3)\rangle\bigg)
\end{align}
Evaluating this explicitly  using $\langle J_1^\mu(p) J_1^\nu(-p)\rangle=p\pi^{\mu\nu}(p)$ and $\langle O(p) O(-p)\rangle=\frac{1}{p}$ yields,
\begin{align}
       \langle k_1\cdot J_3(k_1,z_1)J_1(k_2,z_2) O(k_3)\rangle_{QB}\propto \frac{k_2(z_1\cdot k_2)}{k_3}\left(2\epsilon^{z_1 z_2 k_2}-\epsilon^{z_1 z_2 k_3}\right)
\end{align}
So we see that for QB theory non-conservation indeed contributes and it will lead to parity odd result. This parity odd result can be identified to be given by critical fermionic theory in 3d.

\subsection*{$\langle J_{s_1} J_{s_2} J_{s_3} \rangle$} 
The contribution to the correlator outside the triangle inequality due to the slightly broken higher spin symmetry is most manifest in the case of $\langle J_{s_1}J_{s_2}J_{s_3}\rangle$. For this case, we would require non-conservation expression
in \eqref{noncss1s2}. Now, we look at  $\langle\partial\cdot J_{s_1} J_{s_2}J_{s_3}\rangle$ which leads to
\begin{align}
   & \langle\partial_1.J_{s_1}(x_1)J_{s_2}(x_2)J_{s_3}(x_3)\rangle = \sum_{m=0}^{s_1-s_2-s_3}\epsilon_{\mu\nu-}[a_m \partial^m_{1}\langle J^{\mu}_{s_2}(x_1) J_{s_2}(x_2)\rangle\partial^{p-m}_{1}\langle J^{\nu}_{s_3}(x_1) J_{s_3}(x_3)\rangle\notag\\&+b_m\partial^{m-1}_{1}\partial^{\nu}_{1}\langle J^{\mu}_{s_2}(x_1) J_{s_2}(x_2)\rangle\partial^{p-m}_{1}\langle J_{s_3}(x_1) J_{s_3}(x_3)\rangle+e_m\partial^{m}_{1}\langle J^{\mu}_{s_2}(x_1) J_{s_2}(x_2)\rangle\partial^{p-m-1}_{1}\partial^{\mu}_1\langle J^{\nu}_{s_3}(x_1) J_{s_3}(x_3)\rangle ]
\end{align}
Notice that RHS of above equation is non zero\footnote{Notice how the sum vanishes for $s_1-s_2-s_3 < 0$ i.e. when the correlator obeys triangle inequality. This is consistent with what we have discussed in the previous subsection.} only when spins violate the triangle inequality $s_1>s_2+s_3$. 
 We see that due to presence of epsilon in the RHS of above equation we get a parity-odd contribution to the correlation function.

\subsection*{Example: $\langle J_{4} J_{1} J_{1} \rangle$} 
The covariant non conservation of $J_4$ in $\langle J_4 J_1 J_1\rangle$ is given by,
\begin{align}
  &z_{1\mu}z_{1\nu}z_{1\rho}\partial^\sigma J_{4\sigma}^{\mu\nu\rho}(x)\notag\\&=\epsilon^{a b z_1}z_{1\nu}z_{1\rho} \bigg(4\partial^\nu J_{1a}(x) \partial^\rho J_{1b}(x)-3 \partial^\nu\partial^\rho J_{1a}(x) J_{1b}(x)
     -24 \partial_b J_{1a}(x) \partial^\nu J_1^\rho(x) +18 \partial^\nu\partial_b J_{1a}(x) J_1^\rho(x)\bigg)
\end{align}
The quantity of interest is,
\small
\begin{align}
  &\langle\partial^\sigma J_{4\sigma}(x_1,z_1)J_1(x_2,z_2) J_1(x_3,z_3)\rangle\notag\\&=\epsilon^{a b z_1}z_{1\nu}z_{1\rho}z_{2\alpha}z_{3\beta} \bigg(4\langle\partial^\nu J_{1a}(x_1) \partial^\rho J_{1b}(x_1)J_1^\alpha(x_2) J_1^\beta(x_3)\rangle- 3 \langle\partial^\nu\partial^\rho J_{1a}(x_1) J_{1b}(x_1)J_1^\alpha(x_2) J_1^\beta(x_3)\rangle
     \notag\\&-24\langle \partial_b J_{1a}(x_1) \partial^\nu J_1^\rho(x_1)J_1^\alpha (x_2)J_1^\beta(x_3)\rangle  +18 \langle\partial^\nu\partial_b J_{1a}(x_1) J_1^\rho(x_1) J_1^\alpha(x_2) J_1^\beta(x_3)\rangle\bigg)
\end{align}
\normalsize
Using large-N factorization and  and going to momentum space we get the following result
\small
\begin{align}\label{411wta}
     &\langle k_1\cdot J_4(k_1,z_1)J_1(k_2,z_2)J_1(k_3,z_3)\rangle_{NC}\notag\\&=7(z_1\cdot k_2)\bigg((z_1\cdot k_2)\epsilon^{a b z_1}\langle J_{1a}(-k_2) J_1(k_2,z_2)\rangle\langle J_{1b}(-k_3) J_1(k_3,z_3)\rangle -6\epsilon^{a k_2 z_1}z_{1\rho}\langle J_{1a}(-k_2) J_1(k_2,z_2)\rangle\langle J_1^\rho(-k_3)J_1(k_3,z_3)\rangle \notag\\&+(z_1\cdot k_2)\epsilon^{a b z_1}\langle J_{1a}(-k_3) J_1(k_3,z_3)\rangle\langle J_{1b}(-k_2) J_1(k_2,z_2)\rangle  +6\epsilon^{a k_3 z_1}z_{1\rho}\langle  J_{1a}(-k_3) J_1(k_3,z_3)\rangle\langle J_1^\rho(-k_2)J_1(k_2,z_2)\rangle\bigg)
\end{align}
\normalsize
Using $\langle J_1^\mu(p) J_1^\nu(-p)\rangle=p\pi^{\mu\nu}(p)$ we get the following result
\begin{equation}\label{411nc1} 
\langle k_1\cdot J_4(k_1,z_1)J_1(k_2,z_2)J_1(k_3,z_3)\rangle_{NC}\propto (z_1\cdot k_2)k_2 k_3\left( (z_1\cdot z_2)\epsilon^{z_1 z_3 k_3}-(z_1\cdot z_3)\epsilon^{z_1 z_2 k_2}\right)\,.
\end{equation}
Thus we observe that non conservation indeed contributes and will give rise to parity odd non-homogeneous term for both QB and QF theory.
\subsection{Derivation of non-conservation WT identity from WT identity of exactly conserved current}
The aim of this section is to discuss a remarkable relation between the WT identity for exactly conserved currents and the WT identity obtained for non-conservation discussed in the previous section. We illustrate our point by giving a few explicit examples.
\subsection*{Map between free boson and critical boson WT identities in momentum space}
Higher spin currents in the critical bosonic theory are not conserved \cite{Giombi:2016zwa}. In this section we show that the WT identity obtained from the non-conservation of currents in the critical bosonic theory can be obtained from the free bosonic theory by a simple Legendre transform. To illustrate this fact, let us consider   $\langle J_3 J_1 O\rangle$. In \eqref{WTcb} we calculated the Legendre transform of this WT identity which precisely matches with WT identity derived using non-conservation of $J_3$ current in \eqref{cb1}. It is easy to check that this relation holds for generic $\langle J_{s_1} J_{s_2} O\rangle$.
\subsection*{Map between free fermion and critical boson WT identity in momentum space}
To illustrate the map between free fermion and critical boson WT identities we again consider the example of $\langle J_3 J_1 O\rangle$. The free fermion $J_3$ WT Identity is given by \eqref{WTFf1} and the critical bosonic theory WT identity is given by \eqref{WTcb}. We define an epsilon transform for the free fermionic WT identity as
\begin{align}
   \frac{1}{k_2}\epsilon^{\alpha z_2 k_2}\langle k_1\cdot J_3(k_1,z_1) J_1^\alpha(k_2) O(k_3) \rangle_{FF} 
\end{align}
which gives
\begin{equation}
    (z_1\cdot k_2)(2 (z_1\cdot k_2)(z_2\cdot k_1)+(z_1\cdot z_2)(k_1^2-5k_2^2+6k_2k_3-k_3^2))\,.
\end{equation}
This is precisely what we obtained in \eqref{WTcb} and is same as that of the critical bosonic theory in \eqref{cb1} up to contributions which leads to contact terms as discussed earlier. Thus we have 
\begin{align}\label{rlffcbwt}
 \frac{1}{k_2}\epsilon^{\alpha z_2 k_2}\langle k_1\cdot J_3(k_1,z_1) J_1^\alpha(k_2) O(k_3) \rangle_{FF} =   \langle k_1\cdot J_3(k_1,z_1) J_1(k_2,z_2) O(k_3) \rangle_{CB}
\end{align}
 This discussion can be generalised to $\langle J_{s_1} J_{s_2} O\rangle$.
We conclude that for correlation functions of the form $\langle J_{s_1} J_{s_2} O\rangle$ the free fermion and the critical bosonic WT identity are related by an epsilon transformation. 

\begin{figure}[H]
\centering
  \begin{tikzpicture}
\node (A) at (0,6) {$\langle J_{s_1} J_{s_2} O\rangle_{CF}$};
\node (B) at (6, 6) {$\langle J_{s_1} J_{s_2} O\rangle_{FF}$};
\node (C) at (6, 0) {$\langle J_{s_1} J_{s_2} O\rangle_{CB}$};
\node (D) at (0, 0) {$\langle J_{s_1} J_{s_2} O\rangle_{FB}$};
\draw [<->] (B) -- node [midway,above] {L.T} (A);
\draw [<->] (B) -- node [midway,right] {E.T} (C);
\draw [<->] (D) -- node [midway,below] {L.T} (C);
\draw [<->] (D) -- node [midway,left] {E.T} (A);

\end{tikzpicture}
\caption{\label{ccbfig} A diagram showing how $\langle J_{s_1} J_{s_2} O\rangle$ in different theories are related. Here E.T. implies the epsilon transform and L.T. means Legendre transform.}
\end{figure}
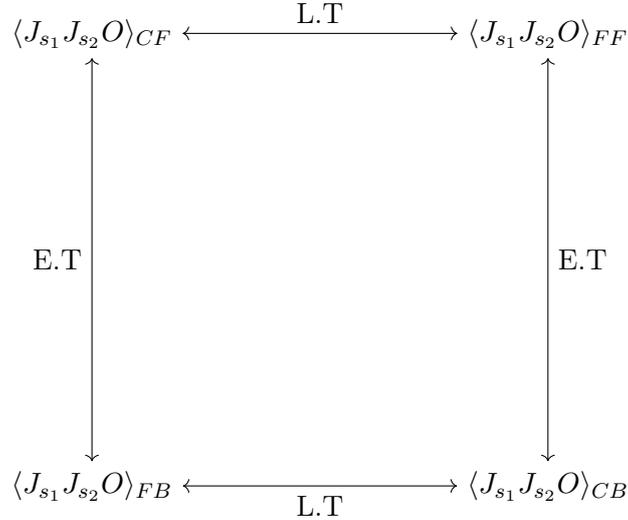
To summarize, let us take WT identity of  free bosonic theory, see \eqref{ffj3j1o},\eqref{WTfb}
\begin{align}
ik_{1\sigma}\langle J^\sigma_{\mu\nu}(k_1)J_{\rho}(k_2)O(k_3)\rangle
&=k_{3\mu}k_{3\nu}k_{3\rho}\langle O(k_3)O(-k_3)\rangle-k_3^2g_{\rho(\mu}k_{3\nu)}\langle O(k_3)O(-k_3)\rangle\cr
&\hspace{.5cm}+k_{2(\mu}\langle J_{\rho}(k_2)J_{\nu)}(-k_2)\rangle
\end{align}
Under Legendre transform the  first line of last equation becomes a contact term and second line roughly gives WT identity proportional to $\langle JJ\rangle \langle OO \rangle$ which matches with the WT identity for critical bosonic theory. Instead if we do epsilon transform, the last line becomes a contact term as epsilon transform of $\langle JJ\rangle_{even}$ gives $\langle JJ\rangle_{odd}$  which is contact term and the first line gives precisely the WT identity for critical fermionic theory. This statement can be explicitly checked in position space as well by appropriate fourier transform. We have summarized this relation in fig \ref{ccbfig}.

\noindent {\bf{Example: $\langle J_4 J_1 J_1\rangle$}}\\

\noindent For $\langle J_4 J_1 J_1\rangle$, the non-conservation of $J_4$ current is discussed in \eqref{411nc1}. One can directly solve this WT identity to calculate the parity-odd contribution to $\langle J_4 J_1 J_1\rangle$ due to non-conservation of $J_4$. However, below we make an interesting observation which helps us to solve the parity-odd correlation function easily. We will show that the RHS of \eqref{411nc1} can be  found from the WT identity in the free theory.
For this let us calculate the difference between the Boson and Fermion WT identities using \eqref{WIEx2b} and \eqref{WIEx2f}. We have,
\begin{align}
     &k_{1\sigma}\langle J^\sigma_4(k_1,z_1)J_1(k_2,z_2)J_1(k_3,z_3)\rangle_{FB-FF}\nonumber\\
     &=\frac{-21}{512}(z_1\cdot k_2)(k_2-k_3) \Bigg(2 (z_1\cdot k_2)^2 (z_2\cdot z_3)+(z_1\cdot k_2) (2 (z_3\cdot k_1) (z_1\cdot z_2)\notag\\&~~~-2 (z_2\cdot k_1) (z_1\cdot z_3))+((k_2-k_3)^2-k_1^2)(z_1\cdot z_2) (z_1\cdot z_3)\Bigg).
\end{align}
Now we apply an epsilon transformation to the above quantity with respect to one of the $J_1$'s and then symmetrize the result under $2\leftrightarrow 3$ exchange. The result is,
\begin{align}
    &\frac{1}{k_2} \epsilon^{\alpha z_2 k_2} k_{1\sigma}\langle J^\sigma_4(k_1,z_1)J_{1\alpha}(k_2)J_1(k_3,z_3)\rangle_{FB-FF}+2\leftrightarrow 3 \nonumber\\
    &=\frac{21}{1024 k_2 k_3}(z_1\cdot k_2)(k_2-k_3)(-((k_1^2-(k_2-k_3)^2)((z_1\cdot z_3)k_3\epsilon^{z_1 z_2 k_2}+(z_1\cdot z_2) k_2 \epsilon^{z_1 z_3 k_3}))\notag\\&+(z_1\cdot k_2)^2(-2k_3\epsilon^{z_2 z_3 k_2}+2 k_2 \epsilon^{z_2 z_3 k_3})+2(z_1\cdot k_2)(k_3((z_3\cdot k_1)\epsilon^{z_1 z_2 k_2}+(z_1\cdot z_3)\epsilon^{z_2 k_1 k_2})\nonumber\\
    &-k_2(z_2\cdot k_1)\epsilon^{z_1 z_3 k_3}+z_1\cdot z_2\epsilon^{z_3 k_1 k_3})))
\end{align}
The RHS appears quite complicated at first glance, but one can show\footnote{This is verified by going to Spinor Helicity variables and checking that both sides of the expression are identical in all helicities.
} that this equals,
\begin{equation}\label{epwt}
    \frac{21}{256}(z_1\cdot k_2)k_2 k_3\left( (z_1\cdot z_2)\epsilon^{z_1 z_3 k_3}-(z_1\cdot z_3)\epsilon^{z_1 z_2 k_2}\right)+\text{Contact Terms}
\end{equation}
where the contact terms appearing in the above expression are given by,
\begin{align}
    \text{Contact Terms}=\frac{21}{1024}(z_1\cdot k_2)((z_1\cdot z_3)(-k_1^2+k_2^2+3 k_3^2)\epsilon^{z_1 z_2 k_2}+(z_1\cdot z_2)(k_1^2-3k_2^2-k_3^2)\epsilon^{z_1 z_3 k_3}\notag\\
    +2(z_1\cdot k_2)^2\epsilon^{z_2 z_3 k_1}+2(z_1\cdot k_2)\Bigg((z_3\cdot k_1)\epsilon^{z_1 z_2 k_2}+(z_2\cdot k_1)\epsilon^{z_1 z_3 k_3}+(z_1\cdot z_3)\epsilon^{z_2 k_1 k_2}+(z_1\cdot z_2)\epsilon^{z_3 k_1 k_3})\Bigg)
\end{align}
where these terms of the WT identity leads to contact term in correlation function as we have discussed in section \ref{smlcon}.
We observe that removing the contact terms from \eqref{epwt} gives the WT identity obtained in \eqref{411nc1} using non conservation of $J_4$ current. Thus we have 
\begin{equation}\label{411rloe}
    \langle k_1\cdot J_4(k_1,z_1)J_1(k_2,z_2)J_1(k_3,z_3)\rangle_{NC,odd}=  \frac{1}{k_2} \epsilon^{\alpha z_2 k_2}\langle k_1\cdot J_4(k_1,z_1)J_{1\alpha}(k_2)J_1(k_3,z_3)\rangle_{FB-FF}+2\leftrightarrow 3
\end{equation}
The above relation can be easily generalised to  general correlators. Let us consider most general correlation function of the form $\langle J_{s_1}  J_{s_2} J_{s_3}\rangle$. Assuming $s_1>s_2+s_3$ it is easy to show that parity-odd $\langle J_{s_1}  J_{s_2} J_{s_3}\rangle$ gets contribution only from non-conservation of $s_1$. With few more examples, it should be easy to generalize \eqref{411rloe} to
\begin{align}
    &\langle k_1\cdot J_{s_1}(k_1,z_1)J_{s_2}(k_2,z_2)J_{s_3}(k_3,z_3)\rangle_{NC,odd}\nonumber\\
    &={\rm{ Epsilon ~transform:}}~\langle k_1\cdot J_{s_1}(k_1,z_1)J_{s_2}(k_2,z_2)J_{s_3}(k_3,z_3)\rangle_{FB-FF}+2\leftrightarrow 3\,.
    \end{align}

\subsubsection*{In spinor-helicity variables}
{\bf{Example: $\langle J_3 J_1 O\rangle$}}\\
Converting \eqref{WTcb} and \eqref{WTFf1} into spinor helicity variables we notice that,
\begin{align}
    \langle k_1\cdot J_3^{-} J_1^{-} O\rangle_{CB}&=i\langle k_1\cdot J_3^{-} J_1^{-} O\rangle_{FF}
\notag\\
    \langle k_1\cdot  J_3^{-} J_1^{+} O\rangle_{CB}&=-i\langle k_1\cdot  J_3^{-} J_1^{+} O\rangle_{FF}
\end{align}
The relations in the remaining helicities are obtained by complex conjugating the given results.
\\

\noindent {\bf{Example: $\langle J_4 J_1 J_1\rangle$}}\\
Taking \eqref{411nc1} and \eqref{WIEx2b}-\eqref{WIEx2f} and converting them into spinor helicity variables, we obtain the following results:
\begin{align}
     \langle k_1\cdot  J_4^{-} J_1^{-} J_1^{-}\rangle_{NC,odd}&=i\langle k_1\cdot  J_4^{-} J_1^{-} J_1^{-}\rangle_{FB-FF}\notag\\
          \langle k_1\cdot  J_4^{+} J_1^{-} J_1^{-}\rangle_{NC,odd}&=i\langle k_1\cdot  J_4^{+} J_1^{-} J_1^{-}\rangle_{FB-FF}
\end{align}
All other non-zero helicity configurations are either obtained by complex conjugates and/or by  a $2\leftrightarrow 3$ exchange of the listed ones are zero. Also as we remarked before, these relations are true upto terms in WT identity which contributes to contact terms.

\subsection*{Implication of this relation in QF and QB theory}

\subsection*{When triangle inequality is satisfied}
Let us consider correlators of the kind $\langle J_{s}J_{s}O\rangle$. For these correlators one has the following WT identity
\begin{align}
k_{1}\cdot \langle J_{s}J_{s}O\rangle=0
\end{align}
Therefore the only contribution to such correlators is homogeneous. Now let us consider correlators of the kind $\langle J_{s_1}J_{s_2}J_{s_3}\rangle$. These correlators are given by
\begin{align}
\langle J_{s_1}J_{s_2}J_{s_3}\rangle=n_b\langle J_{s_1}J_{s_2}J_{s_3}\rangle_{FB}+n_f\langle J_{s_1}J_{s_2}J_{s_3}\rangle_{FF}+n_{odd}\langle J_{s_1}J_{s_2}J_{s_3}\rangle_{odd}
\end{align}
where
\begin{align}
    n_b=\tilde N\frac{\tilde\lambda^2}{1+\tilde\lambda^2},\quad n_f=\tilde N\frac{1}{1+\tilde\lambda^2},\quad n_{odd}=\tilde N\frac{\tilde\lambda}{1+\tilde\lambda^2}
\end{align}
where $\tilde N=2N \frac{\sin \pi\lambda}{\pi\lambda}$ and $\tilde\lambda=\tan\frac{\pi\lambda}{2}$, where these relation are true for both bosonic and fermion theory.

The odd parts of these correlators satisfy a trivial Ward identity 
\begin{align}
k_{1}\cdot \langle J_{s_1}J_{s_2}J_{s_3}\rangle_{odd}=0
\end{align}
The fermionic and bosonic parts satisfy the same Ward identity 
\begin{align}\label{wtfbff}
k_{1}\cdot \langle J_{s_1}J_{s_2}J_{s_3}\rangle_{FB}=k_{1}\cdot \langle J_{s_1}J_{s_2}J_{s_3}\rangle_{FF}
\end{align}
We also observe this fact by looking into the higher spin algebra, see \eqref{WTfbffsm}. One can see that the higher spin algebra for free boson and free fermion is exactly identical when spins satisfy triangle inequality.
The WT identity satisfied $\langle J_{s_1}J_{s_2}J_{s_3}\rangle$ is given by
\begin{align}
k_{1}\cdot \langle J_{s_1}J_{s_2}J_{s_3}\rangle &=\tilde N k_{1}\cdot \langle J_{s_1}J_{s_2}J_{s_3}\rangle_{FB}\nonumber\\
&=\tilde N  k_{1}\cdot \langle J_{s_1}J_{s_2}J_{s_3}\rangle_{FF}
\end{align}

\subsection*{When triangle inequality is violated}
Let us consider correlators of the kind $\langle J_{s_1}J_{s_2}O\rangle$. In the  quasi-fermionic theories these correlators are given by
\begin{align}
\langle J_{s_1}J_{s_2}O\rangle_{QF}=n_{f}\langle J_{s_1}J_{s_2}O\rangle_{FF}+n_{odd}\langle J_{s_1}J_{s_2}O\rangle_{odd}
\end{align}
where \footnote{Here we differ from the convention in \cite{Maldacena:2011jn} by absorbing a factor of $1+\tilde\lambda^2$ in the definition of the scalar operator in the interacting theory.}
\begin{align}
    n_{f}=\tilde N,\quad n_{odd}=\tilde N\tilde\lambda
\end{align}
In spinor-helicity variables we saw that 
\begin{align}
    k_1\cdot\langle J_3JO\rangle_{odd}=i\,k_1\cdot \langle J_3JO\rangle_{FF}
\end{align}
This gives the following WT identity for the correlator in spinor-helicity variables
\begin{align}
\label{any1}
    k_1\cdot\langle J_3JO\rangle_{QF}&=\tilde N(1+i\tilde\lambda)k_1\cdot\langle J_3JO\rangle_{FF}\cr
    &=\frac{2i(1-e^{i\pi\lambda})N}{\pi\lambda}k_1\cdot\langle J_3JO\rangle_{FF}
\end{align}
%
Let us now consider correlators of the kind $\langle J_{s_1}J_{s_2}J_{s_3}\rangle$. 
In this case unlike \eqref{wtfbff}
\begin{align}
    k_1\cdot \langle J_{s_1}J_{s_2}J_{s_3}\rangle_{FB}\neq  k_1\cdot \langle J_{s_1}J_{s_2}J_{s_3}\rangle_{FF}
\end{align}
This fact can also be understood by looking into the higher spin algebra, see  \eqref{WTfbffdif}. One can see that the higher spin algebra for free boson and free fermion is different when spins violate triangle inequality.
These correlators are given by
\begin{align}
\langle J_{s_1}J_{s_2}J_{s_3}\rangle&=\frac{\tilde N\tilde\lambda^2}{1+\tilde\lambda^2}\langle J_{s_1}J_{s_2}J_{s_3}\rangle_{FB}+\frac{\tilde N}{1+\tilde\lambda^2}\langle J_{s_1}J_{s_2}J_{s_3}\rangle_{FF}+\frac{\tilde N\tilde\lambda}{1+\tilde\lambda^2}\langle J_{s_1}J_{s_2}J_{s_3}\rangle_{odd}\cr
&=\frac{\tilde N}{2}(\langle J_{s_1}J_{s_2}J_{s_3}\rangle_{FB}+J_{s_1}J_{s_2}J_{s_3}\rangle_{FF})-\frac{\tilde N}{2}\frac{1-\tilde\lambda^2}{1+\tilde\lambda^2}\left(\langle J_{s_1}J_{s_2}J_{s_3}\rangle_{FB}-J_{s_1}J_{s_2}J_{s_3}\rangle_{FF})\right)\cr
&+\frac{\tilde N\tilde\lambda}{1+\tilde\lambda^2}\langle J_{s_1}J_{s_2}J_{s_3}\rangle_{odd}
\end{align}
We saw in spinor-helicity variables that  in all minus helicity
\begin{align}
    k_1\cdot\langle J_{s_1}^{-}J_{s_2}^{-}J_{s_3}^{-}\rangle_{odd}=i\,k_1\cdot (\langle J_{s_1}^{-}J_{s_2}^{-}J_{s_3}^{-}\rangle_{FB}-\langle J_{s_1}^{-}J_{s_2}^{-}J_{s_3}^{-}\rangle_{FF})
\end{align}
Other helicity can also be written appropriately. This helps us write the WT identity satisfied by the correlator in the following form
\begin{align}
\label{any2}
&k_{1}\cdot \langle J_{s_1}J_{s_2}J_{s_3}\rangle\\
&=\frac{\tilde N}{2}k_{1}\cdot \left(\langle J_{s_1}J_{s_2}J_{s_3}\rangle_{FB}+\langle J_{s_1}J_{s_2}J_{s_3}\rangle_{FF}\right)
+\frac{\tilde N}{2}\frac{\tilde\lambda+i}{\tilde\lambda-i}k_{1}\cdot \left(\langle J_{s_1}J_{s_2}J_{s_3}\rangle_{FB}-\langle J_{s_1}J_{s_2}J_{s_3}\rangle_{FF}\right)\nonumber\\
&=\frac{\tilde N}{2}k_1\cdot\langle J_{s_1}J_{s_2}J_{s_3}\rangle_{FB+FF}
-\frac{\tilde N}{2}e^{-i\pi\lambda}k_1\cdot\langle J_{s_1}J_{s_2}J_{s_3}\rangle_{FB-FF}.
\end{align}
where by $FB+FF$ and $FB-FF$ denote the sum and difference of $\langle J_{s_1}J_{s_2}J_{s_3}\rangle$ in the free boson and free fermion theories.

\subsection{Parity odd CFT correlator outside the triangle inequality}
We have already shown that outside the triangle inequality, the CFT correlators are non-homogeneous and we can not have any homogeneous contribution. Non-homogeneous piece is purely determined in terms of the WT identity. In the previous section, we have mapped the WT identity obtained from non-conservation of higher spin currents to that of WT identity obtained from exactly conserved currents. This implies that we should be able to obtain the correlation function involving weakly broken higher spin symmetry interms of the correlation function of exactly conserved currents. 
\subsubsection{In momentum space}
For example, \eqref{rlffcbwt} implies
\begin{equation}
    \langle J_3(k_1,z_1) J_1(k_2,z_2) O(k_3)\rangle_{CB}=\frac{1}{k_2}\epsilon^{\alpha z_2 k_2} \langle  J_3(k_1,z_1) J_1^\alpha(k_2) O(k_3)\rangle_{FF}.
\end{equation}
Similarly \eqref{411rloe} implies
\begin{equation}
    \langle J_4(k_1,z_1)J_1(k_2,z_2)J_1(k_3,z_3)\rangle_{NC,odd}=  \frac{1}{k_2} \epsilon^{\alpha z_2 k_2}\langle J_4(k_1,z_1)J_{1\alpha}(k_2)J_1(k_3,z_3)\rangle_{FB-FF}+2\leftrightarrow 3
\end{equation}
This relation should easily generalise to correlators involving  arbitrary spin. Notice that in \cite{Jain:2021gwa}, it was shown that the same relation to hold for homogeneous parity-even and parity-odd contribution 
 \begin{align}
      \langle J_{s_1}J_{s_2}J_{s_3} \rangle_{\text{odd}} =  \frac{1}{k_1}\epsilon_{\alpha}^{\,\,k_1 (\mu_1}\langle J_{s_1}^{\mu_2 \cdots \mu_{s_1})\alpha}J_{s_2}J_{s_3}\rangle_{\text{even},\bf{h}}
 \end{align}

\subsubsection{In spinor-helicity variables}
{\bf{Example: $\langle J_3 J_1 O\rangle$}}\\
\begin{align}
    &\langle J^-_3J^-_1O\rangle_{CB} =i\langle J^-_3J^-_1O\rangle_{FF} = f_{-~-}(k_1, k_2, k_3)\langle 12\rangle^4\langle 31\rangle^2\langle \bar{2}\bar{3}\rangle^2\notag\\
    & \langle J^-_3J^+_1O\rangle_{CB} = -i\langle J^-_3J^+_1O\rangle_{FF} =f_{-+}(k_1, k_2, k_3)\langle 12\rangle^2\langle 31\rangle^4\langle \bar{2}\bar{3}\rangle^4
\end{align}
where
\begin{align}
   &f_{-~-}(k_1, k_2, k_3) =\frac{2 E^3+3 E^2 k_1+4 E k_1^2+4 k_1^3}{128 E^6 k_1^3 }\notag\\
    &f_{-~+}(k_1, k_2, k_3) =\frac{2E+k_1}{E^6 k_1^3 }
\end{align}
Other helicity configurations are obtained via complex conjugation of the listed ones.\\

\noindent {\bf{Example: $\langle J_4 J_1 J_1\rangle$}}\\
\begin{align}
     \langle J_4^{-} J_1^{-} J_1^{-}\rangle_{NC,odd}=i\langle J_4^{-} J_1^{-} J_1^{-}\rangle_{FB-FF}&= f_{-~-~-}(k_1,k_2,k_3)\langle 1 2\rangle^4 \langle 3 1\rangle^4  \langle \bar{2}\bar{3}\rangle^ 2\nonumber\\
  \langle J_4^{+} J_1^{-} J_1^{-}\rangle_{NC,odd}=i\langle J_4^{+} J_1^{-} J_1^{-}\rangle_{FB-FF}&= f_{+~-~-}(k_1,k_2,k_3) \langle 2 3\rangle^6  \langle \bar{1} \bar{2}\rangle^4\langle \bar{3}\bar{1}\rangle^ 4
\end{align}
where,
\begin{align}
    &f_{-~-~-}=i\frac{3 E^5+5 E^4 k_1+8 E^3 k_1^2+12 E^2 k_1^3+16 E k_1^4+16 k_1^5}{524288 E^8 k_1^4}\notag\\
    &f_{+~-~-}=i\frac{3E+k_1}{524288 E^8 k_1^4}
\end{align}
All other non-zero helicity configurations are either obtained by complex conjugates and/or by  a $2\leftrightarrow 3$ exchange of the listed ones are zero. Also as we remarked before, these relations are true upto to contact terms.

\subsection*{Correlation functions in  QF and QB theory}
We make use of the WT identities in \eqref{any1} and \eqref{any2} to express the correlators as 
\begin{align}
   \langle J_{s_1}J_{s_2}O\rangle_{QF}=\frac{2i(1-e^{i\pi\lambda})N}{\pi\lambda}\langle J_{s_1}J_{s_2}O\rangle_{FF}
\end{align}
and
\begin{align}
\langle J_{s_1}J_{s_2}J_{s_3}\rangle
=\frac{\tilde N}{2}\langle J_{s_1}J_{s_2}J_{s_3}\rangle_{FB+FF}
-\frac{\tilde N}{2}e^{-i\pi\lambda}\langle J_{s_1}J_{s_2}J_{s_3}\rangle_{FB-FF}.
\end{align}

\subsubsection{In position space}
In position space the odd part of the correlator can be written in terms of  free theory correlators by relations of the following kind 

\begin{align}\label{pso12}
    \langle J_{\mu\nu\rho\sigma}J_{\alpha}J_{\beta}\rangle_{\text{odd}}=\int\frac{1}{|x-x_1|^2}\partial_{x_{\gamma}}\epsilon_{\gamma b\alpha}(\langle J_{\mu\nu\rho\sigma}J^{b}J_{\beta}\rangle_{\text{FF}}-\langle J_{\mu\nu\rho\sigma}J^{b}J_{\beta}\rangle_{\text{FB}})
\end{align}
This can be directly generalised to correlators of the kind $\langle J_{s_1}J_{s_2}J_{s_3}\rangle$, for a similar discussion on correlators inside the triangle inequality see \cite{Jain:2021gwa}. See also \cite{Maldacena:2012sf} for similar but weaker \footnote{In Appendix D of the reference a non-covariant relation was established between derivatives of $\langle TJ_{s_1}J_{s_2}\rangle_{\text{odd}}$ and those of the free theory correlators using higher spin symmetry. Our approach in this paper does not make use of higher spin algebra.} relations derived using higher spin symmetry.

\section{Three-point function for supersymmetric theory}
\label{susy}
In \cite{Aharony:2019mbc,Inbasekar:2019wdw,Nizami:2013tpa} correlation functions in supersymmetric theory were computed. In particular, it was observed in \cite{Nizami:2013tpa} for some generic superconformal theory and later conjectured in \cite{Aharony:2019mbc} that for ${\mathcal N}=1$ theory only two structures exist which are given by
\begin{align}\label{susy1}
    \langle J_{s_1} J_{s_2} J_{s_3} \rangle= \alpha_{s_1 s_2 s_3}   \langle J_{s_1} J_{s_2} J_{s_3} \rangle_{free}+ \beta_{s_1 s_2 s_3}   \langle J_{s_1} J_{s_2} J_{s_3} \rangle_{odd}
\end{align}
This conjecture was checked for some special case in  \cite{Aharony:2019mbc}. Here we present a simple explanation of the conjecture \eqref{susy1}. 

For supersymmetric theories it is natural to demand that the number of bosonic and fermionic degree of freedom should be same which implies $n_f=n_s$. Using this condition in \eqref{fff1} we get
\begin{equation}\label{susy1}
   \langle J_{s_1} J_{s_2}J_{s_3}\rangle = \left(n_s+n_f\right)  \langle J_{s_1} J_{s_2}J_{s_3}\rangle_{\text{nh,even}}+ n_{odd}  \langle J_{s_1} J_{s_2}J_{s_3}\rangle_{odd}
 \end{equation}
So we conclude that for a supersymmetric theory, the two structures that was observed in \cite{Aharony:2019mbc,Nizami:2013tpa} can be interpreted to be non-homogeneous component and parity-odd homogeneous component of the non-super symmetric theory. One intuitive way to understand the result in \eqref{susy1} is that non-homogeneous part is needed to reproduce correct WT identity where as homogeneous part does not contribute to WT identity. Since parity-odd term is always a homogeneous contribution\footnote{For correlators with spin inside the triangle inequality.} as was argued in \cite{Jain:2021vrv}, one of the two contribution in supersymmetric theory should be non-homogeneous parity-even contribution. A more detailed analysis is deferred to future.

\section{Conformal collider bound}
\label{ccb}
Finding out bounds on the three-point function coefficients has been very important topic. There has been a lot of work to put bound on three dimensional CFT as well. In this section we show that splitting of three-point functions into their homogeneous and non-homogeneous parts makes the conformal collider bound natural and it just bounds the coefficient of the homogeneous part. The coefficient of non-homogeneous part is already determined in terms of the two-point function coefficient. Bounds on homogeneous part are best described in terms of the spinor-helicity variables. Homogeneous part of the correlator is zero for mixed helicity. So we concentrate on the all $-$ or all $+$ helicity. Using \eqref{fff1}
and
\begin{align}\label{odh}
    \langle J_{s_1}^{-} J_{s_2}^{-} J_{s_3}^{-}\rangle_{\text{odd}}& =   i \langle J_{s_1}^{-} J_{s_2}^{-} J_{s_3}^{-}\rangle_{\bf{h}}\nonumber\\
    \langle J_{s_1}^{+} J_{s_2}^{+} J_{s_3}^{+}\rangle_{\text{odd}}& = -  i \langle J_{s_1}^{+} J_{s_2}^{+} J_{s_3}^{+}\rangle_{\bf{h}}.
     \end{align}
 we get 
 \begin{align}\label{nhr}
   \langle J_{s_1}^{-}J_{s_2}^{-}J_{s_3}^{-}\rangle = \left(n_f+n_{b} \right)  \langle J_{s_1}^{-}J_{s_2}^{-}J_{s_3}^{-}\rangle_{\text{nh}}+\left(n_f+n_{b} \right) \left(\frac{n_{b}-n_f}{n_f+n_{b}}+ i \frac{n_{odd}}{n_f+n_{b}}\right)\langle J_{s_1}^{-}J_{s_2}^{-}J_{s_3}^{-}\rangle_{\bf{h}} 
\end{align}
and all positive helicity results is just complex conjugate of \eqref{nhr}. Now introducing notation $$N_{s}= n_f+ n_b, ~~~\left(\frac{n_b-n_{f}}{n_f+n_{b}}+  i \frac{n_{odd}}{n_f+n_{b}}\right)= \gamma_{s} e^{-i\pi\theta}$$ we get
\begin{align}
   \langle J_{s_1}^{-}J_{s_2}^{-}J_{s_3}^{-}\rangle &= N_s\left(  \langle J_{s_1}^{-}J_{s_2}^{-}J_{s_3}^{-}\rangle_{\text{nh}}  +  \gamma_s e^{- i \pi \theta_s}   \langle J_{s_1}^{-}J_{s_2}^{-}J_{s_3}^{-}\rangle_{\bf{h}} \right)\nonumber\\[5pt]
     \langle J_{s_1}^{+}J_{s_2}^{+}J_{s_3}^{+}\rangle &= N_s\left(  \langle J_{s_1}^{+}J_{s_2}^{+}J_{s_3}^{+}\rangle_{\text{nh}}  +  \gamma_s e^{ i \pi \theta_s}   \langle J_{s_1}^{+}J_{s_2}^{+}J_{s_3}^{+}\rangle_{\bf{h}} \right).
\end{align} 
In \cite{Gandhi:2021gwn} it was observed that 
\begin{center}\label{hstabl1}
    \begin{tabular}{ | l | l | l | l |}
    \hline
     Theory & $\gamma_s$ & $\theta$ & $\gamma_s e^{-i \pi \theta_s}$  \\ \hline
    Free boson & 1 & 0 & 1 \\ \hline
    Free fermion & 1 & $\pi$ &-1 \\ \hline
     Boson$+$Chen-Simons & 1 & $\pi \lambda_b$ & $ e^{-i\pi \lambda_b}$\\ \hline
     Fermion$+$Chen-Simons & 1 & $\pi \left(1-\lambda_f\right)$ & $-e^{i\pi \lambda_f}$\\ \hline
    \end{tabular}
\end{center}
where $\lambda_b,\lambda_f$ are t'Hooft coupling for theories with boson and fermion coupled to Chern-Simons gauge field respectively defined in \ref{theory1}.
With all the notation and convention setup, we are now ready to discuss the conformal collider bound described in \cite{Chowdhury:2017vel}, \cite{Chowdhury:2018uyv}, \cite{Afkhami-Jeddi:2018own}. We immediately notice that Table (\ref{hstabl1}) and figure (\ref{ccbfig}) implies that  higher spin theories satisfies $\gamma_s=1$ for all $s.$ We show below that conformal collider bound implies that for  generic CFT $\gamma_s\le 1$ with higher-spin theories saturating the bound.

For this purpose following \cite{Chowdhury:2017vel}, \cite{Chowdhury:2018uyv}, \cite{Afkhami-Jeddi:2018own} we consider 
 \begin{align}
 \langle J^-J^-T^-\rangle &= N_J   \langle J^-J^-T^-\rangle_{\text{nh}}+ \gamma_J e^{-i\pi\theta_J}    \langle J^-J^-T^-\rangle_{\bf{h}}\nonumber\\[5pt]
 \langle T^-T^-T^-\rangle &=  N_T   \langle T^-T^-T^-\rangle_{\text{nh}}+   \gamma_T e^{-i\pi\theta_T}  \langle T^-T^-T^-\rangle_{\bf{h}}.
\end{align}    
where $N_J, N_T$ are coefficient of two-point function of $\langle JJ \rangle$ and $\langle TT \rangle$ two-point function.
The bounds in \cite{Chowdhury:2017vel}, \cite{Chowdhury:2018uyv} is just the statement that 
\begin{equation}
\gamma_J \le 1,~~~~~~ \gamma_T \le 1
\end{equation}
for generic CFT\footnote{It is interesting to note that double copy structure found in \cite{Jain:2021qcl} is consistent with conformal collider bound.}. We see that free bosonic, free fermionic theory and Chern-Simons matter theories saturates the bound. In \cite{Afkhami-Jeddi:2018own}, collider bound was found for holographic theories in terms of $\Delta_{gap}$. In our notation it simply implies
\begin{align}
    \gamma_J \le \frac{\ln \Delta_{gap}}{\Delta_{gap}^2}, ~~~~\gamma_T \le \frac{\ln \Delta_{gap}}{\Delta_{gap}^4}
\end{align}
Details of the calculation can be found in appendix \ref{cfba}.
These results can be summarised in the two pictures drawn below.

\begin{center}
\begin{figure}[H]
\begin{tikzpicture}
\draw (0,0) circle (4cm);
\coordinate (O) at (0,0);
\coordinate (N) at (0,4);
\coordinate (S) at (0,-4);

\filldraw[black] (O) circle (2pt);
\node[below =1mm of O] {$\scriptstyle{Free SUSY\,CFT}$};
\draw (0,-4)[thick,red] -- (0,4) ;
\draw (-4,0) -- (0,0) coordinate (b) -- (4,0) coordinate (c);
\filldraw[black] (-4,0) circle (2pt) node[anchor=east] {Free fermion};
\filldraw[black] (4,0) circle (2pt) node[anchor=west] {Free boson};


\path[draw=blue,dashed,stealth-stealth,
  postaction={decorate,decoration={text effects along path,
    text={\ Interacting higher spin theory\ }, text align=center,
    text effects/.cd,
      text along path,
      every character/.style={fill=white, yshift=-0.5ex}}}]
  (-4.4,-.55) arc [start angle=188, end angle=352, radius=4.4];

  \path[draw=blue,dashed,stealth-stealth,
 postaction={decorate,decoration={text effects along path,
   text={\ Interacting higher spin theory\ }, text align=center,
   text effects/.cd,
     text along path,
     every character/.style={fill=white, yshift=-0.5ex}}}]
 (-4.4,.55) arc [start angle=-188, end angle=-352, radius=4.4];


\end{tikzpicture}
\caption{\label{ccbfig} The free bosonic and fermionic theories and the interacting bosonic and fermionic theories lie along the circle as indicated. Supersymmetric theories lie along the line marked in red and at the centre lies free supersymmetric theory.}
\end{figure}
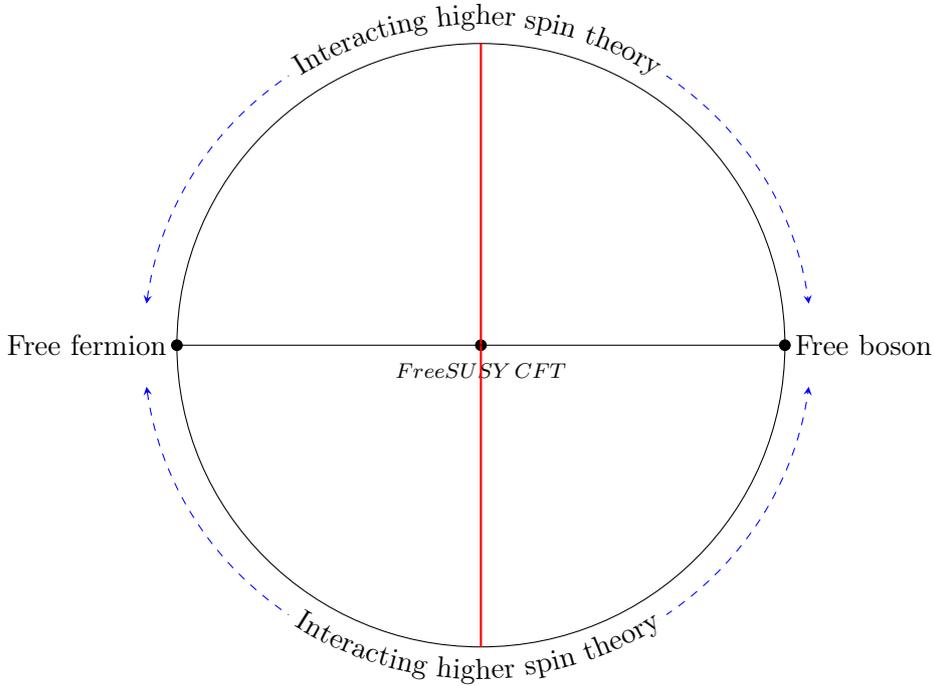
    \end{center}

\begin{center}
\begin{figure}[H]
\begin{tikzpicture}
\draw (0,0) circle (4cm);
\coordinate (O) at (0,0);
\coordinate (N) at (0,4);
\coordinate (S) at (0,-4);

\draw[fill=cyan, opacity=.5] (0,0) circle (3cm);
\draw (0,-4) -- (0,4) ;
\draw (-4,0) -- (0,0) coordinate (b) -- (4,0) coordinate (c);
\draw[dashed] (0,0) -- (3,2.655)  coordinate (a)   node[midway,above]{$\gamma_s$}
pic["$\pi\theta_s$",draw=black,<->,angle eccentricity=1.2,angle radius=1cm] {angle=c--b--a};;
\filldraw[black] (-4,0) circle (2pt) node[anchor=east] {Free fermion};
\filldraw[black] (4,0) circle (2pt) node[anchor=west] {Free boson};
\node at (-0.3,1) {\footnotesize Holographic CFT};


\path[draw=blue,dashed,stealth-stealth,
  postaction={decorate,decoration={text effects along path,
    text={\ Interacting higher spin theory\ }, text align=center,
    text effects/.cd,
      text along path,
      every character/.style={fill=white, yshift=-0.5ex}}}]
  (-4.4,-.55) arc [start angle=188, end angle=352, radius=4.4];

  \path[draw=blue,dashed,stealth-stealth,
 postaction={decorate,decoration={text effects along path,
   text={\ Interacting higher spin theory\ }, text align=center,
   text effects/.cd,
     text along path,
     every character/.style={fill=white, yshift=-0.5ex}}}]
 (-4.4,.55) arc [start angle=-188, end angle=-352, radius=4.4];


\end{tikzpicture}
\caption{\label{ccbfig} This is a circle of unit radius. Exactly conserved or weakly broken at large N  higher spin theories lies on the circle of radius $\gamma_s=1$.   Generic CFT lies inside the disc $\gamma_s\le 1$.  Holographic CFT lies inside the blue region.}
\end{figure}
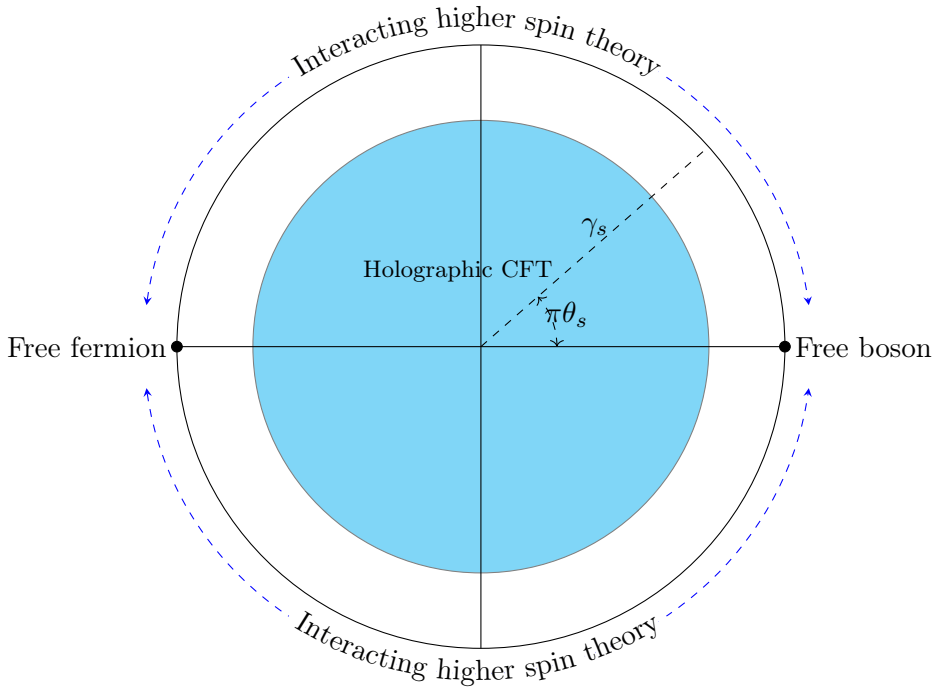
    \end{center}

\section{Summary and Discussion}
\label{summary}
Understanding CFT correlation functions in momentum space is an important topic. In this paper we have explored the general structure of 3d CFT correlation functions. For our analysis it is important to split up the correlation function into  homogeneous and non-homogeneous parts. We observed a very interesting interplay of homogeneous and non-homogeneous structures. We showed that when the spins satisfy triangle inequality, both homogeneous and  non-homogeneous structures are allowed. In particular, we got one homogeneous even, one homogeneous odd and one non-homogeneous even structure. However, when the spins violate the triangle inequality, we showed that homogeneous parity even and odd pieces had bad poles which implies they are not consistent with position space OPE limit. This in particular disallowed the parity-odd structure outside the triangle inequality for exactly conserved currents when all the spins of three-point function is non-zero. We found two  non-homogeneous parity-even structures outside the triangle inequality. However, in the flat space limit, these disallowed parity-even and odd homogeneous pieces give correct flat space amplitude.  For correlators of the kind 
$\langle J_{s_1} J_{s_2} O\rangle$ with $s_1\neq s_2 \neq 0$, parity-odd structure can exist but it is a non-homogeneous piece whereas for $\langle J_{s} J_{s} O\rangle$, which is inside the triangle inequality, parity-odd correlator is a homogeneous piece. Summary of the results are presented in the following table for correlators comprising exactly conserved currents.

\begin{center}
\begin{table}[h!]
\begin{tabular}{ | l | l | l |}
\hline
    \multicolumn{3}{|l|}{\,\,\,\,\,\,\,\,\,\,\,\,\,\,\,\,\,\,\,\,\,\,\,\,\,\,\,\,\,\,\,\,\,\,\,\,\,\,\,\,\,\,\,\,\,\,\,\,\,\,\,\,{\bf {Correlators comprising exactly conserved currents}}}\\
    \hline
    Correlator & Inside triangle inequality & Outside triangle inequality \\ \hline
    $\langle J_{s} O O \rangle$ & $\cross$ & one NH even  \\ \hline
     $\langle J_{s} J_{s} O \rangle$ & 1  H even and 1 H odd  &  $\cross$   \\ \hline
     $\langle J_{s_1} J_{s_2} O \rangle$ with $s_1\neq s_2$ &  $\cross$  &  1 NH even, 1 NH odd  \\ \hline
    $\langle J_{s_1} J_{s_2} J_{s_3} \rangle$ with all non-zero spin & 1 H even, 1 H odd and
    1 NH even & 2 NH even\\
    \hline
      \end{tabular}
     \caption{\label{tab:table-name1}In this table we summarise the structure of various 3-point correlators comprising exactly conserved currents. Here H denotes homogenous contribution and NH denotes non-homogeneous contribution.}
    \end{table}
\end{center}
We then analysed theories with weakly broken higher spin symmetry. The results are summarised in the table below. 
\begin{center}
\begin{table}[h!]
\begin{tabular}{ | l | l | l |}
    \hline
     \multicolumn{3}{|l|}{\,\,\,\,\,\,\,\,\,\,\,\,\,\,\,\,\,\,\,\,\,\,\,\,\,\,\,\,\,\,\,\,\,\,\,\,\,\,\,\,\,\,\,\,\,\,\,\,\,\,\,\,{\bf {Correlators comprising weakly broken currents}}}\\
    \hline
    Correlator & Inside triangle inequality & Outside triangle inequality \\ \hline
    $\langle J_{s} O O \rangle$ & $\cross$ & 1 NH even  \\ \hline
     $\langle J_{s} J_{s} O \rangle$ & 1  H even and one H odd  &  $\cross$   \\ \hline
     $\langle J_{s_1} J_{s_2} O \rangle$ with $s_1\neq s_2$ &  $\cross$  &  1 NH even, 1 NH odd  \\ \hline
    $\langle J_{s_1} J_{s_2} J_{s_3} \rangle$ with all non-zero spin & 1 H even, 1 H odd and
    1 NH even & 2 NH even, 1 NH odd.\\
    \hline
    \end{tabular}
     \caption{\label{tab:table-name2}In this table we summarise the structure of various 3-point correlators comprising weakly broken currents. Here H denotes homogenous contribution and NH denotes non-homogeneous contribution.}
    \end{table}
\end{center}
We observe that results in the two tables are the same except in the last row where for weakly broken symmetry a parity-odd term is allowed. We also found a remarkable relation between WT identity derived using weakly broken higher spin currents and
WT identity derived using exactly conserved currents. This identification also gives a precise prescription for calculating parity-odd correlation function outside the triangle inequality starting from parity-even free bosonic and free fermionic results. It will be interesting to understand the general structure of WT identity and why there is  nice mapping between the WT identity obtained from non-conservation of currents and that obtained from exactly conserved currents. Let us mention that, we found the WT identity derivation in general in position space to be very complicated. Also for calculating position space correlation function WT identity for exactly conserved currents were not essential by using the fact that we can work at separated points. However in momentum space we need the WT identity. 

In previous papers, \cite{Gandhi:2021gwn,Jain:2021gwa}, we have shown that parity-odd homogeneous results inside the triangle inequality can be obtained from free bosonic and free theory in the same manner. Let us also recall that in general parity even results can be obtained from free theories. However, parity-odd results are obtained in interacting theories such as Chern-Simons matter theories. Our results suggest that there might be a free field theory representation of low energy Hilbert space of interacting theories such as Chern-Simons matter theories. Low energy because we have been able to reproduce parity odd correlation function of conserved currents which are like bound states.
We also discussed the case of supersymmetric theory briefly. We found there are one parity-even and one parity-odd structures. It would be interesting to explore this further. We also found that in momentum space and spinor-helicity variables, the conformal collider bounds becomes natural and easier to understand. It would be interesting to derive these bounds directly in momentum space which might be much more simpler than in position space analysis. It would also be interesting to explore the general structure of four point functions. See \cite{Bzowski:2019kwd,Bzowski:2020kfw} for recent development, but in general this would be a challenging computation.

\section*{Acknowledgments}
The work of S.J and R.R.J is supported by the Ramanujan Fellowship. We acknowledge our debt to the people of India for their steady support of research in basic sciences. We thank S. D. Chowdhury, A. Gadde, N. Kundu, G. Mandal, S. Minwalla,  A. A. Nizami, A. Suresh and S. Trivedi for useful discussions. We thank S. Prakash for comments on the draft. S.J would like to thank participants of the informal discussion in TIFR journal club members for discussions and critical comments where part of the work was presented. AM would like to acknowledge the support of CSIR-UGC.
(JRF) fellowship (09/936(0212)/2019-EMR-I).

\appendix

\section{Details of theories considered in this paper}\label{theory1}
In this section we give the Lagrangian and comment on the operator spectrum of the theories that we study.
\subsection{Free Bosonic Theory}
Let $\phi$ be a free massless scalar field in the fundamental representation of SU($N$). The Lagrangian is given by
\begin{align}
    S=\int d^3x\,\partial^\mu\bar\phi\,\partial_\mu\phi
\end{align}
The operator spectrum of single trace primary operators in the theory consists of a scalar primary $O$ with scaling dimension 1 and conserved spin-1 and spin-2 currents with dimensions 2 and 3 respectively. In addition to these, it contains an infinite tower of exactly conserved higher spin currents. For a spin-$s>3$ higher spin current the conformal dimension is $s+1$.

\subsection{Bosonic Theory with Chern-Simons field}
The bosonic theory coupled to $SU(N_b)$ Chern-Simons gauge field has the following action
\begin{align}
S=\int d^3x\left[D_\mu\bar\phi D^\mu\phi+i\epsilon^{\mu\nu\rho}\frac{\kappa_F}{4\pi}\text{Tr}(A_\mu\partial_\nu A_\rho-\frac{2i}{3}A_\mu A_\nu A_\rho)+\frac{(2\pi)^2}{\kappa_B^2}(x_6^B+1)(\bar\phi\phi)^3\right]    \end{align}
The scalar primary operator has conformal dimension $\Delta=1+\mathcal O\left(\frac 1N\right)$. The spin-1 and spin-2 conserved currents have dimensions 2 and 3 respectively. The theory also has an infinite tower of higher spin currents $J_s$ with spin $s>2$ that are weakly broken with conformal dimension $\Delta=s+1+\mathcal O\left(\frac 1N\right)$.
At large $N_b$ and $\kappa_b$ the t'Hooft coupling is defined as
\begin{align}
   \lambda_b =  \lim_{N_b,\kappa_b\rightarrow \infty}\frac{N_b}{k_b}.
\end{align}

\subsection{Critical Bosonic Theory}

The critical bosonic theory obtained by adding an interaction of the kind $\sigma_B\bar\phi\phi$ where $\sigma_B$ is an auxiliary field  has the following action
\begin{align}
    S=\int d^3x\left[\partial_\mu\bar\phi \partial^\mu\phi+\sigma_B\bar\phi\phi\right]
\end{align}
The conformal dimension of the scalar primary operator is $\Delta=2+\mathcal O\left(\frac{1}{N}\right)$. The scaling dimensions of the spin-1 and spin-2 conserved currents are 2 and 3 respectively. The theory has an infinite tower of slightly broken higher spin currents. The conformal dimension of the spin $s>2$ current $J_s$ is $\Delta=s+1+\mathcal O\left(\frac{1}{N}\right)$.

\subsection{Critical Bosonic Theory with Chern-Simons field}

The critical bosonic theory 
coupled to Chern-Simons gauge field obtained by adding an interaction of the kind $\sigma_B\bar\phi\phi$ where $\sigma_B$ is an auxiliary field  has the following action
\begin{align}
    S=\int d^3x\left[D_\mu\bar\phi D^\mu\phi+i\epsilon^{\mu\nu\rho}\frac{\kappa_B}{4\pi}\text{Tr}(A_\mu\partial_\nu A_\rho-\frac{2i}{3}A_\mu A_\nu A_\rho)+\sigma_B\bar\phi\phi\right]
\end{align}
The conformal dimension of the scalar primary operator is $\Delta=2+\mathcal O\left(\frac{1}{N}\right)$. The scaling dimensions of the spin-1 and spin-2 conserved currents are 2 and 3 respectively. The theory has an infinite tower of slightly broken higher spin currents. The conformal dimension of the spin $s>2$ current $J_s$ is $\Delta=s+1+\mathcal O\left(\frac{1}{N}\right)$.

\subsection{Free Fermionic Theory}
Let $\psi$ be a free massless fermionic field in the fundamental representation of SU($N$). The Lagrangian is given by
\begin{align}
S=\int d^3x\,\bar\psi\gamma^\mu\partial_{\mu}\psi
\end{align}
The scalar primary operator $O$ has scaling dimension 2 and conserved spin-1 and spin-2 currents with dimensions 2 and 3 respectively. There is an infinite tower of higher spin currents that are conserved. The spin-$s>3$ higher spin current has dimension $s+1$.

\subsection{Fermionic Theory with Chern-Simons field}
The fermionic theory coupled to $SU(N_f)$ Chern-Simons gauge field has the following action
\begin{align}
S=\int d^3x\left[\bar\psi\gamma_\mu D^\mu\psi+i\epsilon^{\mu\nu\rho}\frac{\kappa_f}{4\pi}\text{Tr}(A_\mu\partial_\nu A_\rho-\frac{2i}{3}A_\mu A_\nu A_\rho)\right]    \end{align}
The scalar primary operator has conformal dimension $\Delta=2+\mathcal O\left(\frac 1N\right)$. The spin-1 and spin-2 conserved currents have dimensions 2 and 3 respectively. The theory also has an infinite tower of higher spin currents $J_s$ with spin $s>2$ that are weakly broken with conformal dimension $\Delta=s+1+\mathcal O\left(\frac 1N\right)$.
At large $N_f$ and $\kappa_f$ the t'Hooft coupling is defined as
\begin{align}
   \lambda_f =  \lim_{N_f,\kappa_f\rightarrow \infty}\frac{N_f}{k_f}.
\end{align}

\subsection{Critical Fermionic Theory}
The critical fermionic theory has the following action
\begin{align}
S=\int d^3x\left[\bar\psi\gamma_\mu \partial^\mu\psi-\frac{4\pi}{\kappa_F}\xi\bar\psi\psi+\frac{(2\pi)^2}{\kappa_F^2}x_6^F\xi^3\right]    
\end{align}
The conformal dimension of the scalar primary operator is $\Delta=1+\mathcal O\left(\frac{1}{N}\right)$. The scaling dimensions of the spin-1 and spin-2 conserved currents are 2 and 3 respectively. The infinite tower of slightly broken higher spin currents $J_s$ with spin $s$ has conformal dimension  $\Delta=s+1+\mathcal O\left(\frac{1}{N}\right)$.

\subsection{Critical Fermionic Theory with Chern-Simons field}
The critical fermionic theory coupled to Chern-Simons gauge field has the following action
\begin{align}
S=\int d^3x\left[\bar\psi\gamma_\mu D^\mu\psi+i\epsilon^{\mu\nu\rho}\frac{\kappa_F}{4\pi}\text{Tr}(A_\mu\partial_\nu A_\rho-\frac{2i}{3}A_\mu A_\nu A_\rho)-\frac{4\pi}{\kappa_F}\xi\bar\psi\psi+\frac{(2\pi)^2}{\kappa_F^2}x_6^F\xi^3\right]    
\end{align}
The conformal dimension of the scalar primary operator is $\Delta=1+\mathcal O\left(\frac{1}{N}\right)$. The scaling dimensions of the spin-1 and spin-2 conserved currents are 2 and 3 respectively. The infinite tower of slightly broken higher spin currents $J_s$ with spin $s$ has conformal dimension  $\Delta=s+1+\mathcal O\left(\frac{1}{N}\right)$.

We refer to the regular bosonic (regular fermionic) theory coupled to Chern-Simons field and the critical fermionic (critical bosonic) theory coupled to Chern-Simons field as quasi-bosonic (fermionic) theories.

\section{Derivation of WT identity}\label{WT-identity}

Consider an infinitesimal transformation of the space-time coordinates. Under this transformation the action changes as :
\begin{align}
\delta S=\int d^3x\,\partial_\mu J^\mu\,\omega
\end{align}
where $J^\mu$ is the current associated with the infinitesimal transformation and $\omega$ is the infinitesimal parameter of transformation.

Under the infinitesimal transformation, an operator $\Phi(x)$ transforms as :
\begin{align}
\Phi'(x)=\Phi(x)-i\,\omega\,G\Phi(x)
\end{align}
where $G$ is the generator of the transformation. 

Let us now compute the change in the correlation function under the infinitesimal transformation. For sake of brevity we denote by $X$ the product $\Phi(x_1)\ldots\Phi(x_n)$ 
\begin{align}
\langle X\rangle&\equiv\langle\Phi(x_1)\ldots\Phi(x_n)\rangle\cr
&=\frac{1}{Z}\int[d\Phi]Xe^{-S[\Phi]}
\end{align}
where $Z$ is the functional integral without the operator insertions. Let us now express the functional integral in terms of $\Phi'$ \footnote{We assume that the measure is invariant under the infinitesimal transformation, i.e. $[d\Phi']=[d\Phi]$.}. Let us denote by $(X+\delta X)$ the product of the transformed fields, i.e.
\begin{align}
X+\delta X=\Phi'(x_1)\ldots\Phi'(x_n)
\end{align}
The functional integral $\langle X\rangle$ after the change of variables is then given by :
\begin{align}
\langle X\rangle=\frac{1}{Z}\int[d\Phi](X+\delta X)e^{-\left(S[\Phi]+\int d^3x\,\partial_\mu J^\mu\,\omega\right)}
\end{align}
Working to first order in the transformation parameter we obtain :
\begin{align}
\langle\delta X\rangle=\int dx\,\partial_{\mu}\langle J^\mu(x) X\rangle\omega(x)
\end{align}
By computing the explicit change in each of the operator insertions the change in the functional is given by :
\begin{align}
\delta X
=-i\int dx\,\omega(x)\sum_{i=1}^n(\Phi(x_1)\ldots G\Phi(x_i)\ldots\Phi(x_n))\delta(x-x_i)
\end{align}
Equating the two expressions derived for $\langle\delta X\rangle$ we obtain :
\begin{align}
\label{WIdiff}
\partial_{\mu}\langle J^\mu(x) \Phi(x_1)\ldots\Phi(x_n)\rangle=-i\,\sum_{i=1}^n\langle\Phi(x_1)\ldots G\Phi(x_i)\ldots\Phi(x_n)\rangle\delta(x-x_i)
\end{align}
%
%
Let us now consider the charge associated to the current defined as :
\begin{align}
Q=\int d^{d-1}x\,J^0(x)
\end{align}
The integrated form of \eqref{WIdiff} gives the following
%
%
\begin{align}
[Q,\Phi]=-i\,G\,\Phi
\end{align}
This leads to 
\begin{align}
\partial_{\mu}\langle J^\mu(x) \Phi(x_1)\ldots\Phi(x_n)\rangle=
\sum_{i=1}^n\langle\Phi(x_1)\ldots[Q,\Phi(x_i)]\ldots\Phi(x_n)\rangle\delta(x-x_i)
\end{align}
For the $n=2$ case 
\begin{align}
\partial_{\mu}\langle J^\mu(x) \Phi(x_1)\Phi(x_2)\rangle=\langle[Q,\Phi(x_1)]\Phi(x_2)\rangle\delta(x-x_1)+\langle\Phi(x_1)[Q,\Phi(x_2)]\rangle\delta(x-x_2)
\end{align}
A simple Fourier transform gives :
\begin{align}
\label{WIn=2}
ip_\mu\,\langle J^\mu(p) \Phi(k_1)\Phi(k_2)\rangle=\langle[Q,\Phi(k_1+p)]\Phi(k_2)\rangle+\langle\Phi(k_1)[Q,\Phi(k_2+p)]\rangle
\end{align}

\subsection{$\langle J_3J_1O_1\rangle$ : Bosonic theory}
The algebra is given by
\begin{align}
[Q_{\mu\nu},O]&=-\partial_{(\mu}J_{\nu)}\cr
[Q_{\mu\nu},J_{\rho}]&=\partial_\rho T_{\mu\nu}+\partial_{(\mu} T_{\nu)\rho}+g_{\rho(\mu}\partial_{\nu)} T+g_{\mu\nu}\partial_\rho T+\partial_\mu\partial_\nu\partial_\rho O+g_{\mu\nu}\Box\partial_\rho O+g_{\rho(\mu}\Box\partial_{\nu)}O\cr
\end{align}
Thus the Ward identity takes the form 
\begin{align}
\label{ffj3j1o}
ik_{1\sigma}\langle J^\sigma_{\mu\nu}(k_1)J_{\rho}(k_2)O(k_3)\rangle
&=i\langle[Q_{\mu\nu},J_{\rho}(-k_3)]O(k_3)\rangle+i\langle J_{\rho}(k_2)[Q_{\mu\nu},O(-k_2)]\rangle\cr
&=k_{3\mu}k_{3\nu}k_{3\rho}\langle O(k_3)O(-k_3)\rangle-k_3^2g_{\rho(\mu}k_{3\nu)}\langle O(k_3)O(-k_3)\rangle\cr
&\hspace{.5cm}+k_{2(\mu}\langle J_{\rho}(k_2)J_{\nu)}(-k_2)\rangle
\end{align}
Dotting with polarization vectors $z_{1\mu},z_{1\nu},z_{2\rho}$ we get 
\begin{align}
-\frac{1}{k_3}(k_2\cdot z_1)^2(k_1\cdot z_2)+2k_2(k_2\cdot z_1)(z_1\cdot z_2)-2k_3(k_2\cdot z_1)(z_1\cdot z_2)
\end{align}
This agrees with the Ward identity obtained from explicit computations in the free theory.
\subsection{$\langle J_4J_1J_1\rangle$ : Bosonic theory}
The algebra is given by
\begin{align}
[Q_{\mu\nu\rho},J_{\alpha}]&=a\,\partial_{\mu}\partial_{\nu}\partial_{\rho}J_{\alpha}+b\,\partial_{\alpha}\partial_{\mu}\partial_{\nu}J_{\rho}+\partial_{\alpha}J_{\mu\nu\rho}+g_{\mu\nu}\partial_{\rho}\Box J_{\alpha}+g_{\mu\nu}\partial_{\alpha}\Box J_{\rho}\cr
&\hspace{.5cm}+cg_{\mu\alpha}\partial_{\nu}\Box J_{\rho}+cg_{\mu\alpha}\partial_{\rho}\Box J_{\nu}
\end{align}
Thus in the free theory we have
\begin{align}
\label{ffj3j1o}
ik_{1\sigma}\langle J^\sigma_{\mu\nu\rho}(k_1)J_{\alpha}(k_2)J_{\beta}(k_3)\rangle
&=i\langle[Q_{\mu\nu\rho},J_{\alpha}(-k_3)]J_{\beta}(k_3)\rangle+i\langle J_{\alpha}(k_2)[Q_{\mu\nu\rho},J_{\beta}(-k_2)]\rangle
\end{align}
Dotting with polarization vectors $z_{1\mu},z_{1\nu},z_{1\rho},z_{2\alpha},z_{3\beta}$ we get 
\begin{align}
&a(k_2\cdot z_1)^3(z_2\cdot z_3)k_2-a(k_2\cdot z_1)^3(z_2\cdot z_3)k_3+2c(k_3^3-k_2^3)(k_2\cdot z_1)(z_1\cdot z_2)(z_1\cdot z_3)\cr
&+b(k_1\cdot z_2)(k_2\cdot z_1)^2(z_1\cdot z_3)k_3 -b(k_2\cdot z_3)(k_2\cdot z_1)^2(z_1\cdot z_2)k_2  
\end{align}
where $a=7,b=9$ and $c=3$. This agrees with the Ward identity obtained from explicit computations in the free theory.

\section{Counting structures}
\label{countingapp}
The action of the charge corresponding to the spin $s$ current on a spin $s'$ operator is given by \cite{Maldacena:2011jn}
\begin{align}
\label{generalalgebra}
[Q_s,j_{s'}]=(2s-2)!\sum_{r=0}^{s+s'-2}c_{s,s'}^{r}\partial^{s+s'-r-1}j_r
\end{align}
where the coefficients $c_{s,s'}^r$ are given by :
\begin{align}
c_{s,s'}^r=(1+(-1)^{s+s'+r})\left(\frac{1}{\Gamma[r+s-s']\Gamma[s+s'-r]}\pm \frac{1}{\Gamma[r+s+s']\Gamma[s-s'-r]}\right)
\end{align}
and the signs $+$ and $-$ for the second term inside the bracket correspond to the boson and fermion theories respectively.

Let us now consider the Ward identity \eqref{WIn=2} which is roughly given by
\begin{align}
\label{WIQs}
ik_{1\mu_1}\,\langle J^{\mu_1\ldots\mu_{s_1}}(k_1) J^{\nu_1\ldots\nu_{s_2}}(k_2)J^{\rho_1\ldots\rho_{s_3}}(k_3)\rangle&\sim\langle[Q_{s_1},J^{\nu_1\ldots\nu_{s_2}}(-k_3)]J^{\rho_1\ldots\rho_{s_3}}(k_3)\rangle\cr
&\hspace{0.5cm}+\langle J^{\nu_1\ldots\nu_{s_2}}(k_2)[Q_{s_1},J^{\rho_1\ldots\rho_{s_3}}(-k_2)]\rangle
\end{align}
Each term on the RHS of the above equation is non-vanishing when the algebra \eqref{generalalgebra} results in a term that leads to a non-zero 2-point function. When this occurs the Ward identity \eqref{WIQs} takes the form 
\begin{align}
&ik_{1\mu_1}\,\langle J^{\mu_1\ldots\mu_{s_1}}(k_1) J^{\nu_1\ldots\nu_{s_2}}(k_2)J^{\rho_1\ldots\rho_{s_3}}(k_3)\rangle\cr
&=c_{s_1,s_2}^{s_3}k_{3\mu_1}k_{3\mu_{s_3+1}}\ldots k_{3,\mu_{s_1}}k_{3\nu_1}\ldots k_{3\nu_{s_2}}\langle J^{\rho_1\ldots\rho_{s_3}}(k_3)J^{\mu_1\ldots\mu_{s_3}}(-k_3)\rangle\cr
&\hspace{1cm}+c_{s_1,s_3}^{s_2}k_{2\mu_1}k_{2\mu_{s_2+1}}\ldots k_{2,\mu_{s_1}}k_{2\rho_1}\ldots k_{3\rho_{s_3}}\langle J^{\nu_1\ldots\nu_{s_2}}(k_2)J^{\mu_1\ldots\mu_{s_2}}(-k_2)\rangle
\end{align}
Let us now consider the form of the Ward identity in the boson and fermion theories when the spins in the three-point function satisfy triangle inequality.
\subsection{Inside the triangle inequality}
When the spins in the correlation function satisfy triangle inequality we have
\begin{align}
s_2-s_3\le s_1\le s_2+s_3,\quad
s_1-s_3\le s_2\le s_1,\quad
s_1-s_2\le s_3\le s_2
\end{align}
In this case for the bosonic as well as the fermionic theory we have 
\begin{align}
c_{s_1,s_2}^{s_3}=c_{s_1,s_3}^{s_2}=\frac{(1+(-1)^S)(2s_1-2)!}{\Gamma(s_1+s_2-s_3)\Gamma(s_1-s_2+s_3)}
\end{align}
where $S=s_1+s_2+s_3$. The coefficients for the two theories match as the second term which distinguishes the two theories vanishes when triangle inequality is satisfied by the spins. A similar analysis for the transverse Ward identities involving $k_2$ and $k_3$ will give 
\begin{align}
c_{s_2,s_1}^{s_3}&=c_{s_2,s_3}^{s_1}=\frac{(1+(-1)^S)(2s_2-2)!}{\Gamma(s_1+s_2-s_3)\Gamma(-s_1+s_2+s_3)}\cr
c_{s_3,s_1}^{s_2}&=c_{s_3,s_2}^{s_1}=\frac{(1+(-1)^S)(2s_3-2)!}{\Gamma(s_1-s_2+s_3)\Gamma(-s_1+s_2+s_3)}
\end{align}
for the bosonic theory as well as the fermionic theory. 
Thus we see that when the triangle inequality is satisfied the bosonic and fermionic  theories have the same Ward identities. That is 
\begin{align}\label{WTfbffsm}
k_{1\mu_1}\,\langle J^{\mu_1\ldots\mu_{s_1}}(k_1) J^{\nu_1\ldots\nu_{s_2}}(k_2)J^{\rho_1\ldots\rho_{s_3}}(k_3)\rangle_{FF}
=k_{1\mu_1}\,\langle J^{\mu_1\ldots\mu_{s_1}}(k_1) J^{\nu_1\ldots\nu_{s_2}}(k_2)J^{\rho_1\ldots\rho_{s_3}}(k_3)\rangle_{FB}.
\end{align}

Let us now consider the case when triangle inequality is violated.
\subsection{Outside the triangle inequality}
When the spins in the correlation function violate triangle inequatlity we have
\begin{align}
s_1> s_2+s_3, s_2\ge s_3
\end{align}
In this case one can check that the coefficients in the algebra differ for the boson and fermion theories.
In this case we have the following coefficients 
\begin{align}
c_{s_1,s_2}^{s_3}&=c_{s_1,s_3}^{s_2}=\left[\frac{(1+(-1)^S)(2s_1-2)!}{\Gamma(s_1+s_2-s_3)\Gamma(s_1-s_2+s_3)}\pm \frac{(1+(-1)^S)(2s_1-2)!}{\Gamma(s_1-s_2-s_3)\Gamma(s_1+s_2+s_3)}\right]\cr
c_{s_2,s_1}^{s_3}&=c_{s_2,s_3}^{s_1}=0\cr
%
c_{s_3,s_1}^{s_2}&=c_{s_3,s_2}^{s_1}=0
\end{align}
that is $c_{s_1,s_2,Boson}^{s_3}\neq c_{s_1,s_2,Fermion}^{s_3},$
Thus when the spins do not satisfy triangle inequality the bosonic and fermionic  theories have different Ward identities
\begin{align}\label{WTfbffdif}
k_{1\mu_1}\,\langle J^{\mu_1\ldots\mu_{s_1}}(k_1) J^{\nu_1\ldots\nu_{s_2}}(k_2)J^{\rho_1\ldots\rho_{s_3}}(k_3)\rangle_{FF}
\neq k_{1\mu_1}\,\langle J^{\mu_1\ldots\mu_{s_1}}(k_1) J^{\nu_1\ldots\nu_{s_2}}(k_2)J^{\rho_1\ldots\rho_{s_3}}(k_3)\rangle_{FB}.
\end{align}




\subsection{No homogeneous contribution outside the triangle inequality}\label{outTH}
In this section we will prove that when the spins of the operator insertion are such that they violate triangle inequality there is no homogeneous contribution to the correlator. We show this by looking at three classes of correlators : $\langle J_{s_1}O_{\Delta_1}O_{\Delta_2}\rangle$, $\langle J_{s_1}J_{s_2}O_{\Delta}\rangle$ and $\langle J_{s_1}J_{s_2}J_{s_3}\rangle$.
\subsubsection{$\langle J_{s_1}O_{\Delta_1}O_{\Delta_2}\rangle$} \label{outTHjsoo}
We consider the following ansatz for the correlator 
\begin{align}
\label{ansatzjsoogen}
\langle J^{h_1}_{s_1}O_{\Delta_1}O_{\Delta_2}\rangle = f_{h_1s_1}(k_1, k_2, k_3)\left(\frac{\langle 12\rangle\langle 31\rangle}{\langle 23\rangle}\right)^{-h_1s_1}
\end{align}
where $h_1$ is the sign of the helicity. The action of the generator of special conformal transformations on $\langle J^{h_1}_{s_1}O_{\Delta_1}O_{\Delta_2}\rangle$ gives the following Ward identity for the correlator
\begin{align}
\sum_{a = 1}^3b\cdot K^a \frac{\langle J^{h_1}_{s_1}O_{\Delta_1}O_{\Delta_2}\rangle}{k^{s_1-1}_1k^{\Delta_1-2}_2k^{\Delta_2-2}_3} = \frac{b\cdot k_2}{k^{s_1-1}_1k^{\Delta_1}_2k^{\Delta_2-2}_3}(\Delta_1-2)(\Delta_1-1)\langle J_{s_1}O_{\Delta_1}O_{\Delta_2}\rangle + (2 \leftrightarrow 3)
\end{align}
where $b$ is an arbitrary vector. If we choose $b^i = \lambda_2\sigma^i\lambda_3+\lambda_3\sigma^i\lambda_2$, we obtain
\begin{align}
\label{jsoofirst}
(k_2-k_3)\left(\frac{\partial^2\widetilde{f}}{\partial^2k_3}-\frac{\partial^2\widetilde{f}}{\partial^2k_1}\right)&-k_2\left(\frac{\partial^2\widetilde{f}}{\partial^2k_3}-\frac{\partial^2\widetilde{f}}{\partial^2k_2}\right) \cr
&\hspace{.5cm}= -\frac{\widetilde{f}}{k_2}(\Delta_1-2)(\Delta_1-1)+\frac{\widetilde{f}}{k_3}(\Delta_2-2)(\Delta_2-1)
\end{align}
where 
\begin{align}
\widetilde{f} = \frac{f}{k^{s_1-1}_1k^{\Delta_1-2}_2k^{\Delta_2-2}_3}
\end{align}
For $b^i =  \lambda_1\sigma^i\lambda_3+\lambda_3\sigma^i\lambda_1$, we obtain
\begin{align}
\label{jsoosecond}
&(k_3-k_1)\left(\frac{\partial^2\widetilde{f}}{\partial^2k_1}-\frac{\partial^2\widetilde{f}}{\partial^2k_2}\right)-k_3\left(\frac{\partial^2\widetilde{f}}{\partial^2k_1}-\frac{\partial^2\widetilde{f}}{\partial^2k_3}\right)+2h_1s_1\left(\frac{\partial\widetilde{f}}{\partial k_2}-\frac{\partial\widetilde{f}}{\partial k_1}\right) \notag\\&= \frac{(k_3-k_1)\widetilde{f}}{k^2_2}(\Delta_1-1)(\Delta_1-2)-\frac{\widetilde{f}}{k_3}(\Delta_2-1)(\Delta_2-2)
\end{align}
Let us consider the following ansatz for $\widetilde f$ 
\begin{align}
\label{ansatzjsoo}
\widetilde{f} = \frac{1}{k^{\tau}_1}I_{\alpha, \{\beta_1, \beta_2, \beta_3\}}
\end{align}
If we insert this into \eqref{jsoofirst} we get \begin{align}
  & \frac{ -\tau(1+\tau)k_2(k_2-k_3)k_3+k^2_1[(2-\Delta_1)(1-\Delta_1)k_2-(2-\Delta_2)(1-\Delta_2)k_3]}{k^2_1k_2k_3}I_{\alpha, \{\beta_1, \beta_2, \beta_3\}}\nonumber\\[5pt]&+(-1+2\beta_1-2\tau)(k_2-k_3)I_{1+\alpha, \{\beta_1-1, \beta_2, \beta_3\}}+(1-2\beta_2)k_2I_{1+\alpha, \{\beta_1, \beta_2-1, \beta_3\}}\nonumber\\[5pt]
  &+(1-2\beta_3)k_2I_{1+\alpha, \{\beta_1, \beta_2, \beta_3-1\}} = 0
\end{align}
If we take $I_{\alpha+1, \{\beta_i-\delta_{in}\}}$ and $I_{\alpha, \{\beta_i\}}$ as independent, we obtain the following constrains on the parameters that appear in the triple-K integral in \eqref{ansatzjsoo}.
\begin{align}
    \beta_1 = \beta_2 = \beta_3 = \frac{1}{2}, \quad \tau = 0  \label{sol22}
\end{align}
It will also require the conformal dimensions of the scalar insertions to be 
\begin{align}
\label{sol22a}
\Delta_i = 1, 2, \quad\quad i = 1, 2
\end{align}
This solves the second equation \eqref{jsoosecond} as well. Therefore, to have a nontrivial simultaneous solution to conformal Ward identities \eqref{jsoofirst} and \eqref{jsoosecond}, we must have \eqref{sol22} and \eqref{sol22a}. When $\Delta_1 \neq \Delta_2 \neq 1, 2$, we only have the trivial solution i.e. $\widetilde{f} = 0$. For  $\Delta_i = 1, 2$ for $i=1,2$, the values of the parameters in \eqref{sol22} in the ansatz \eqref{ansatzjsoo} gives
\begin{align}
\widetilde{f} = \frac{1}{E^{s_1}}
\end{align}
However, due to the bad pole because of the spinor-helicity bracket $\langle ij \rangle$ in the denominator of \eqref{ansatzjsoogen}, this solution is unacceptable. This proves that there is no homogeneous contribution to $\langle J_{s_1}O_{\Delta_1}O_{\Delta_2}\rangle$.
\subsubsection{$\langle J_{s_1} J_{s_2} O_{\Delta}\rangle$}
\label{Js1Js2Odeltaapp}
Let us now consider correlators of the kind $\langle J_{s_1} J_{s_2} O_{\Delta}\rangle$.
In spinor-helicity variables, we have the following ansatz for the correlator \footnote{The notation for the current used is $J^{\pm}_{s} = (2k)^{s+1}z^{\pm, i_1}z^{\pm, i_2}\cdots z^{\pm, i_s} J_{i_1i_2\cdots i_s} $}
\begin{align}
&\langle J^{h_1}_{s_1}J^{h_2}_{s_2}O_{\Delta}\rangle = f_{h_1, h_2}(k_1, k_2, k_3)\langle12\rangle^{-h_1s_1-h_2s_2}\langle23\rangle^{h_1s_1-h_2s_2}\langle31\rangle^{h_2s_2-h_1s_1}
\end{align}
where $h_i$'s are the signs of the helicities i.e.  $h_i = \pm $. For scalars the helicity is  zero. 
The action of the generator of special conformal transformations gives 
\begin{align}
\sum_{a=1}^{3}{b}\cdot\widetilde{K}_a\frac{\langle J^{h_1}_{s_1}J^{h_2}_{s_2}O_{\Delta}\rangle}{k^{\Delta-2}_3k^{s_1-1}_1k^{s_2-1}_2} &= \frac{{b}\cdot k_3}{k^{\Delta}_3 k^{s_1-1}_1k^{s_2-1}_2}(\Delta-2)(\Delta-1)\langle J^{h_1}_{s_1}J^{h_2}_{s_2}O_{\Delta}\rangle\notag\\&\hspace{0.2cm}+N_1({b}\cdot z^{h_1}_1)\frac{k^{i_1}_1z^{h_1,i_2}_1\cdots z^{h_1,i_{s_1}}_1\langle J_{i_1\mu_2\cdots i_{s_1}}J^{h_2}_{s_2}O_{\Delta}\rangle}{k^{\Delta-2}_3k^{s_1+1}_1k^{s_2-1}_2}\notag\\&\hspace{0.2cm}+(1 \leftrightarrow 2)\label{WISH}
\end{align}
where $b$ is an arbitrary vector and 
\begin{align}
    z^{-, i} = \frac{\lambda\sigma^i\lambda}{2k} \quad  z^{+, i} = \frac{\bar{\lambda}\sigma^i\bar{\lambda}}{2k}
\end{align}
and $N_2$ denotes a numerical coefficient. Let us split the solution into the homogeneous and non-homogeneous pieces
\begin{align}
    \langle J^{h_1}_{s_1}J^{h_2}_{s_2}O_{\Delta}\rangle=\langle J^{h_1}_{s_1}J^{h_2}_{s_2}O_{\Delta}\rangle_\text{h}+\langle J^{h_1}_{s_1}J^{h_2}_{s_2}O_{\Delta}\rangle_{\text{nh}}.
\end{align}
and analyse the homogeneous contribution.
\subsection*{Homogeneous solution}
 We will show that there is no solution for $s_1\ne s_2$ when $\Delta \ne 1,2$.  If we choose ${b} = \lambda_2{\sigma}^i\lambda_3+\lambda_3{\sigma}^i\lambda_2$ in \eqref{WISH}, we obtain the following equation for the homogeneous part
\begin{align}
&(k_2-k_3)\left(\frac{\partial^2 \widetilde{f}}{\partial k^2_3}-\frac{\partial^2 \widetilde{f}}{\partial k^2_1}\right)-k_2\left(\frac{\partial^2 \widetilde{f}}{\partial k^2_3}-\frac{\partial^2 \widetilde{f}}{\partial k^2_2}\right)+\left(2h_2s_2 \frac{\partial \widetilde{f}}{\partial k_1}- 2h_2 s_2\frac{\partial \widetilde{f}}{\partial k_2}\right)= -\frac{1}{k_3}(\Delta-2)(\Delta-1)\widetilde{f}\label{hgp}
\end{align}
where
\begin{align}
\widetilde{f} = \frac{f}{k^{s_1-1}_1k^{s_2-1}_2k^{\Delta-2}_3}
\end{align}
Let us consider the following ansatz for $\widetilde f$
\begin{align}
\widetilde{f} =  \frac{1}{k^\tau_3} I_{\alpha,\{\beta_1, \beta_1, \beta_3\}} \label{soln}
\end{align}
The conformal Ward identity \eqref{hgp} then takes the form 
\begin{align}
   &-(-2+\Delta-\tau)(-1+\Delta+\tau)I_{\alpha, \{\beta_1, \beta_2, \beta_3\}}+k_3[(2h_2s_2 k_1+k_2-2\beta_1k_2+(2\beta_1-1)k_3)I_{1+\alpha, \{\beta_1-1, \beta_2, \beta_3\}}\notag\nonumber\\[5pt]&+(-1-2h_2s_2+2\beta_2)k_2 I_{1+\alpha,\{\beta_1, \beta_2-1, \beta_3\}}+(1-2\beta_3+2\tau)k_3 I_{1+\alpha, \{\beta_1, \beta_2, \beta_3-1\}}] = 0
\end{align}
We consider $I_{\alpha, \{\beta_i\}}$ and $I_{1+\alpha, \{\beta_i-\delta_{i, n}\}}$ as independent. To satisfy the above equation we require the coefficients of these triple-$K$ integrals in this equation to be zero
\begin{align}
   & \tau = \Delta -2 \implies \beta_3 = \Delta-\frac{3}{2}\\ 
    &\tau = 1-\Delta \implies \beta_3 =  \frac{3}{2}-\Delta\label{cons1}
\end{align}
When $\beta_1$ and $\beta_2$ are unrelated then we can assume that $I_{\alpha, \{\beta_i-\delta_{in}\}}$'s are independent of each other. This would then imply from the coefficients of $I_{1+\alpha, \{\beta_1-1, \beta_2, \beta_3\}}$ that $s_2 = 0$, which contradicts our initial assumption that $s_1$ and $s_2$ are non-zero. Instead, we require that the coefficient of $s_2$ vanishes i.e.
\begin{align}
    k_1I_{1+\alpha, \{\beta_1-1, \beta_2, \beta_3\}}-k_2I_{1+\alpha, \{\beta_1, \beta_2-1, \beta_3\}} = 0
\end{align}
which is satisfied only when \footnote{Here we have made use of the identity $I_{\alpha, \{\beta_1, \beta_2, \beta_3-1\}}= k^{-2\beta_3}_3I_{\alpha, \{\beta_1, \beta_2, \beta_3\}}$.}
\begin{align}
    \beta_1 = \beta_2 = \frac{1}{2}\label{cons2}
\end{align}
For ${b} = \lambda_1{\sigma}\lambda_3+\lambda_3{\sigma}\lambda_1$ and $b=\lambda_1{\sigma}\lambda_2+\lambda_2{\sigma}\lambda_1$, respectively, we get the following equations
\begin{align}
    &(k_3-k_1)\left(\frac{\partial^2 \widetilde{f}}{\partial k^2_1}-\frac{\partial^2 \widetilde{f}}{\partial k^2_2}\right)-k_3\left(\frac{\partial^2 \widetilde{f}}{\partial k^2_1}-\frac{\partial^2 \widetilde{f}}{\partial k^2_3}\right)+\left(-2s_1h_1 \frac{\partial \widetilde{f}}{\partial k_2}+ 2s_1h_1 s_2\frac{\partial \widetilde{f}}{\partial k_1}\right) \notag\\&\hspace{0.5cm}= \frac{1}{k_3}(\Delta-2)(\Delta-1)\widetilde{f}\label{hgp11}
    \end{align}
    and \begin{align}
    &(k_1-k_2)\left(\frac{\partial^2 \widetilde{f}}{\partial k^2_2}-\frac{\partial^2 \widetilde{f}}{\partial k^2_3}\right)-k_1\left(\frac{\partial^2 \widetilde{f}}{\partial k^2_2}-\frac{\partial^2 \widetilde{f}}{\partial k^2_1}\right)+\left(2(s_1h_1-s_2h_2) \frac{\partial \widetilde{f}}{\partial k_3}-2s_1h_1\frac{\partial \widetilde{f}}{\partial k_1}-2s_2h_2\frac{\partial \widetilde{f}}{\partial k_2}\right) \notag\\&\hspace{0.5cm}= -\frac{k_1-k_2}{k^2_3}(\Delta-2)(\Delta-1)\widetilde{f}\label{hgp12}
\end{align}
If we use the ansatz \eqref{soln} along with \eqref{cons1} in the above equations,  \eqref{hgp11} is solved automatically. On the other hand  \eqref{hgp12} gives
\begin{align}
  &4(s_1h_1-s_2h_2)\left[k_3(\Delta-2) I_{\alpha,\{\frac{1}{2}, \frac{1}{2},\Delta-\frac{3}{2}\}}+k^2_3(k_3 I_{\alpha,\{\frac{1}{2}, \frac{1}{2},\Delta-\frac{5}{2}\}}-I_{\alpha,\{\frac{1}{2}, \frac{1}{2},\Delta-\frac{3}{2}\}})\right] = 0
  \end{align}
  for $(\tau, \beta_3) = (\Delta-2, \Delta-\frac{3}{2})$ and 
  \begin{align}
  &4(s_1h_1-s_2h_2)\left[k_3(\Delta-1) I_{\alpha,\{\frac{1}{2}, \frac{1}{2},\frac{3}{2}-\Delta\}}+k^2_3(-k_3 I_{1+\alpha,\{\frac{1}{2}, \frac{1}{2},\frac{1}{2}-\Delta\}}+I_{1+\alpha,\{\frac{1}{2}, \frac{1}{2},\frac{3}{2}-\Delta\}})\right] = 0 \end{align}
for $(\tau, \beta_3) = (1-\Delta, \frac{3}{2}-\Delta)$
both of which imply that
\begin{align}
    s_1 = s_2~~\text{or}~~\Delta = 1, 2
\end{align}
This shows that no homogeneous solution exists when $s_2\neq s_1$, $\Delta \neq 1, 2$.
When $\Delta = 1, 2$, there is another solution that is given by
$$\widetilde{f} = \left(k_{1}+k_{2}-k_{3}\right)^{s_1+s_2}\left(k_{2}+k_{3}-k_{1}\right)^{-s_1+s_2}\left(k_{1}+k_{3}-k_{2}\right)^{-s_2+s_1}$$
However, $\langle O J_{s_2} J_{s_3}\rangle_h$ has an unphysical pole of the form $k_i - k_j - k_k$ either due to the form factor $\widetilde{f}$ or due to $\langle j k\rangle^{-1}$. Therefore, all correlators with one scalar operator depending on the dimension of the scalar operator either has an unphysical homogeneous solution or has no homogeneous solution.
\subsubsection{$\langle J_{s_1} J_{s_2}J_{s_3}\rangle$}
Let us now consider correlators of the kind $\langle J_{s_1}J_{s_2}J_{s_3}\rangle$. The ansatz for the correlator takes the form 
\begin{align}
&\langle J^{h_1}_{s_1}J^{h_2}_{s_2}J^{h_3}_{s_3}\rangle = f_{h_1, h_2, h_3}(k_1, k_2, k_3)\langle12\rangle^{h_3-h_1-h_2}\langle23\rangle^{h_1-h_2-h_3}\langle31\rangle^{h_2-h_3-h_1}\label{ansatz}
\end{align}
The action of the generator of special conformal transformations on this correlator gives 
\begin{align}
\sum_{a=1}^{3}b\cdot\widetilde{K}_a\frac{\langle J^{h_1}_{s_1}J^{h_2}_{s_2}J^{h_3}_{s_3}\rangle}{k^{s_1-1}_1k^{s_2-1}_2k^{s_3-1}_3} &\sim (b\cdot z^{h_1}_1)\frac{k^{i_1}_1z^{h_1i_2}_1\cdots z_1^{h_1i_{s_1}}}{k^{s_1+1}_1 k_2^{s_2-1}k_3^{s_3-1}}\langle J_{i_1i_2\cdot i_{s_1}} J^{h_2}_{s_2} J^{h_3}_{s_3}\rangle \notag\\&+\hspace{.5cm} \text{cyclic permutation in 1, 2, 3}\label{WISH2}
\end{align}
Using the ansatz (\ref{ansatz}) in the above equation, the LHS takes the following form
\begin{align}
&\langle 12\rangle^{a}\langle 23\rangle^{b}\langle 31\rangle^cb^i[\sum_{i}\widetilde{K}_a^i\widetilde{f} +(a+c)\frac{k^i_1}{k_1}\frac{\partial\widetilde{f}}{\partial k_1}+(b+a)\frac{k^i_2}{k_2}\frac{\partial\widetilde{f}}{\partial k_2}+(c+b)\frac{k^i_3}{k_3}\frac{\partial\widetilde{f}}{\partial k_3}]\notag\\
&-(b\cdot z_{1-})\langle 12\rangle^{a-1}\langle 23\rangle^{b+1}\langle 31\rangle^{c-1}[a(k_2-k_3-k_1)-c(k_3-k_2-k_1)]\frac{\partial\widetilde{f}}{\partial k_1}
\notag\\&-(b\cdot z_{2-})\langle 12\rangle^{a-1}\langle 23\rangle^{b-1}\langle 31\rangle^{c+1}[b(k_3-k_1-k_2)-a(k_1-k_3-k_2)]\frac{\partial\widetilde{f}}{\partial k_2}
\notag\\&-(b\cdot z_{3-})\langle 12\rangle^{a+1}\langle 23\rangle^{b-1}\langle 31\rangle^{c-1}[c(k_1-k_2-k_3)-b(k_2-k_1-k_3)]\frac{\partial\widetilde{f}}{\partial k_3}\label{WIU}
\end{align}
where $\widetilde{K}^i_a\widetilde{f} = k_a^i\frac{\partial^2\widetilde{f}}{\partial k^2_a}$.
For $b= \lambda_1\sigma\lambda_1$, the above expression for the RHS takes the following form 
\begin{align}
&L_1\equiv (k_1-k_2-k_3)\left[\frac{\partial^2\widetilde{f}}{\partial k^2_2}-\frac{\partial^2\widetilde{f}}{\partial k^2_3}+\left(\frac{a+b}{k_2}\frac{\partial\widetilde{f}}{\partial k_2}-\frac{c+b}{k_3}\frac{\partial\widetilde{f}}{\partial k_3}\right)\right]\notag\\&\hspace{.2cm}+\left[b(k_3-k_1-k_2)-a(k_1-k_3-k_2)\right]\frac{1}{k_2}\frac{\partial\widetilde{f}}{\partial k_2}+\left[c(k_1-k_2-k_3)-b(k_2-k_1-k_3)\right]\frac{1}{k_3}\frac{\partial\widetilde{f}}{\partial k_3}
\end{align}
while other choices of $b^i = \lambda_2\sigma^i\lambda_2, \lambda_3\sigma^i\lambda_3$ give cyclically permuted expressions $L_2$ and $L_3$ of $R_1$ in $(a, b, c)$ and $(k_1, k_2, k_3)$.
The solutions to the homogeneous part of the above equations i.e. $R_1 = R_2 = R_3 = 0$ are given by
\begin{align}
\widetilde{f} &= \frac{1}{E^n}\nonumber\\[5pt]
\widetilde{f} &= (k_1+k_2-k_3)^{-a}(k_2+k_3-k_1)^{-b}(k_1+k_3-k_2)^{-c}
\end{align}
This proves the existence of homogeneous solutions to the most general spinning correlator. However, for correlators that violate triangle inequality we see that the above solutions have a pole of the form $k_i+k_j-k_k$ either through $\langle ij\rangle^{-1}$ or through one of the solutions of $\widetilde{f}$, which is unphysical. Thus when triangle inequaltiy is violated there are no physical homogeneous solutions to the correlators. Thus we conclude that
\begin{equation}
    \langle J_{s_1} J_{s_2}J_{s_3}\rangle_{\textbf{h,even}}= \langle J_{s_1} J_{s_2}J_{s_3}\rangle_{\textbf{h,odd}}=0.
\end{equation}

\subsection*{Inside the triangle inequality: $\langle TTT\rangle$}
The spinor-helicity structures of various helicities are given by
\begin{align}
  &  \langle T_{-}T_{-}T_{-}\rangle = \frac{k_1k_2k_3}{E^6} \langle 12\rangle^2\langle 23\rangle^2\langle 31\rangle^2\notag\\
  & \langle T_{-}T_{-}T_{-}\rangle =
  \frac{k_1 k_2 k_3}{(k_1+k_2-k_3)^2(k_1+k_3-k_2)^2(k_3+k_2-k_1)^2}\langle 12\rangle^2\langle 23\rangle^2\langle 31\rangle^2  \notag\\
  &\langle T_{-}T_{-}T_{+}\rangle =  \frac{k_1k_2k_3}{E^2} \frac{\langle 12\rangle^6}{\langle 23\rangle^2\langle 13\rangle^2} = \frac{k_1k_2k_3}{E^6}\frac{\langle 12\rangle^6\langle \bar{2}\bar{3}\rangle^2\langle \bar{1}\bar{3}\rangle^2}{(k_2+k_3-k_1)^2(k_1+k_3-k_2)^2} \notag\\&\langle T_{-}T_{-}T_{+}\rangle =  \frac{k_1 k_2 k_3(k_1+k_3-k_2)^{2}(k_3+k_2-k_1)^{2}}{(k_1+k_2-k_3)^6}\frac{\langle 12\rangle^6}{\langle 23\rangle^2\langle 13\rangle^2}
\end{align}
We see that only the first solution in $---$ helicity is acceptable as it lacks any bad poles. We see that $\langle TTT\rangle$ has a homogeneous solution and it exists only in the $---$ helicity.
\subsection*{Outside the triangle inequality: $\langle J_4J_1J_1\rangle$}
The spinor-helicity structures of various helicities are given by
\begin{align}
  &  \langle J^{-}_{4}J^{-}_{1}J^{-}_{1}\rangle = \frac{k^3_1}{E^6} \frac{\langle 12\rangle^4\langle 23\rangle^4}{\langle 23\rangle^2} =\frac{k^3_1}{E^8}\frac{\langle 12\rangle^4\langle 23\rangle^4\langle\bar{2}\bar{3}\rangle^2}{(k_2+k_3-k_1)^2}\notag\\
  & \langle J^{-}_{4}J^{-}_{1}J^{-}_{1}\rangle =\frac{k^3_1(k_1+k_3-k_2)^{2}}{(k_1+k_2-k_3)^4(k_3+k_2-k_1)^4}\frac{\langle 12\rangle^4\langle 23\rangle^4}{\langle 23\rangle^2}\notag\\
  &\langle J^{-}_{4}J^{-}_{1}J^{+}_{1}\rangle =  \frac{k^3_1}{E^4} \frac{\langle 12\rangle^6}{\langle 23\rangle^4\langle 13\rangle^2} = \frac{k^3_1}{E^{10}}\frac{\langle 12\rangle^6\langle\bar{2}\bar{3}\rangle^4\langle\bar{1}\bar{3}\rangle^2}{(k_2+k_3-k_1)^4(k_1+k_3-k_2)^2} \notag\\
  &\langle J^{-}_{4}J^{-}_{1}J^{+}_{1}\rangle = \frac{k^3_1}{(k_1+k_2-k_3)^6(k_1+k_3-k_2)^{-2}(k_3+k_2-k_1)^{-4}}\frac{\langle 12\rangle^6}{\langle 23\rangle^4\langle 13\rangle^2} \notag\\
  &\langle J^{-}_{4}J^{+}_{1}J^{+}_{1}\rangle =  \frac{k^3_1}{E^2} \frac{\langle 12\rangle^4\langle 31\rangle^4}{\langle 23\rangle^6} = \frac{k^3_1}{E^8} \frac{\langle 12\rangle^4\langle 31\rangle^4\langle\bar{2}\bar{3}\rangle^6}{(k_2+k_3-k_1)^6} \notag\\
  &\langle J^{-}_{4}J^{+}_{1}J^{+}_{1}\rangle =  \frac{k^3_1}{(k_1+k_2-k_3)^4(k_1+k_3-k_2)^{4}(k_3+k_2-k_1)^{-6}}\frac{\langle 12\rangle^4\langle 31\rangle^4}{\langle 23\rangle^6}
  \end{align}
We see that none of the solutions are acceptable as each of them has bad poles. Therefore, $\langle J_4J_1J_1\rangle$ has no homogeneous solution.

\section{A relation between 2-point function coefficients}
\label{2pt1}
In this section we will argue that conformal invariance along with the WT identity implies a relation between various 2-point function coefficients. In spinor-helicity variables conformal invariance of a 3-point correlator is a non-homogeneous differential equation whose R.H.S is given by the sum of Ward identities associated to each of the momentum insertions. For instance, for the case of $\langle JJJ\rangle$, the conformal Ward identity takes the form :
\begin{align}\label{Kkappajjj}
\widetilde{K}^{\kappa} \langle J^- J^- J^- \rangle &= 2\left(z_1^{-\kappa}\frac{k_{1\mu}}{k_1^2}\langle J^{\mu} J^- J^- \rangle+z_2^{-\kappa}\frac{k_{2\mu}}{k_2^2}\langle J^- J^{\mu} J^- \rangle+z_3^{-\kappa}\frac{k_{3\mu}}{k_3^2}\langle J^- J^- J^{\mu} \rangle\right)
\end{align}
For a generic 3-point correlator $\langle J_{s_1}J_{s_2}J_{s_3}\rangle$, this implies that the R.H.S of the above equation is proportional to two-point function coefficients $c_{s_1}, c_{s_2}$ and $c_{s_3}$. 
The equation of special conformal invariance \eqref{Kkappajjj} is a vector equation and one can dot it with appropriate vectors such that it projects the Ward identity with respect to only one of the momenta. With an appropriate choice of such a vector to dot with we can obtain equations which have only one of $c_{s_1}$, $c_{s_2}$ or $c_{s_3}$ on the R.H.S. 

We consider two scalar differential equations that correspond to one having say $c_{s_1}$ on the RHS and the other having $c_{s_2}$ on the RHS. The correlator that is obtained after solving the two differential equations should match. Requiring this consistency condition leads to a relation between the two-point function coefficients  $c_{s_1}$ and  $c_{s_2}$. We see that a combination of conformal invariance and transverse Ward identities leads to a relation between two-point function coefficients.

Let us consider the correlator $\langle J_3J_1O\rangle$. The action of special conformal generator in spinor-helicity variables takes the form 
\begin{align}
\widetilde{K}^{\kappa} \langle J_3 J_1 O \rangle &= 2\left(z_1^{-\kappa}\frac{k_{1\mu}}{k_1^2}\langle J^{\mu}_{\nu\rho} J^- O \rangle+z_2^{-\kappa}\frac{k_{2\mu}}{k_2^2}\langle J^- J^{\mu} O\rangle\right)
\end{align}
One can contract this vector equation with two different vectors that project out separately the two WT identities on the RHS. With suitable projections we get the following equations
\begin{align}\label{kk31ap}
\vec b_1\cdot \widetilde{K}^{\kappa} \langle J_3 J_1 O \rangle&= k_1\cdot \langle J_3J_1O\rangle\cr
&\propto c_O\langle OO\rangle+c_J\langle J_1J_1\rangle\cr
\vec b_2\cdot \widetilde{K}^{\kappa} \langle J_3 J_1 O \rangle&=k_2\cdot \langle J_3J_1O\rangle\cr 
&=0
\end{align}
Let us assume that the solution of of first equation is given by $f(c_O,c_J,k_i)$. And since homogeneous solution for this correlation function does not exist, plugging this solution in the second equation in \eqref{kk31ap} would give a relation between coefficients $c_O$ and $c_J$. This is exactly the statement that higher spin symmetry relates all the two-point function coefficient. This can formally be shown by using higher spin algebra\cite{Maldacena:2011jn,Maldacena:2012sf}.

    \section{Outside the triangle inequality: Exactly conserved current - some details}\label{hotngap}
In this section we give details of the results presented in section \ref{hotng}.
\\

\noindent{\bf{Bosonic $\langle J_3J_1O\rangle$}}\\
The $J_3$ higher spin current is defined as
\begin{align}
    J_3(z, x) = \frac{1}{6}\left(-\bar{\phi}(z\cdot\partial)^3\phi+15(z\cdot\partial)\bar{\phi}(z\cdot\partial)^2\phi-15(z\cdot\partial)^2\bar{\phi}(z\cdot\partial)\phi+ (z\cdot\partial)^3\bar{\phi} \phi\right)
\end{align}
where we have contracted the free indices with the null polarization Tensors. In momentum space taking $z_i.p_i=0,$ we get
\begin{align}
    J_3(z, k) = \int d^3l \bar{\phi}(l)\phi(l-p)(z\cdot l)^3
\end{align}
By directly computing the correlator we obtain
\begin{align}
    \langle J_3(z_1, k_1)J_1(z_2, k_2)O(k_3)\rangle = A (z_1\cdot k_2 )^3(z_2\cdot k_1 )+B (z_1\cdot k_2 )^2(z_1\cdot z_2)
\end{align}
where the form factors $A, B$ are given by
\begin{align}
   A = -\frac{E^2 + 2k_1(E+k_1)}{E^4 k_3}\quad B = \frac{3}{E}+\frac{k_1(3E + 2k_1)}{E^3}
\end{align}
The WT Identity corresponding to the $J_3$  and $J_1$ current of this correlator is given by,
\begin{align}
 k_{1 \mu}\left\langle J_{\nu \sigma}^{\mu}\left(k_{1}\right) J_{\rho}\left(k_{2}\right) O\left(k_{3}\right)\right\rangle &= k_{3 \mu} k_{3 \nu} k_{3 \rho}\left\langle O\left(k_{3}\right) O\left(-k_{3}\right)\right\rangle-k_{3}^{2} g_{\rho(\mu} k_{3 \nu)}\left\langle O\left(k_{3}\right) O\left(-k_{3}\right)\right\rangle \nonumber\\
  k^{2 \rho}\left\langle J_{\nu \sigma}^{\mu}\left(k_{1}\right) J_{\rho}\left(k_{2}\right) O\left(k_{3}\right)\right\rangle &= k_{2\mu} k_2^\nu k_2^\sigma \langle O(-k_3)O(k_3)\rangle 
\end{align}
After contracting with the transverse polarizations and evaluating the RHS we obtain
\begin{align}\label{wt310b}
    \left\langle k_1\cdot J_3\left(z_1, k_{1}\right) J_1\left(z_2, k_{2}\right) O\left(k_{3}\right)\right\rangle &= -\frac{15}{k_{3}}\left(z_{1} \cdot k_{2}\right)^{2}\left(z_{2} \cdot k_{1}\right)+30k_{2}\left(z_{1} \cdot k_{2}\right)\left(z_{1} \cdot z_{2}\right)-30k_{3}\left(z_{1} \cdot k_{2}\right)\left(z_{1} \cdot z_{2}\right)\nonumber\\
   k^{2 \rho}\left\langle J_3\left(z_1,k_{1}\right) J_{\rho}\left(z_2,k_{2}\right) O\left(k_{3}\right)\right\rangle &=-\frac{5}{k_3}(z_1\cdot k_2)^3
\end{align}
where we have used shorthand notation $k_1\cdot J_3= k_{1\mu} J_3^{\mu\alpha\beta}$. We will be using the same notation below quite often.
Converting the results in spinor-helicity variables we get
\begin{align}
    &\langle J^-_3J^-_1O\rangle = f_{-~-}(k_1, k_2, k_3)\langle 12\rangle^4\langle 31\rangle^2\langle \bar{2}\bar{3}\rangle^2\notag\\
    & \langle J^-_3J^+_1O\rangle = f_{-+}(k_1, k_2, k_3)\langle 12\rangle^2\langle 31\rangle^4\langle \bar{2}\bar{3}\rangle^4
\end{align}
Comparing with the explicit result we get,
\begin{align}
    &f_{-~-}(k_1, k_2, k_3) =\frac{-5 E^4+4 E^3 k_2+6 E^2 k_1 k_2+8 E k_1^2 k_2+8 k_1^3 k_2}{256 E^6 k_1^3 k_2 k_3}\notag\\
    &f_{-~+}(k_1, k_2, k_3) = \frac{-5 E^2+4 E k_2+2 k_1 k_2}{256 E^6 k_1^3 k_2 k_3}
\end{align}
The other helicity configurations are just complex conjugates of the ones we have listed.
Also, notice that neither the momentum space expression nor the spinor-helicity expression have any unphysical poles. 

We note, however that the above correlator has a contribution due to $J_1$  WT Identity
\eqref{wt310b}. We make use of the following redefinition of the correlator to remove the $J_1$ contribution 
\begin{equation}
    \langle J_3 J_1 O\rangle_{here}= \langle J_3 J_1 O\rangle_{there}-\frac{5}{2k_3}(z_1\cdot z_2)(z_1\cdot k_2)^2
\end{equation}
The redefined correlator in momentum space is given by \eqref{res1}. See \eqref{WTfb} for answer in spinor helicity variables.
\\

\noindent {\bf{Example: Bosonic $\langle J_4 J_1 J_1\rangle$}}\\
The $J_4$ current is given by
\begin{align}
    J_4(z, x) = \frac{1}{128}\left(\bar{\phi}(z\cdot\partial)^4\phi -28 (z\cdot\partial)\bar{\phi}(z\cdot\partial)^3\phi+70 (z\cdot\partial)^2\bar{\phi}(z\cdot\partial)^2\phi+28 (z\cdot\partial)^3\bar{\phi}(z\cdot\partial)\phi+(z\cdot\partial)^4\bar{\phi} \phi \right)
\end{align}
which in momentum space is given by
\begin{align}
    J_4(z, k) = \int d^3l \bar{\phi}(l)\phi(l-k)(z\cdot l)^4
\end{align}
We now write an ansatz for the correlator $\langle J_4 J_1 J_1\rangle$ after appropriate contractions with transverse polarizations
\begin{align}\label{411anst1}
   &\langle J_4(z_1, k_1)J_1(z_2, k_2)J_1(z_3, k_3)\rangle = A (z_1\cdot k_2 )^4 z_2\cdot z_3+B (z_1\cdot k_2 )^4z_2\cdot k_1  z_3\cdot k_1  \notag\\&+ C (z_1\cdot k_2 )^3(z_1\cdot z_2)(z_3\cdot k_1 )- C(k_2 \leftrightarrow k_3)(z_1\cdot k_2 )^3(z_1\cdot z_3)(z_2\cdot k_1 )+ D (z_1\cdot k_2 )^2(z_1\cdot z_2)(z_1\cdot z_3) 
\end{align}
Using a momentum space degeneracy identity\footnote{We use$\bigg(-4 k_1^2 (z_1\cdot k_2)^2 (z_2\cdot z_3)+4 ((z_1\cdot k_2))^2(z_2\cdot k_1) (z_3\cdot k_1)+2(z_1\cdot k_2) ((k_1^2+k_2^2-k_3^2) (z_2\cdot k_1) (z_1\cdot z_3)$$ $$-(k_1^2-k_2^2+k_3^2) (z_3\cdot k_1) (z_1\cdot z_2))+(k_1^4-2 (k_2^2+k_3^2) k_1^2+(k_2^2-k_3^2)^2) (z_1\cdot z_2) (z_1\cdot z_3)\bigg)=0$}, we can restrict our ansatz to the following.
\begin{align}
\label{res1bans}
   &\langle J_4(z_1, k_1)J_1(z_2, k_2)J_1(z_3, k_3)\rangle = A (z_1\cdot k_2 )^4 z_2\cdot z_3+ C (z_1\cdot k_2 )^3(z_1\cdot z_2)(z_3\cdot k_1 )\notag\\&- C(k_2 \leftrightarrow k_3)(z_1\cdot k_2 )^3(z_1\cdot z_3)(z_2\cdot k_1 )+ D (z_1\cdot k_2 )^2(z_1\cdot z_2)(z_1\cdot z_3) 
\end{align}
where the form factors $A, C, D$ can be obtained by explicit computations in the free bosonic theory. After an appropriate redefinition\footnote{ We have $\langle J_4 J_1 J_1\rangle_{here} = \langle J_4 J_1 J_1\rangle_{there}+\frac{7}{1024}(z_1\cdot z_2)(z_1\cdot z_3)(z_1\cdot k_2 )^2(k_2+k_3)$} of the correlator to make the $J_1$ WT identity $0$ we have,
\small
\begin{align}
   & A =\frac{\left(E-k_1\right) \left(7 E^4+14 k_1 E^3+ 18 k_1^2E^2+16 k_1^3E+8k_1^4\right)}{2048 E^6}\notag\\& C = \frac{7 E^5+7 k_1E^4+2 k_1^2 \left(E-2 k_1\right) \left(2 E^2+3 E k_1+2 k_1^2\right)- 2k_2E^4-4 k_1 k_2 E^3-2k_1^2 k_2 \left(3 E^2+4 k_1 \left(E+k_1\right)\right)}{2048 E^6}\notag\\
   &C(k_2\leftrightarrow k_3) = -\frac{5 E^5+5 k_1E^4+2 k_1^2 E \left(E^2-2 k_1 \left(E+2 k_1\right)\right)+2 k_2E^4+2k_1k_2 \left(2 E^3+k_1 \left(3 E^2+4 k_1 \left(E+k_1\right)\right)\right)}{2048 E^6}\notag\\
   &D =\frac{5 E^6+20k_2 E^5+4k_1^2k_2 \left(E^3+2 E^2 k_1-8 k_1^3\right)-5 k_1E^5-8 k_1^2 E \left(E^3+E^2 k_1-2 k_1^3\right)}{4096 E^5}\notag\\&-\frac{20 k_2^2 E^4+4k_1k_2^2 \left(5 E^3+6 E^2 k_1+8 k_1^2 \left(E+k_1\right)\right)}{4096 E^5}
  \label{res1b}
\end{align}
\normalsize
The WT identity  and answer in spinor helicity variables can be found in \eqref{WIEx2b}, \eqref{411bhlc}.
\\

\noindent{\bf{Fermionic $\langle J_4 J_1 J_1\rangle$}}\\
The $J_4$ current is given by
\begin{align}
    J_4(z, x) = -\bar{\psi}(z\cdot\gamma)(z\cdot\partial)^3\psi+7(z\cdot\partial)\bar{\psi}(z\cdot\gamma)(z\cdot\partial)^2\psi-7(z\cdot\partial)^2\bar{\psi}(z\cdot\gamma)(z\cdot\partial)\psi+(z\cdot\partial)^3\bar{\psi}(z\cdot\gamma)\psi
\end{align}
which in momentum space is given by
\begin{align}
    J_4(z, k) = \int d^3l \bar{\psi}(l)(z\cdot\gamma)\psi(l-k)(z\cdot l)^3
\end{align}
The ansatz for the correlator $\langle J_4 J_1 J_1\rangle$ is exactly the same as \eqref{411anst1} with 
\small
\begin{align}
   & A =\frac{13E^5+13 k_1E^4+12k_1^2 E^3+10k_1^3 E^2+8 k_1^4E+8k_1^5}{2048 E^6}\notag\\&C=\frac{E^4 \left(13 E+2 k_2\right)+E^3 k_1 \left(13 E+4 k_2\right)+6 E^2 k_1^2 \left(2 E+k_2\right)+8 k_1^4 \left(E+k_2\right)+2 E k_1^3 \left(5 E+4 k_2\right)+8 k_1^5}{2048 E^6}\notag\\
   &C(k_2\leftrightarrow k_3) =-\frac{15 E^5+15 k_1 E^4+2 k_1^2 E \left(7 E^2+6 E k_1+4 k_1^2\right)-2 k_2 E^4-4 k_1 k_2E^3-2k_1^2k_2 \left(3 E^2+4 k_1 \left(E+k_1\right)\right)}{2048 E^6}\notag\\
   &D = \frac{15 E^6-15 k_1 E^5-16 k_1^2 E \left(E+k_1\right) \left(E^2+k_1^2\right)-20 k_2E^5-4k_1^2k_2 \left(E^3+2 E^2 k_1-8 k_1^3\right)}{4096 E^5}\notag\\+&\frac{20 k_2^2E^4+4k_1k_2^2 \left(5 E^3+6 E^2 k_1+8 k_1^2 \left(E+k_1\right)\right)}{4096 E^5}\label{res1f}
\end{align}
\normalsize
where the form factors $A, C, D$ are obtained by explicit computations in free fermionic theory. We also have done a redefinition of the correlator\footnote{We have $\langle J_4 J_1 J_1\rangle_{here} = \langle J_4 J_1 J_1\rangle_{there}-\frac{7}{4096}(z_1\cdot z_2)(z_1\cdot z_3)(z_1\cdot k_2 )^2(k_2+k_3)$}, to make it transverse with respect to $J_1$.
The WT identity is given by 
\begin{align}\label{WIEx2f}
    \left\langle k_1.J_4\left(z_1, k_{1}\right) J_1\left(z_2, k_{2}\right) J_1\left(z_3, k_{3}\right)\right\rangle &= \frac{7}{512}z_1\cdot k_2 [13 (z_1\cdot k_2 )^2 z_2\cdot z_3 (k_2-k_3)\notag\\&+2z_1\cdot z_2 z_1\cdot z_3 (k_2-k_3)(-2k^2_1+2k^2_2+k_2k_3+2k^2_3)\notag\\&+z_1\cdot k_2 (z_1\cdot z_3 z_3\cdot k_1 (8k_2-13k_3)+z_1\cdot z_3 z_2\cdot k_1 (-13k_2+8k_3))] \nonumber\\
    k_{2 \rho}\left\langle J_{4}\left(z_1, k_{1}\right) J_1^{\rho}\left( k_{2}\right) J_1\left(z_3, k_{3}\right)\right\rangle &=0.
\end{align}
Again notice any lack of bad poles in both momentum Space and in spinor-helicity variables, see \eqref{411fhlc}.

\section{Conformal bounds: some details}\label{cfba}
Let us define the correlation functions as follows
\begin{align}
\langle JJT\rangle&=n_b\langle JJT\rangle_{\text{FB}}+n_f\langle JJT\rangle_{\text{FF}}+n_{\text{odd}}\langle JJT\rangle_{\text{odd}}\cr
\langle TTT\rangle&=n_b\langle TTT\rangle_{\text{FB}}+n_f\langle TTT\rangle_{\text{FF}}+n_{\text{odd}}\langle TTT\rangle_{\text{odd}}
\end{align}
In \cite{Chowdhury:2017vel}, \cite{Chowdhury:2018uyv} for a generic CFT it was shown that 
\begin{align}
\frac{(n_b-n_f)^2}{c_J^2}+\frac{n_{\text{odd}}^2}{c_J^2}&\le 1\cr
\frac{(n_b-n_f)^2}{c_T^2}+\frac{n_{\text{odd}}^2}{c_T^2}&\le 1
\end{align}
For holographic CFT it was shown in \cite{Afkhami-Jeddi:2018own} that
\begin{align}
\frac{n_b-n_f}{c_J}\le\frac{\ln\Delta}{\Delta^2},\quad\frac{n_{\text{odd}}}{c_J}\le\frac{\ln\Delta}{\Delta^2}\cr
\frac{n_b-n_f}{c_T}\le\frac{\ln\Delta}{\Delta^4},\quad\frac{n_{\text{odd}}}{c_T}\le\frac{\ln\Delta}{\Delta^4}
\end{align}
where $c_J$ and $c_T$ are the two-point functions of the spin-one and spin-two currents.
These inequalities imply the following 
\begin{align}
\frac{(n_b-n_f)^2}{c_J^2}+\frac{n_{\text{odd}}^2}{c_J^2}&\le 2\left(\frac{\ln\Delta}{\Delta^2}\right)^2\cr
\frac{(n_b-n_f)^2}{c_T^2}+\frac{n_{\text{odd}}^2}{c_T^2}&\le 2\left(\frac{\ln\Delta}{\Delta^4}\right)^2
\end{align}
In our language this is simply the statement
\begin{align}
\gamma_J\le\frac{\ln\Delta}{\Delta^2},~~~~~~~
\gamma_T\le\frac{\ln\Delta}{\Delta^4}
\end{align}
A similar analysis was performed in \cite{Chowdhury:2017vel}, \cite{Chowdhury:2018uyv}. One can translate between the notations in that paper and our paper using \footnote{$$ 2\pi^4p_J=n_{\text{odd}},\quad n_f^J+n_s^J=c_J,\quad \frac{2\pi^4p_T}{3}=n_{\text{odd}},\quad n_f^T+n_s^T=c_T$$ where \cite{Chowdhury:2017vel,Chowdhury:2018uyv} followed the notation
	$$\langle JJT\rangle=n_s^J\langle JJT\rangle_{\text{FB}}+n_f^J\langle JJT\rangle_{\text{FF}}+p_J\langle JJT\rangle_{\text{odd}},\quad
	\langle TTT\rangle=n_s^T\langle TTT\rangle_{\text{FB}}+n_f^T\langle TTT\rangle_{\text{FF}}+p_T\langle TTT\rangle_{\text{odd}}$$
}.

\providecommand{\href}[2]{#2}\begingroup\raggedright
\bibliography{refs}

\providecommand{\href}[2]{#2}\begingroup\raggedright\begin{thebibliography}{10}

\bibitem{Giombi:2011rz}
S.~Giombi, S.~Prakash and X.~Yin, \emph{{A Note on CFT Correlators in Three
  Dimensions}}, \href{https://doi.org/10.1007/JHEP07(2013)105}{\emph{JHEP}
  {\bfseries 07} (2013) 105} [\href{https://arxiv.org/abs/1104.4317}{{\ttfamily
  1104.4317}}].

\bibitem{Costa:2011mg}
M.~S. Costa, J.~Penedones, D.~Poland and S.~Rychkov, \emph{{Spinning Conformal
  Correlators}}, \href{https://doi.org/10.1007/JHEP11(2011)071}{\emph{JHEP}
  {\bfseries 11} (2011) 071} [\href{https://arxiv.org/abs/1107.3554}{{\ttfamily
  1107.3554}}].

\bibitem{Costa:2011dw}
M.~S. Costa, J.~Penedones, D.~Poland and S.~Rychkov, \emph{{Spinning Conformal
  Blocks}}, \href{https://doi.org/10.1007/JHEP11(2011)154}{\emph{JHEP}
  {\bfseries 11} (2011) 154} [\href{https://arxiv.org/abs/1109.6321}{{\ttfamily
  1109.6321}}].

\bibitem{Maldacena:2011jn}
J.~Maldacena and A.~Zhiboedov, \emph{{Constraining Conformal Field Theories
  with A Higher Spin Symmetry}},
  \href{https://doi.org/10.1088/1751-8113/46/21/214011}{\emph{J. Phys. A}
  {\bfseries 46} (2013) 214011}
  [\href{https://arxiv.org/abs/1112.1016}{{\ttfamily 1112.1016}}].

\bibitem{Maldacena:2012sf}
J.~Maldacena and A.~Zhiboedov, \emph{{Constraining conformal field theories
  with a slightly broken higher spin symmetry}},
  \href{https://doi.org/10.1088/0264-9381/30/10/104003}{\emph{Class. Quant.
  Grav.} {\bfseries 30} (2013) 104003}
  [\href{https://arxiv.org/abs/1204.3882}{{\ttfamily 1204.3882}}].

\bibitem{Giombi:2016zwa}
S.~Giombi, V.~Gurucharan, V.~Kirilin, S.~Prakash and E.~Skvortsov, \emph{{On
  the Higher-Spin Spectrum in Large N Chern-Simons Vector Models}},
  \href{https://doi.org/10.1007/JHEP01(2017)058}{\emph{JHEP} {\bfseries 01}
  (2017) 058} [\href{https://arxiv.org/abs/1610.08472}{{\ttfamily
  1610.08472}}].

\bibitem{Giombi:2011kc}
S.~Giombi, S.~Minwalla, S.~Prakash, S.~P. Trivedi, S.~R. Wadia and X.~Yin,
  \emph{{Chern-Simons Theory with Vector Fermion Matter}},
  \href{https://doi.org/10.1140/epjc/s10052-012-2112-0}{\emph{Eur. Phys. J. C}
  {\bfseries 72} (2012) 2112}
  [\href{https://arxiv.org/abs/1110.4386}{{\ttfamily 1110.4386}}].

\bibitem{Aharony:2011jz}
O.~Aharony, G.~Gur-Ari and R.~Yacoby, \emph{{d=3 Bosonic Vector Models Coupled
  to Chern-Simons Gauge Theories}},
  \href{https://doi.org/10.1007/JHEP03(2012)037}{\emph{JHEP} {\bfseries 03}
  (2012) 037} [\href{https://arxiv.org/abs/1110.4382}{{\ttfamily 1110.4382}}].

\bibitem{Gerasimenko:2021sxj}
P.~Gerasimenko, A.~Sharapov and E.~Skvortsov, \emph{{Slightly Broken Higher
  Spin Symmetry: General Structure of Correlators}},
  \href{https://arxiv.org/abs/2108.05441}{{\ttfamily 2108.05441}}.

\bibitem{Skvortsov:2018uru}
E.~Skvortsov, \emph{{Light-Front Bootstrap for Chern-Simons Matter Theories}},
  \href{https://doi.org/10.1007/JHEP06(2019)058}{\emph{JHEP} {\bfseries 06}
  (2019) 058} [\href{https://arxiv.org/abs/1811.12333}{{\ttfamily
  1811.12333}}].

\bibitem{Aharony:2012nh}
O.~Aharony, G.~Gur-Ari and R.~Yacoby, \emph{{Correlation Functions of Large N
  Chern-Simons-Matter Theories and Bosonization in Three Dimensions}},
  \href{https://doi.org/10.1007/JHEP12(2012)028}{\emph{JHEP} {\bfseries 12}
  (2012) 028} [\href{https://arxiv.org/abs/1207.4593}{{\ttfamily 1207.4593}}].

\bibitem{GurAri:2012is}
G.~Gur-Ari and R.~Yacoby, \emph{{Correlators of Large N Fermionic Chern-Simons
  Vector Models}}, \href{https://doi.org/10.1007/JHEP02(2013)150}{\emph{JHEP}
  {\bfseries 02} (2013) 150} [\href{https://arxiv.org/abs/1211.1866}{{\ttfamily
  1211.1866}}].

\bibitem{Bedhotiya:2015uga}
A.~Bedhotiya and S.~Prakash, \emph{{A test of bosonization at the level of
  four-point functions in Chern-Simons vector models}},
  \href{https://doi.org/10.1007/JHEP12(2015)032}{\emph{JHEP} {\bfseries 12}
  (2015) 032} [\href{https://arxiv.org/abs/1506.05412}{{\ttfamily
  1506.05412}}].

\bibitem{Jain:2020puw}
S.~Jain, R.~R. John and V.~Malvimat, \emph{{Constraining momentum space
  correlators using slightly broken higher spin symmetry}},
  \href{https://doi.org/10.1007/JHEP04(2021)231}{\emph{JHEP} {\bfseries 04}
  (2021) 231} [\href{https://arxiv.org/abs/2008.08610}{{\ttfamily
  2008.08610}}].

\bibitem{Aharony:2019mbc}
O.~Aharony and A.~Sharon, \emph{{Large N renormalization group flows in 3d $
  \mathcal{N} $ = 1 Chern-Simons-Matter theories}},
  \href{https://doi.org/10.1007/JHEP07(2019)160}{\emph{JHEP} {\bfseries 07}
  (2019) 160} [\href{https://arxiv.org/abs/1905.07146}{{\ttfamily
  1905.07146}}].

\bibitem{Inbasekar:2019wdw}
K.~Inbasekar, S.~Jain, V.~Malvimat, A.~Mehta, P.~Nayak and T.~Sharma,
  \emph{{Correlation functions in ${\cal N}=2$ Supersymmetric vector matter
  Chern-Simons theory}},
  \href{https://doi.org/10.1007/JHEP04(2020)207}{\emph{JHEP} {\bfseries 04}
  (2020) 207} [\href{https://arxiv.org/abs/1907.11722}{{\ttfamily
  1907.11722}}].

\bibitem{Jain:2020rmw}
S.~Jain, R.~R. John and V.~Malvimat, \emph{{Momentum space spinning correlators
  and higher spin equations in three dimensions}},
  \href{https://doi.org/10.1007/JHEP11(2020)049}{\emph{JHEP} {\bfseries 11}
  (2020) 049} [\href{https://arxiv.org/abs/2005.07212}{{\ttfamily
  2005.07212}}].

\bibitem{Aharony:2018pjn}
O.~Aharony, S.~Jain and S.~Minwalla, \emph{{Flows, Fixed Points and Duality in
  Chern-Simons-matter theories}},
  \href{https://doi.org/10.1007/JHEP12(2018)058}{\emph{JHEP} {\bfseries 12}
  (2018) 058} [\href{https://arxiv.org/abs/1808.03317}{{\ttfamily
  1808.03317}}].

\bibitem{Li:2019twz}
Z.~Li, \emph{{Bootstrapping conformal four-point correlators with slightly
  broken higher spin symmetry and $3D$ bosonization}},
  \href{https://doi.org/10.1007/JHEP10(2020)007}{\emph{JHEP} {\bfseries 10}
  (2020) 007} [\href{https://arxiv.org/abs/1906.05834}{{\ttfamily
  1906.05834}}].

\bibitem{Silva:2021ece}
J.~A. Silva, \emph{{Four point functions in CFT\textquoteright{}s with slightly
  broken higher spin symmetry}},
  \href{https://doi.org/10.1007/JHEP05(2021)097}{\emph{JHEP} {\bfseries 05}
  (2021) 097} [\href{https://arxiv.org/abs/2103.00275}{{\ttfamily
  2103.00275}}].

\bibitem{Kalloor:2019xjb}
R.~R. Kalloor, \emph{{Four-point functions in large $N$ Chern-Simons fermionic
  theories}}, \href{https://doi.org/10.1007/JHEP10(2020)028}{\emph{JHEP}
  {\bfseries 10} (2020) 028}
  [\href{https://arxiv.org/abs/1910.14617}{{\ttfamily 1910.14617}}].

\bibitem{Vasiliev:1992av}
M.~A. Vasiliev, \emph{{More on equations of motion for interacting massless
  fields of all spins in (3+1)-dimensions}},
  \href{https://doi.org/10.1016/0370-2693(92)91457-K}{\emph{Phys. Lett. B}
  {\bfseries 285} (1992) 225}.

\bibitem{Vasiliev:1995dn}
M.~A. Vasiliev, \emph{{Higher spin gauge theories in four-dimensions,
  three-dimensions, and two-dimensions}},
  \href{https://doi.org/10.1142/S0218271896000473}{\emph{Int. J. Mod. Phys. D}
  {\bfseries 5} (1996) 763}
  [\href{https://arxiv.org/abs/hep-th/9611024}{{\ttfamily hep-th/9611024}}].

\bibitem{Vasiliev:1999ba}
M.~A. Vasiliev, \emph{{Higher spin gauge theories: Star product and AdS
  space}},  \href{https://arxiv.org/abs/hep-th/9910096}{{\ttfamily
  hep-th/9910096}}.

\bibitem{Vasiliev:2003ev}
M.~A. Vasiliev, \emph{{Nonlinear equations for symmetric massless higher spin
  fields in (A)dS(d)}},
  \href{https://doi.org/10.1016/S0370-2693(03)00872-4}{\emph{Phys. Lett. B}
  {\bfseries 567} (2003) 139}
  [\href{https://arxiv.org/abs/hep-th/0304049}{{\ttfamily hep-th/0304049}}].

\bibitem{Giombi:2010vg}
S.~Giombi and X.~Yin, \emph{{Higher Spins in AdS and Twistorial Holography}},
  \href{https://doi.org/10.1007/JHEP04(2011)086}{\emph{JHEP} {\bfseries 04}
  (2011) 086} [\href{https://arxiv.org/abs/1004.3736}{{\ttfamily 1004.3736}}].

\bibitem{Giombi:2016ejx}
S.~Giombi, \emph{{Higher Spin \textemdash{} CFT Duality}},  in
  \emph{{Theoretical Advanced Study Institute in Elementary Particle Physics}:
  {New Frontiers in Fields and Strings}}, pp.~137--214, 2017,
  \href{https://arxiv.org/abs/1607.02967}{{\ttfamily 1607.02967}},
  \href{https://doi.org/10.1142/9789813149441_0003}{DOI}.

\bibitem{Bzowski:2013sza}
A.~Bzowski, P.~McFadden and K.~Skenderis, \emph{{Implications of conformal
  invariance in momentum space}},
  \href{https://doi.org/10.1007/JHEP03(2014)111}{\emph{JHEP} {\bfseries 03}
  (2014) 111} [\href{https://arxiv.org/abs/1304.7760}{{\ttfamily 1304.7760}}].

\bibitem{Coriano:2013jba}
C.~Coriano, L.~Delle~Rose, E.~Mottola and M.~Serino, \emph{{Solving the
  Conformal Constraints for Scalar Operators in Momentum Space and the
  Evaluation of Feynman's Master Integrals}},
  \href{https://doi.org/10.1007/JHEP07(2013)011}{\emph{JHEP} {\bfseries 07}
  (2013) 011} [\href{https://arxiv.org/abs/1304.6944}{{\ttfamily 1304.6944}}].

\bibitem{sissathesis}
L.~de~Souza., \emph{{CFT’s, contact terms and anomalies}},
  \href{https://arxiv.org/abs/PhD Thesis}{{\ttfamily PhD Thesis}}.

\bibitem{Bzowski:2015pba}
A.~Bzowski, P.~McFadden and K.~Skenderis, \emph{{Scalar 3-point functions in
  CFT: renormalisation, beta functions and anomalies}},
  \href{https://doi.org/10.1007/JHEP03(2016)066}{\emph{JHEP} {\bfseries 03}
  (2016) 066} [\href{https://arxiv.org/abs/1510.08442}{{\ttfamily
  1510.08442}}].

\bibitem{Bzowski:2017poo}
A.~Bzowski, P.~McFadden and K.~Skenderis, \emph{{Renormalised 3-point functions
  of stress tensors and conserved currents in CFT}},
  \href{https://doi.org/10.1007/JHEP11(2018)153}{\emph{JHEP} {\bfseries 11}
  (2018) 153} [\href{https://arxiv.org/abs/1711.09105}{{\ttfamily
  1711.09105}}].

\bibitem{Bzowski:2018fql}
A.~Bzowski, P.~McFadden and K.~Skenderis, \emph{{Renormalised CFT 3-point
  functions of scalars, currents and stress tensors}},
  \href{https://doi.org/10.1007/JHEP11(2018)159}{\emph{JHEP} {\bfseries 11}
  (2018) 159} [\href{https://arxiv.org/abs/1805.12100}{{\ttfamily
  1805.12100}}].

\bibitem{Isono:2019ihz}
H.~Isono, T.~Noumi and T.~Takeuchi, \emph{{Momentum space conformal three-point
  functions of conserved currents and a general spinning operator}},
  \href{https://doi.org/10.1007/JHEP05(2019)057}{\emph{JHEP} {\bfseries 05}
  (2019) 057} [\href{https://arxiv.org/abs/1903.01110}{{\ttfamily
  1903.01110}}].

\bibitem{Gillioz:2019lgs}
M.~Gillioz, \emph{{Conformal 3-point functions and the Lorentzian OPE in
  momentum space}},
  \href{https://doi.org/10.1007/s00220-020-03836-8}{\emph{Commun. Math. Phys.}
  {\bfseries 379} (2020) 227}
  [\href{https://arxiv.org/abs/1909.00878}{{\ttfamily 1909.00878}}].

\bibitem{Bzowski:2019kwd}
A.~Bzowski, P.~McFadden and K.~Skenderis, \emph{{Conformal $n$-point functions
  in momentum space}},
  \href{https://doi.org/10.1103/PhysRevLett.124.131602}{\emph{Phys. Rev. Lett.}
  {\bfseries 124} (2020) 131602}
  [\href{https://arxiv.org/abs/1910.10162}{{\ttfamily 1910.10162}}].

\bibitem{Bzowski:2020kfw}
A.~Bzowski, P.~McFadden and K.~Skenderis, \emph{{Conformal correlators as
  simplex integrals in momentum space}},
  \href{https://doi.org/10.1007/JHEP01(2021)192}{\emph{JHEP} {\bfseries 01}
  (2021) 192} [\href{https://arxiv.org/abs/2008.07543}{{\ttfamily
  2008.07543}}].

\bibitem{Jain:2021wyn}
S.~Jain, R.~R. John, A.~Mehta, A.~A. Nizami and A.~Suresh, \emph{{Momentum
  space parity-odd CFT 3-point functions}},
  \href{https://arxiv.org/abs/2101.11635}{{\ttfamily 2101.11635}}.

\bibitem{Jain:2021qcl}
S.~Jain, R.~R. John, A.~Mehta, A.~A. Nizami and A.~Suresh, \emph{{Double copy
  structure of parity-violating CFT correlators}},
  \href{https://doi.org/10.1007/JHEP07(2021)033}{\emph{JHEP} {\bfseries 07}
  (2021) 033} [\href{https://arxiv.org/abs/2104.12803}{{\ttfamily
  2104.12803}}].

\bibitem{Coriano:2020ees}
C.~Corian\`o and M.~M. Maglio, \emph{{Conformal Field Theory in Momentum Space
  and Anomaly Actions in Gravity: The Analysis of 3- and 4-Point Functions}},
  \href{https://arxiv.org/abs/2005.06873}{{\ttfamily 2005.06873}}.

\bibitem{Mata:2012bx}
I.~Mata, S.~Raju and S.~Trivedi, \emph{{CMB from CFT}},
  \href{https://doi.org/10.1007/JHEP07(2013)015}{\emph{JHEP} {\bfseries 07}
  (2013) 015} [\href{https://arxiv.org/abs/1211.5482}{{\ttfamily 1211.5482}}].

\bibitem{Maldacena:2011nz}
J.~M. Maldacena and G.~L. Pimentel, \emph{{On graviton non-Gaussianities during
  inflation}}, \href{https://doi.org/10.1007/JHEP09(2011)045}{\emph{JHEP}
  {\bfseries 09} (2011) 045} [\href{https://arxiv.org/abs/1104.2846}{{\ttfamily
  1104.2846}}].

\bibitem{Gillioz:2021sce}
M.~Gillioz, \emph{{From Schwinger to Wightman: all conformal 3-point functions
  in momentum space}},  \href{https://arxiv.org/abs/2109.15140}{{\ttfamily
  2109.15140}}.

\bibitem{Farrow:2018yni}
J.~A. Farrow, A.~E. Lipstein and P.~McFadden, \emph{{Double copy structure of
  CFT correlators}}, \href{https://doi.org/10.1007/JHEP02(2019)130}{\emph{JHEP}
  {\bfseries 02} (2019) 130}
  [\href{https://arxiv.org/abs/1812.11129}{{\ttfamily 1812.11129}}].

\bibitem{Lipstein:2019mpu}
A.~E. Lipstein and P.~McFadden, \emph{{Double copy structure and the flat space
  limit of conformal correlators in even dimensions}},
  \href{https://doi.org/10.1103/PhysRevD.101.125006}{\emph{Phys. Rev. D}
  {\bfseries 101} (2020) 125006}
  [\href{https://arxiv.org/abs/1912.10046}{{\ttfamily 1912.10046}}].

\bibitem{Caron-Huot:2021kjy}
S.~Caron-Huot and Y.-Z. Li, \emph{{Helicity basis for three-dimensional
  conformal field theory}},  \href{https://arxiv.org/abs/2102.08160}{{\ttfamily
  2102.08160}}.

\bibitem{Jain:2021gwa}
S.~Jain and R.~R. John, \emph{{Relation between parity-even and parity-odd CFT
  correlation functions in three dimensions}},
  \href{https://arxiv.org/abs/2107.00695}{{\ttfamily 2107.00695}}.

\bibitem{Gandhi:2021gwn}
Y.~Gandhi, S.~Jain and R.~R. John, \emph{{Anyonic correlation functions in
  Chern-Simons matter theories}},
  \href{https://arxiv.org/abs/2106.09043}{{\ttfamily 2106.09043}}.

\bibitem{Arkani-Hamed:2018kmz}
N.~Arkani-Hamed, D.~Baumann, H.~Lee and G.~L. Pimentel, \emph{{The Cosmological
  Bootstrap: Inflationary Correlators from Symmetries and Singularities}},
  \href{https://doi.org/10.1007/JHEP04(2020)105}{\emph{JHEP} {\bfseries 04}
  (2020) 105} [\href{https://arxiv.org/abs/1811.00024}{{\ttfamily
  1811.00024}}].

\bibitem{Baumann:2019oyu}
D.~Baumann, C.~Duaso~Pueyo, A.~Joyce, H.~Lee and G.~L. Pimentel, \emph{{The
  cosmological bootstrap: weight-shifting operators and scalar seeds}},
  \href{https://doi.org/10.1007/JHEP12(2020)204}{\emph{JHEP} {\bfseries 12}
  (2020) 204} [\href{https://arxiv.org/abs/1910.14051}{{\ttfamily
  1910.14051}}].

\bibitem{Baumann:2020dch}
D.~Baumann, C.~Duaso~Pueyo, A.~Joyce, H.~Lee and G.~L. Pimentel, \emph{{The
  Cosmological Bootstrap: Spinning Correlators from Symmetries and
  Factorization}},
  \href{https://doi.org/10.21468/SciPostPhys.11.3.071}{\emph{SciPost Phys.}
  {\bfseries 11} (2021) 071}
  [\href{https://arxiv.org/abs/2005.04234}{{\ttfamily 2005.04234}}].

\bibitem{Kundu:2015xta}
N.~Kundu, A.~Shukla and S.~P. Trivedi, \emph{{Ward Identities for Scale and
  Special Conformal Transformations in Inflation}},
  \href{https://doi.org/10.1007/JHEP01(2016)046}{\emph{JHEP} {\bfseries 01}
  (2016) 046} [\href{https://arxiv.org/abs/1507.06017}{{\ttfamily
  1507.06017}}].

\bibitem{Kundu:2014gxa}
N.~Kundu, A.~Shukla and S.~P. Trivedi, \emph{{Constraints from Conformal
  Symmetry on the Three Point Scalar Correlator in Inflation}},
  \href{https://doi.org/10.1007/JHEP04(2015)061}{\emph{JHEP} {\bfseries 04}
  (2015) 061} [\href{https://arxiv.org/abs/1410.2606}{{\ttfamily 1410.2606}}].

\bibitem{Ghosh:2014kba}
A.~Ghosh, N.~Kundu, S.~Raju and S.~P. Trivedi, \emph{{Conformal Invariance and
  the Four Point Scalar Correlator in Slow-Roll Inflation}},
  \href{https://doi.org/10.1007/JHEP07(2014)011}{\emph{JHEP} {\bfseries 07}
  (2014) 011} [\href{https://arxiv.org/abs/1401.1426}{{\ttfamily 1401.1426}}].

\bibitem{Pajer:2020wxk}
E.~Pajer, \emph{{Building a Boostless Bootstrap for the Bispectrum}},
  \href{https://doi.org/10.1088/1475-7516/2021/01/023}{\emph{JCAP} {\bfseries
  01} (2021) 023} [\href{https://arxiv.org/abs/2010.12818}{{\ttfamily
  2010.12818}}].

\bibitem{Jazayeri:2021fvk}
S.~Jazayeri, E.~Pajer and D.~Stefanyszyn, \emph{{From locality and unitarity to
  cosmological correlators}},
  \href{https://doi.org/10.1007/JHEP10(2021)065}{\emph{JHEP} {\bfseries 10}
  (2021) 065} [\href{https://arxiv.org/abs/2103.08649}{{\ttfamily
  2103.08649}}].

\bibitem{Bonifacio:2021azc}
J.~Bonifacio, E.~Pajer and D.-G. Wang, \emph{{From Amplitudes to Contact
  Cosmological Correlators}},
  \href{https://arxiv.org/abs/2106.15468}{{\ttfamily 2106.15468}}.

\bibitem{Jain:2021vrv}
S.~Jain, R.~R. John, A.~Mehta, A.~A. Nizami and A.~Suresh, \emph{{Higher spin
  3-point functions in 3d CFT using spinor-helicity variables}},
  \href{https://doi.org/10.1007/JHEP09(2021)041}{\emph{JHEP} {\bfseries 09}
  (2021) 041} [\href{https://arxiv.org/abs/2106.00016}{{\ttfamily
  2106.00016}}].

\bibitem{Baumann:2021fxj}
D.~Baumann, W.-M. Chen, C.~Duaso~Pueyo, A.~Joyce, H.~Lee and G.~L. Pimentel,
  \emph{{Linking the Singularities of Cosmological Correlators}},
  \href{https://arxiv.org/abs/2106.05294}{{\ttfamily 2106.05294}}.

\bibitem{Conde:2016izb}
E.~Conde, E.~Joung and K.~Mkrtchyan, \emph{{Spinor-Helicity Three-Point
  Amplitudes from Local Cubic Interactions}},
  \href{https://doi.org/10.1007/JHEP08(2016)040}{\emph{JHEP} {\bfseries 08}
  (2016) 040} [\href{https://arxiv.org/abs/1605.07402}{{\ttfamily
  1605.07402}}].

\bibitem{Nizami:2013tpa}
A.~A. Nizami, T.~Sharma and V.~Umesh, \emph{{Superspace formulation and
  correlation functions of 3d superconformal field theories}},
  \href{https://doi.org/10.1007/JHEP07(2014)022}{\emph{JHEP} {\bfseries 07}
  (2014) 022} [\href{https://arxiv.org/abs/1308.4778}{{\ttfamily 1308.4778}}].

\bibitem{Chowdhury:2017vel}
S.~D. Chowdhury, J.~R. David and S.~Prakash, \emph{{Constraints on parity
  violating conformal field theories in $d=3$}},
  \href{https://doi.org/10.1007/JHEP11(2017)171}{\emph{JHEP} {\bfseries 11}
  (2017) 171} [\href{https://arxiv.org/abs/1707.03007}{{\ttfamily
  1707.03007}}].

\bibitem{Chowdhury:2018uyv}
S.~D. Chowdhury, J.~R. David and S.~Prakash, \emph{{Bootstrap and collider
  physics of parity violating conformal field theories in d = 3}},
  \href{https://doi.org/10.1007/JHEP04(2019)023}{\emph{JHEP} {\bfseries 04}
  (2019) 023} [\href{https://arxiv.org/abs/1812.07774}{{\ttfamily
  1812.07774}}].

\bibitem{Afkhami-Jeddi:2018own}
N.~Afkhami-Jeddi, S.~Kundu and A.~Tajdini, \emph{{A Conformal Collider for
  Holographic CFTs}},
  \href{https://doi.org/10.1007/JHEP10(2018)156}{\emph{JHEP} {\bfseries 10}
  (2018) 156} [\href{https://arxiv.org/abs/1805.07393}{{\ttfamily
  1805.07393}}].

\end{thebibliography}\endgroup
\bibliographystyle{JHEP}
\endgroup

\end{document}